\documentclass[fleqn,usenatbib]{mnras}

\usepackage{amsmath}
\usepackage{amssymb}
\usepackage{amsfonts}
\usepackage{mathtools}
\usepackage{etoolbox}

\usepackage{graphicx}
\usepackage{booktabs}
\usepackage{multirow}
\usepackage{overpic}
\usepackage{wrapfig}
\usepackage{float}
\usepackage{natbib}
\usepackage{needspace}

\usepackage{nicefrac}
\usepackage{siunitx}
\usepackage{cancel}

\usepackage{tikz}
\usepackage{xcolor}

\usepackage[T1]{fontenc}
\usepackage{aecompl}
\usepackage{calligra}

\usepackage{comment}

\usepackage{lineno}

\DeclareMathAlphabet{\mathcalligra}{T1}{calligra}{m}{n}
\DeclareFontShape{T1}{calligra}{m}{n}
    {<->s*[2.2]callig15}{}

\usepackage{hyperref}

\hypersetup{
    pdfnewwindow=true,
    colorlinks=true,
    linkcolor=blue,
    citecolor=teal,
    filecolor=blue,
    urlcolor=teal
}

\newcommand{\eg}{e.g.\ }
\newcommand{\ie}{i.e.\ }
\newcommand{\cf}{cf.\ }

\title[SMBHBs in Retrograde CBDs ]
{Electromagnetic Emission and Orbital Evolution of Eccentric Supermassive Black Hole Binaries in Retrograde Disks}

\author[D. O'Neill et al.]{
David O'Neill$^{1}$\thanks{E-mail: david.o'neill@ist.ac.at},
Mark J. Avara$^{1}$,
Christopher Tiede$^{2}$,
Daniel J. D'Orazio$^{2,3,4}$,
\newauthor
\hspace{0.5mm} Zolt\'an Haiman$^{1,5,6}$
and Andrew MacFadyen$^{7}$
\\
$^{1}$Institute of Science and Technology Austria (ISTA),
Am Campus 1, Klosterneuburg 3400, Austria\\
$^{2}$Niels Bohr International Academy, Niels Bohr Institute,
Blegdamsvej 17, 2100 Copenhagen, Denmark\\
$^{3}$Space Telescope Science Institute,
3700 San Martin Drive, Baltimore, MD 21218, USA\\
$^{4}$Department of Physics and Astronomy,
Johns Hopkins University, 3400 North Charles Street,
Baltimore, MD 21218, USA\\
$^{5}$Department of Astronomy, Columbia University,
550 W. 120th Street, New York, NY 10027, USA\\
$^{6}$Department of Physics, Columbia University,
550 W. 120th Street, New York, NY 10027, USA\\
$^{7}$Center for Cosmology and Particle Physics,
Physics Department, New York University, New York, NY 10003, USA
}

\begin{document}

\maketitle

\begin{abstract}
Circumbinary disks around supermassive black hole binaries (SMBHBs) are expected to form across a broad range of inclinations, with retrograde configurations potentially a common occurrence. Here we present the first grid-based hydrodynamical simulations of retrograde circumbinary disks around eccentric SMBHBs, solving an energy equation that balances viscous and shock heating against blackbody radiative cooling. We investigate different initial disk Mach numbers $\mathcal{M}_a \in \{10, 20, 40\}$ and consider binary eccentricities $e_\mathrm{b}\in[0.0,\,0.8]$, finding that multiple stable states exist for the same binary eccentricity and Mach number. These states differ by the sense of rotation of their minidisks; both retrograde ($\downarrow\downarrow$), both prograde ($\uparrow\uparrow$), or one of each ($\uparrow\downarrow$), a property set by the initial conditions. Our findings indicate that each state produces qualitatively distinct orbital evolution: $\downarrow\downarrow$ and $\uparrow\downarrow$ minidisks drive circular inspirals whereas $\uparrow\uparrow$ minidisks drive eccentric inspirals with potentially observable eccentricity in the LISA band. We measure the electromagnetic emission produced by the disk, finding that a binary of mass $M_\mathrm{b}=8\times10^6\mathrm{M}_\odot$ at redshift $z=1$ would be detectable by both current and upcoming optical and UV instruments. We demonstrate that the time- and azimuthally-averaged disk profiles are well described by 1D models, which naturally set a cavity radius within which angular momentum transport is dominated by Reynolds stresses rather than viscosity.
\end{abstract}




\section{Introduction}
\needspace{3\baselineskip}
Gas accretion onto supermassive black hole binaries (SMBHBs) has important astrophysical applications for their growth, orbital evolution and electromagnetic emission. The interaction between a binary and a circumbinary disk can shape the demographic properties of SMBHBs on a population level, but can also provide identifying characteristics for binaries on an individual level. To this end, electromagnetic (EM) searches \citep[\eg with LSST, ZTF, eROSITA][]{LSST19, ZTF2019, ERosita2021} can use distinct predictions for the EM signatures of SMBHBS in accretion disks. These efforts are complemented by current \citep[][]{NanoGrav2023, InPTA_EPTA, ParkesPTA} and future \citep[TianQin and LISA;][]{TinQian2021, AmaroSeoane23}  generations of low-frequency gravitational wave (GW) observatories, capable of making multimessenger characterisation of these systems an increasingly realistic prospect.

The degree of anisotropy associated with episodes of accretion onto SMBHs in galactic nuclei has been studied from analytical, numerical and observational perspectives. Analytically, \cite{King2006} argue that SMBH growth at high redshift is most efficient when accretion proceeds through a sequence of randomly oriented episodes, each too small to coherently spin up the black hole. Correspondingly, \cite{Levine2010} and \cite{Hopkins2011} model the angular momentum transport of gas from the galactic scale to a central SMBH and find that by the time gas reaches sub-parsec scales its angular momentum orientation has been effectively randomised leading to highly chaotic, variable gas inflow rates \citep[see also][]{Hobbs2011}. At larger scales, hydrodynamical simulations of gas-rich galaxy mergers \citep{Mayer2007, Chapon2013} find gravoturbulent nuclear disk formation, where subsequent accretion episodes may carry no preferred angular momentum. 
Observationally, by analysing jet directions in active galactic nuclei (AGN), \cite{Kinney2000} and \cite{Schmitt2002} find evidence for misalignment between the accretion disk and the host galaxy disk, assuming that the jet is perpendicular to the accretion disk. Studies of narrow-line-region outflow kinematics find no correlation between the host galaxy and AGN disk \citep{Fischer2013, Fischer2014}, while dynamical modelling of the broad-line region finds the broad-line-region inclination to be similarly uncorrelated with the host galaxy disk inclination \citep{Bennert2026}. Beyond these axis- and inclination-based tests, \cite{Gigi2026} use broad-line spectropolarimetry to constrain the \emph{sense} of AGN rotation, finding equal numbers of AGN that co- and counter-rotate with respect to their host galaxy, suggesting that the relative spin orientations are effectively random. Taken together, these studies support the expectation that circumbinary disks form with a random initial inclination relative to the binary orbital plane \citep[as previously suggested by][]{Nixon12_StableRetrograde}.\\ 

Viscously driven alignment dictates that a circumbinary disk can attain one of three stable orientations around a fixed binary; namely prograde, retrograde or polar \citep[assuming that the disk does not tear, \cf][]{King2005, Nixon12_StableRetrograde, Martin2017, Lubow2018}. Consequently, a low-mass, viscous disk with random initial inclination $\iota_0$ (with $\iota$ being the relative inclination between disk and the orbital plane of the binary) will ultimately settle into one of these three configurations. We denote their respective probabilities $f_\rightrightarrows$ (prograde), $f_\rightleftarrows$ (retrograde) $f_\perp$ (polar). The exact value of each probability is determined by non-trivial dependencies on the mass ratio, binary eccentricity $e_\mathrm{b}$, binary angular momentum magnitude $J_\mathrm{b}$ and disk angular momentum magnitude $J_\mathrm{d}$. While we do not discuss in detail the functional forms of $f_\rightrightarrows, f_\rightleftarrows, f_\perp$, we note that the polar disk fraction depends strongly on the binary eccentricity and vanishes for circular binaries $f_\perp\to0$ as $e_\mathrm{b}\to0$ \citep[][]{Martin2017}. Moreover, if we assume that for eccentric binaries ($e_\mathrm{b}>0$) the polar basin is roughly symmetric around $\iota_0 =\pi/2$ we may assume that the ratio of prograde to retrograde disks is approximately given by (\citealt{King2005}, \citealt{Nixon12_StableRetrograde}):
\begin{equation}
    \frac{f_\rightleftarrows}{f_\rightleftarrows + f_\rightrightarrows} = \frac{1}{2}\left(1-\frac{J_\mathrm{d}}{J_\mathrm{b}}\right).
    \label{RetrogradeFraction}
\end{equation}
Under the assumption that the initial gas angular momentum is small with respect to the binary angular momentum -- or equivalently, the binary mass far exceeds the disk mass out to some characteristic radius \citep[\eg the propagating warp radius as suggested by][]{Nixon12_StableRetrograde} -- Eq.~(\ref{RetrogradeFraction}) predicts a similar abundance of prograde and retrograde disks around SMBHBs. Therefore, based on viscously driven alignment, isotropic feeding of gas onto a binary with larger angular momentum than the gas, will lead to comparable fractions of prograde and retrograde circumbinary disks.\\

Despite this expectation, the vast majority of circumbinary disk studies have focused on the prograde configuration \citep[\eg][]{MacFadyen2008, Cuadra2009, Noble2012, Shi2012, DOrazio2013, Gold2014, Shi2015, Farris15, Miranda2017, Moody2019, Tiede2020, Dorazio2021, Zrake2020, RyanWS2021, Dittmann2022, Franchini2023, Dittmann2023, Avara2024, Tiede2025, DeLaurentiis2025, Betancourt2026, Dittmann2026}, while retrograde disks are comparatively less well studied \citep{Nixon2011, Roedig2014, Bankert2015, Schnittman2015, AmaroSeoane2016, Tiede2024, Bourne2024, ONeill2025}. The latter studies suggest that binaries in retrograde disks admit many potentially interesting properties, such as stronger torques (\citealt{Nixon2011}, \citealt{Tiede2024}), eccentricity growth (\citealt{Roedig2014}, \citealt{Schnittman2015}, \citealt{AmaroSeoane2016}, \citealt{Tiede2024}) and different resonant structures than their prograde counterparts, making retrograde disks efficient paths towards gravitational wave-driven coalescence. Finally, we note that recent studies of realistically thin, prograde circumbinary disks have revealed potentially long periods of suppressed accretion \citep{Ragusa2016, Tiede2020, Dittmann2022, Tiede2025, Tiede2026}, which may be a consequence of colder and narrower tidal streams becoming less efficient at feeding material onto the binary components. In contrast, the dynamics of accretion is different in retrograde disks, potentially bypassing the suppressed accretion problem and making retrograde binaries consistently luminous electromagnetic sources.\\

In this work, we present the first grid-based simulations of eccentric retrograde binaries with blackbody cooling. This paper is outlined as follows. In Section \ref{sec:Methods} we present our analytical and numerical methods followed by our results in Section~\ref{sec:Results} and discussion in Section~\ref{sec:Discussion}. Finally, we summarise our conclusions in Section~\ref{sec:Conclusions}.

\section{Methods}
\label{sec:Methods}
\subsection{Binary Model}
We consider an equal-mass binary ($q=1$) with total mass $M_\mathrm{b}=8\times 10^6 \mathrm{M}_\odot$ at an orbital semi-major axis of $a_\mathrm{b}=2530 r_\mathrm{g}\approx 10^{-3}\mathrm{pc}$. We choose these orbital parameters to ensure that the binary is a suitable target for both EM and GW surveys, with a year-long orbital period and a mass within the range where LISA \citep{AmaroSeoane2016} will have near-optimal sensitivity. We treat both binary components as Plummer potentials with softening radii $r_\mathrm{soft} = 0.05a_\mathrm{b}$ moving on prescribed Keplerian orbits. To investigate the effects of binary eccentricity, we perform a set of adiabatic eccentricity sweeps \citep[see][]{Dorazio2021, Tiede2024} where we linearly increase the binary eccentricity as a function of time,
\begin{equation}
    e(t) = \begin{dcases*}
            \makebox[\widthof{$\left(\dfrac{e_\mathrm{f}(t-t_\mathrm{s})}{2\pi n}\right)$}][c]{$0$} & if $t < t_\mathrm{s}$\,, \\[1ex]
            \frac{e_\mathrm{f}(t-t_\mathrm{s})\Omega_\mathrm{b}}{2\pi n_P} & if $t > t_\mathrm{s}$\,.
            \end{dcases*}
\end{equation}
where $t_\mathrm{s}$ is the sweep start time, $e_\mathrm{f} = 0.8$ is the final eccentricity, and $n_P$ is duration of the sweep measured in units of binary orbital periods. We set $n_P = 4000\,(\mathcal{M}_\mathrm{a}/10)^2$, where $\mathcal{M}_\mathrm{a}$ is the disk Mach number at $r = a_\mathrm{b}$, chosen such that the eccentricity increases by $\Delta e \sim 0.015$ per viscous time at $r = a_\mathrm{b}$. In addition to these sweep simulations, we perform a complementary set of fixed-eccentricity runs to benchmark our results.

\subsection{Disk Model}
The steady-state surface density $\Sigma_0$ and vertically integrated pressure $P_0$ for a gas-pressure and electron-scattering-opacity dominated disk around a single central black hole are given by \cite{SS73} as,
\begin{align}
            \frac{\Sigma_0}{M_\mathrm{b}a_\mathrm{b}^{-2}} &\approx 1.1\times10^{-4}~\lambda_\mathrm{Edd}^{\tfrac{3}{5}}\left(\frac{\alpha}{0.1}\right)^{-\tfrac{4}{5}} \left(\frac{M}{M_\mathrm{b}}\right)^{\tfrac{1}{5}} \left(\frac{r}{a_\mathrm{b}}\right)^{-\tfrac{3}{5}},\nonumber\\
            \frac{P_0}{M_\mathrm{b}\Omega_\mathrm{b}^2} &\approx 2.2\times 10^{-9}~\lambda_\mathrm{Edd}\left(\frac{\alpha}{0.1}\right)^{-1} \left(\frac{M}{M_\mathrm{b}}\right)^{\tfrac{1}{2}} \left(\frac{r}{a_\mathrm{b}}\right)^{-\tfrac{3}{2}},
        \label{SS73Solution}
\end{align}
\citep[see also][]{FrankKingBook2002, Goodman2003, Haiman2009}. Here, $\lambda_\mathrm{Edd}$ is the Eddington accretion fraction, $\Omega_\mathrm{b}$ is the angular frequency of the binary and $\alpha$ is a parameterisation of the efficiency of angular momentum transport \eg due to MRI-driven turbulence \citep{BalbusHawley91}.\\

We perform simulations of circumbinary disks with three different Mach numbers evaluated at $r=a_\mathrm{b}$, $\mathcal{M}_a \in \left[10, 20, 40\right]$. We choose these values based on numerical limitations; larger Mach number disks become less tractable due to increased spatial resolution and simulation timescale requirements \cite[see][for recent high Mach prograde disk simulations]{Tiede2025, Tiede2026}. It is important to note that these values imply hyper-Eddington accretion rates,
\begin{equation}
    \lambda_\mathrm{Edd} \approx 1.55\times 10^6\left(\frac{\alpha}{0.1}\right)^{\tfrac{1}{2}}\left(\frac{M}{M_\mathrm{b}}\right)^{\tfrac{7}{4}}\left(\frac{a}{a_\mathrm{b}}\right)^{-\tfrac{1}{4}}\left(\frac{\mathcal{M}_a}{10}\right)^{-5}.
    \label{Eddington}
\end{equation}
The hyper-Eddington accretion rates implied by the Mach numbers we have chosen (1) result in  electromagnetic disk emission orders of magnitude in excess of realistic systems (with Eddington fractions of order unity) and (2) violate our assumptions of radiatively efficient, gas-pressure dominated disk solutions. 
To compensate for this, in post-processing we perform a remapping for the disk surface density and pressure by rescaling their values to that of a target disk with Eddington fraction $\hat \lambda_\mathrm{Edd}$, 
\citep{RyanWS2021},
\begin{align}
    \Sigma \to \hat\Sigma &= \Sigma \left(\frac{\hat \lambda_\mathrm{Edd}}{\lambda_\mathrm{Edd}}\right)^\frac{3}{5},\\
    {P}\to \hat{P} &= {P}\left(\frac{\hat\lambda_\mathrm{Edd}}{\lambda_\mathrm{Edd}}\right).
\end{align}
We chose a target accretion rate of $\hat \lambda_\mathrm{Edd}=0.1$ for the following reasons: (1) Observed Eddington ratio distributions are broadly distributed around $0.1$ \citep{Schulze2010, Ananna2022}. (2) The transition from the radiation pressure to gas pressure dominated region occurs at $r_\mathrm{rad} \approx 0.1 a_\mathrm{b}$ \citep{SS73} for a $\hat\lambda_\mathrm{Edd}=0.1$ disk for our chosen black hole mass, justifying our neglect of radiation pressure across the majority of the spatial domain ($r_\mathrm{rad} \sim \lambda_\mathrm{Edd}^{16/21}$). (3) Radiatively efficient, geometrically thin disks are unstable at super-Eddington accretion rates, and a slim disk treatment would be required \citep{Abramowicz1988}, which is beyond the scope of this work. Therefore our choice of Eddington fraction is most consistent with the underlying disk model and the astrophysical observations of accreting SMBHs.

\subsection{An Analytical Model of Binary-Disk Interactions}
In this section, we revisit analytical models of retrograde binary–disk interactions, providing a description that spans test-particle orbital stability (Section~\ref{sec:MinidiskStability}) to the density and pressure profiles of the circumbinary disk (Section~\ref{sec:CBDTorques}).

\subsubsection{The Stability of Retrograde Minidisks}
\label{sec:MinidiskStability}
A retrograde circumsingle "minidisk" is the material that is bound to a single black hole with negative specific angular momentum in the frame corotating with the binary. Both \cite{Tiede2024} and \cite{Nixon2011} identify retrograde minidisks in simulations of retrograde circumbinary disks, while \cite{Overton2024} and \cite{Overton2025} study isolated retrograde circumprimary disks in Be/X-ray binary systems, finding them unstable to tilting and, when the binary is eccentric, subject to forced eccentricity growth and disk breaking. In this subsection, we employ the restricted 3-body approach to obtain an estimate for the maximum radius out to which orbits in the minidisk are stable in the context of an equal-mass, circular binary.\\

We first note that the Jacobi constant of a test particle in the corotating frame is insensitive to the orbital direction of the particle -- only the magnitude of its velocity squared enters. Therefore, assuming all other factors are kept constant, simply reversing the direction of motion of the test particle changes neither the Jacobi constant nor the zero-velocity curves. Instead, the dynamical distinction between prograde and retrograde is set by the direction-dependent Coriolis force.\\

One key difference between prograde and retrograde (circular) orbits is how strongly they resonate with the binary orbit. Resonances can coherently transfer angular momentum from the binary to the disk, potentially leading to the ejection of material. In a viscous disk, the tidal truncation radius of the minidisks is set where resonant torques balance viscous torques \citep{Papaloizou1977, Artymowicz1994}\footnote{For an analysis of circumbinary disk stability, see \cite{Mahesh2024}, who attributes orbital instability to radial epicyclic instabilities acting on orbital timescales, and \cite{Ragusa2026}, who connects the dynamical stability of the circumbinary disk to its instantaneous orbital properties -- in contrast to the viscous-timescale resonant torque balance invoked here.} For prograde circumsingle disks around one component of a circular binary, the $(l,m)=(2,2)$ outer Lindblad resonance (which is the outermost of the $m=l$ resonance sequence) sets the outer truncation boundary \citep{Artymowicz1994}, giving a truncation radius of $r_\mathrm{tr} \approx 0.3\,a_b$ for an equal-mass binary \citep{Artymowicz1994}. As noted by \cite{Nixon2011, Ivanov2015} and \cite{Overton2024}, retrograde minidisks around one component of a circular binary are free of low-order ordinary Lindblad resonances \citep{Nixon2015}. Equivalently, retrograde mean-motion resonances (which coincide with Lindblad resonances in a Keplerian disk) exist but are of higher order than their prograde counterparts, and hence more strongly suppressed \citep{Morais2012}. Therefore, retrograde resonances exert very weak torques and are unlikely to limit the radius of circumsingle minidisks.\\

\begin{figure}
    \centering
    \includegraphics[width=\linewidth]{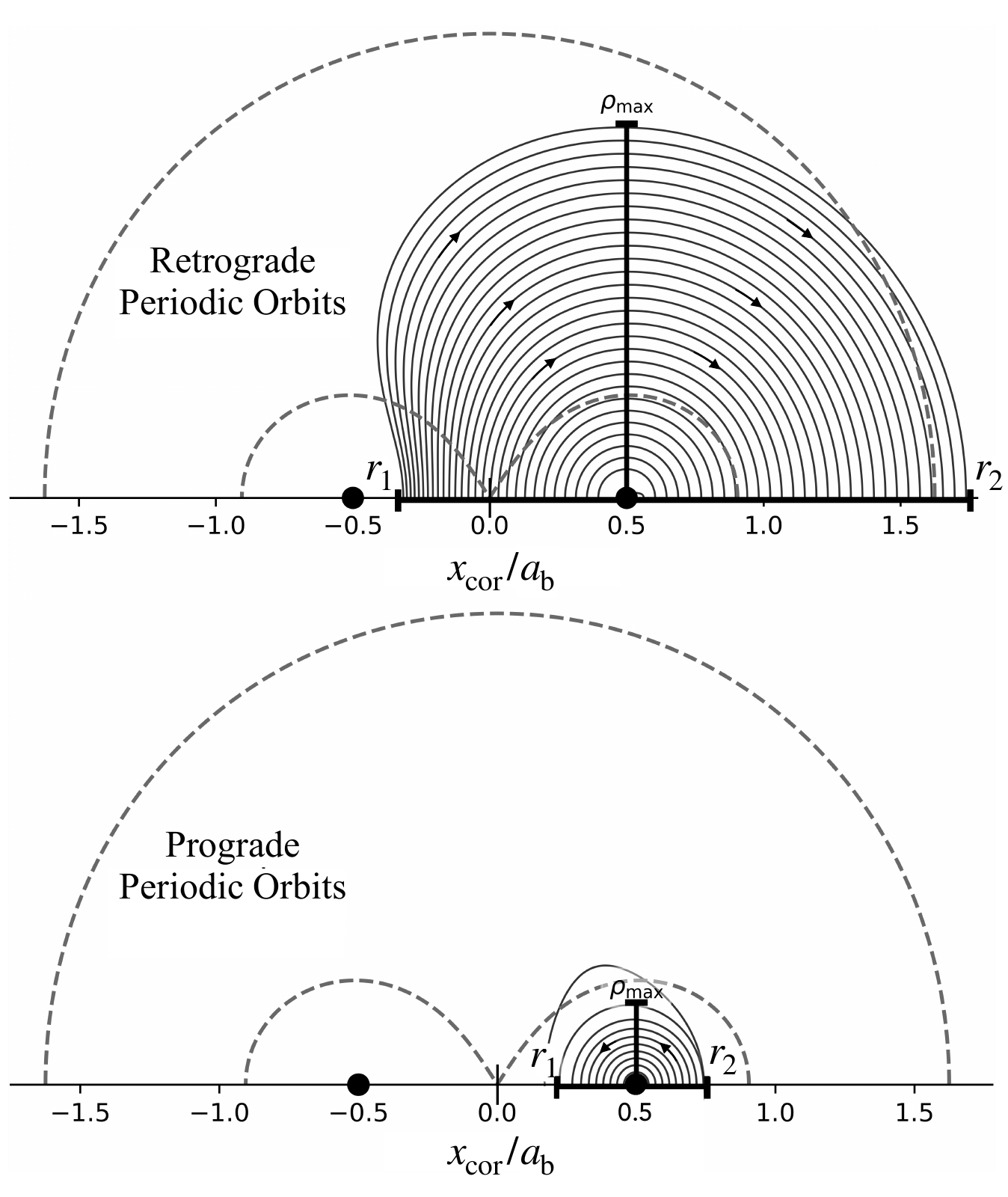}
    \caption{Strictly periodic orbits in the corotating frame around one of the binary components. The retrograde family of orbits (\emph{top}) is non-intersecting within a maximum radius $\rho_\mathrm{max} \approx 1.36 a_\mathrm{b}$. In contrast, the prograde family of orbits (\emph{bottom}) is non-intersecting within a much smaller maximum radius of only $\rho_\mathrm{max} \approx 0.3 a_\mathrm{b}$. The dashed lines denote zero velocity curves of test particle orbits, for a Jacobi constant $C_J=4$.
    }
    \label{fig:StrictlyPeriodicRetrograde}
\end{figure}

An alternative approach to derive the size of minidisks comes from the self-intersection of periodic streamlines. \cite{Paczynski1977} showed that strictly periodic, prograde orbits in the corotating frame will experience self-intersections beyond a critical radius $r_* \approx 0.302\,a_b$ \citep[see Table 1 for $\mu=0.5$ of][]{Paczynski1977} as shown in the bottom panel of Fig.~\ref{fig:StrictlyPeriodicRetrograde}. In a disk, intersecting streamlines cause shocks and ultimately disrupt the disk structure. Therefore, the outermost non-self-intersecting periodic orbit sets a maximum disk radius. In Fig.~\ref{fig:StrictlyPeriodicRetrograde} we demonstrate that the outermost non-self-intersecting retrograde orbit around a single component extends to radii $r_1 \approx 0.82 a_\mathrm{b}$ and $r_2 \approx1.23a_\mathrm{b}$ along the binary axis, with a maximum extent $\rho_\mathrm{max}\approx1.36 a_\mathrm{b}$.\footnote{We note that for an eccentric binary, the self-intersecting streamline approach was generalised by \cite{Pichardo2005} from requiring strictly periodic orbits to stable invariant loops.} This maximum retrograde minidisk radius is significantly larger than the maximum prograde radius ($\approx 0.302\,a_\mathrm{b}$) and since the retrograde orbit extends to $r_2 \approx 1.23\,a_\mathrm{b}$ along the binary axis, two such minidisks cannot simultaneously fill their maximum stable extent without overlapping. In the full hydrodynamic problem the two would necessarily interact, suggesting their behaviour may be more complex than in the prograde case. We further note that pressure and shocks in a collisional disk may reduce the physical minidisk radius below these single-particle estimates.\\

\subsubsection{Angular Momentum Flux in the Circumbinary Disk}
\label{sec:CBDTorques}
For the discussion presented below, we will consider the circumbinary disk to be all of the disk material that is not bound to an individual component of the binary. Instead, circumbinary disk material orbits about the binary barycenter and, to good approximation, with a Keplerian velocity profile.\\

In steady-state, binary accretion necessitates an interaction between material in the circumbinary disk and the minidisks (ensuring a constant mass flux at all radii). This interaction mediates an exchange of angular momentum between the binary and the disk, causing the binary to respond through a change in its orbital elements and the disk to respond through internal stresses that communicate an angular momentum flux. For a viscous hydrodynamical fluid, the radial transport of angular momentum can be decomposed following \citet[][Eq.~C4]{Noble2012}, neglecting the radiative flux $\mathcal{F}_\phi$. In their magnetized flow, angular momentum is transported by the sum of the Maxwell and turbulent Reynolds stresses generated by the MRI. Here, we impose an explicit viscous stress $F_{J,\nu}$ designed to emulate this MRI-driven turbulent transport. The viscous stress does not, however, capture the large-scale, resolved shocks and streams in our simulations, which also transport angular momentum; we measure this contribution through the Reynolds stress $F_{J,\mathrm{Rey}}$. Radially integrating Eq.~C4 of \citet{Noble2012} and grouping the advective, viscous, and Reynolds contributions, the total hydrodynamical angular momentum flux is
\begin{equation}
    F_{J,\mathrm{hydro}} = F_{J,\mathrm{adv}} + F_{J,\nu} + F_{J,\mathrm{Rey}}.
    \label{TotalAngularMomentumFlux}
\end{equation}
In steady state, the net rate of change of angular momentum vanishes, and therefore the angular momentum flux supplied by the binary, $F_{J,\mathrm{bin}}$, must balance the hydrodynamic angular momentum flux, $F_{J,\mathrm{hydro}}$, such that $\dot J = F_{J,\mathrm{bin}} - F_{J,\mathrm{hydro}} = 0$. We directly measure the total angular momentum flux $F_{J,\mathrm{hydro}}$ at a given snapshot by reconstructing the fluxes at cell faces using the solver's own face-reconstruction routine, identical to that used during evolution. The viscous component $F_{J,\nu} \equiv 2\pi r^2 \langle \Pi_{r\phi}\rangle_\phi$ is obtained by azimuthally averaging the shear tensor $\Pi_{r\phi}$ rather than assuming $F_{J,\nu} = 3\pi\nu\Sigma l$; the two agree closely except very close to the binary, where the rotation profile departs from Keplerian. The advected angular momentum flux $F_{J,\mathrm{adv}}\equiv \langle\dot M\rangle_\phi \langle l\rangle_\phi$ is the mean-field angular momentum flux, and $F_{J,\mathrm{Rey}}\equiv 2\pi r \langle \Sigma \delta v_r \delta l\rangle_\phi$ is the Reynolds stress which describes correlated-fluctuations in the medium such as shocks and waves which transport angular momentum. We have verified that the binary torque is typically dominated by the gravitational component, $F_{J,\mathrm{bin}}\approx F_{J,\mathrm{grav}}$ which is measured directly from each snapshot by integrating the specific torque exerted by the binary outward from the origin, $F_\mathrm{J,grav}\equiv \int_0^r dr'\, 2\pi r'\,\mathcal{T}(r')$ where $\mathcal{T}(r')$ is the azimuthally-averaged gravitational torque density.\\

The interaction between the binary and the disk will either deposit or remove angular momentum from the inner regions of the disk. For a time-independent inner boundary torque, $T_{\mathrm{b}}$, the steady-state profile of the disk will be modified to accommodate the constant angular momentum flux sourced by the binary \citep[\cf][]{Tanaka2010, Rafikov2013}, 
\begin{align}
    \Sigma &= \Sigma_0 \left(1 + l_0\sqrt{\frac{a_\mathrm{b}}{r}}\right)^{3/5},\nonumber \\
    P      &= P_0 \left(1 + l_0\sqrt{\frac{a_\mathrm{b}}{r}}\right),
    \label{TorquedProfiles}
\end{align}
where we define $l_0 \equiv T_{\mathrm{b}}/({\dot{M}_\mathrm{b}a_\mathrm{b}^2\Omega_\mathrm{b}})$ as the dimensionless torque parameter, a quantity that is typically negative for retrograde binary disk interactions (\ie the binary removes angular momentum from the disk).\footnote{Note the sign difference in the definition of $\ell_0$ compared to, \eg, \citet{LaiMunoz:Review:2022, Duffell2024, Zrake2025, Tiede2026}} A negative torque parameter ($l_0<0$) will impose a central cavity of radius $r_\mathrm{cav} \equiv l_0^2 a_\mathrm{b}$, within which viscous fluxes alone cannot transport the angular momentum flux deposited by the binary. Within the cavity, viscous transport becomes inefficient, so the angular momentum flux is instead carried by Reynolds stresses (see Eq.~\ref{TotalAngularMomentumFlux}) to preserve a constant mass accretion rate at all radii. We anticipate that a disk initialised with net-zero angular momentum flux (given in Eq.~\ref{SS73Solution}) will relax into the steady-state profile given by Eq.~\ref{TorquedProfiles}.

\subsection{Numerical Approach}
We perform numerical hydrodynamical simulations using \texttt{Sailfish} (see \citealt{Zrake24} for details, \citealt{Duffell2024} for a code comparison), a GPU-accelerated Godunov code designed to solve the vertically integrated Navier-Stokes equation over a uniform Cartesian grid in 2D. We implement an $\alpha$-viscosity prescription of the form,
\begin{equation}
    \nu = \alpha c_\mathrm{s}^2 \left(\frac{GM_\mathrm{b}}{2R_1^2} + \frac{GM_\mathrm{b}}{2R_2^2}\right)^{-\frac{1}{2}},
\end{equation}
where $R_1$ and $R_2$ are the radial distances to the binary components. We solve the energy equation, balancing viscous and shock heating against radiative cooling
\begin{equation}
    \dot{Q} = -\frac{8}{3}\frac{\sigma T^4}{\kappa_\mathrm{es} \Sigma},
\end{equation}
through a semi-implicit scheme \citep[see Section 2.2 of][]{RyanMacFadyen2017}.Here $T = (m_\mathrm{p}/k_\mathrm{b})P/\Sigma$ is the midplane temperature and $\kappa_\mathrm{es} = 0.4\mathrm{~cm^2 g^{-1}}$ is the electron-scattering opacity. The gas is cooled everywhere, however, when measuring the electromagnetic emission from the disk, we disregard all cells that have an effective optical depth less than unity\footnote{
The fraction of optically thin cells is Mach dependent: roughly $\sim 10^{-4}$ of the entire domain at Mach 10, rising to $\sim 10^{-2}$ at Mach 20 and Mach 40.} 
\ie $\tau_\mathrm{eff}=\sqrt{\tau_\mathrm{abs}(\tau_\mathrm{abs}+\tau_\mathrm{es})}<1$ where $\tau_\mathrm{abs}$ is the optical depth due to free-free absorption \citep[see Eq.~5.20 in][]{RybickiLightman1986}. This avoids counting any thermalised emission coming from optically thin regions.
For optically thick cells, we assume that they emit as a blackbody with effective temperature \begin{equation}
    T_\mathrm{eff}^4 = \frac{4}{3}\frac{T^4}{\kappa_\mathrm{es}\Sigma}\left(\frac{\hat\lambda_\mathrm{Edd}}{\lambda_\mathrm{Edd}}\right).
\end{equation}
To obtain the multiband emission from each cell, we integrate the Planck spectrum across the infrared, optical, UV and X-ray bands (defined in Table~\ref{table:BandWavelengths}) before summing the contribution from each cell giving the luminosity in each band. We use an adiabatic index $\gamma = 5/3$, as appropriate for a non-relativistic, monatomic, ideal gas. \\

\begin{table}
    \centering
    \caption{Wavelength ranges of the four bands over which the Planck spectrum is integrated to obtain the multiband emission from each cell.}
    \label{table:BandWavelengths}
    \begin{tabular*}{\columnwidth}{@{\extracolsep{\fill}}lcc@{}}
        \hline\hline
        Band & $\lambda_\mathrm{min}$ (cm) & $\lambda_\mathrm{max}$ (cm) \\
        \hline
        IR      & $7\times10^{-5}$ & $10^{-1}$        \\
        Optical & $4\times10^{-5}$ & $7\times10^{-5}$ \\
        UV      & $10^{-6}$        & $4\times10^{-5}$ \\
        X-ray   & $10^{-9}$        & $10^{-6}$        \\
        \hline
    \end{tabular*}
\end{table}

\begin{figure*}
    \centering
    \includegraphics[width=\linewidth]{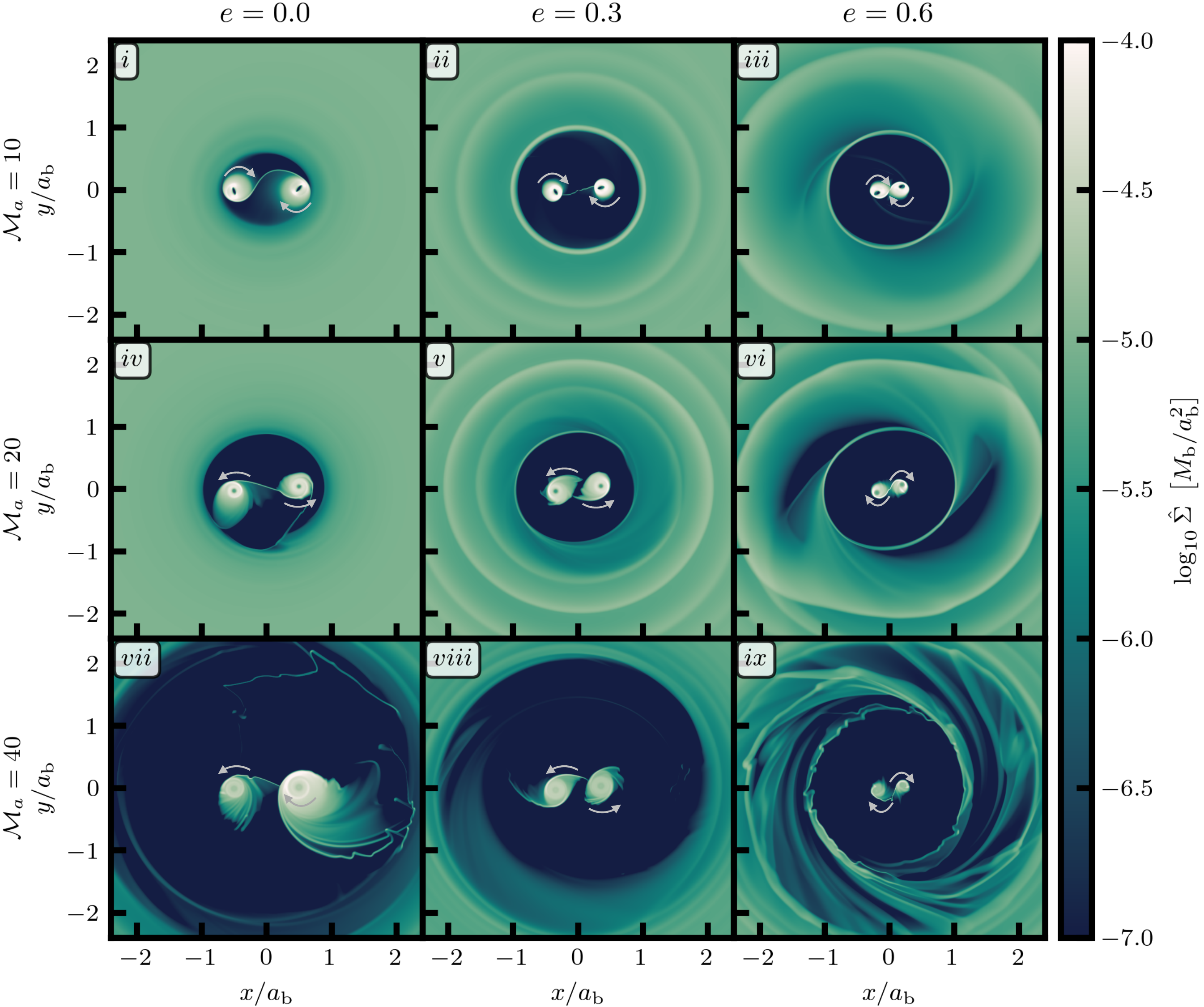}
    \caption{Target surface density maps spanning orbital eccentricity $e=0.0,\,0.3,\,0.6$ (\emph{columns}) and initial disk Mach number $\mathcal{M}_\mathrm{a}=10,\,20,\,40$ (\emph{rows}). The binary orbit is counter-clockwise in all panels, and the grey arrows indicate minidisk orbital directions.
    }
    \label{fig:DensityMaps}
\end{figure*}

We perform our fixed eccentricity simulations over a spatial domain of width $L_\mathrm{fix} = 15a_\mathrm{b}$, with resolution $dx_\mathrm{fix}=0.0067 a_\mathrm{b}$, while our eccentricity sweeps have a spatial domain width $L_\mathrm{swp} = 24a_\mathrm{b}$ and resolution $dx_\mathrm{swp}=0.0075 a_\mathrm{b}$. Since the innermost stable circular orbit (ISCO) is far smaller than our grid scale($r_\mathrm{isco} \approx 0.1 dx_\mathrm{fix}$), we employ a sink prescription as a subgrid model for the inner accretion flow of the form 
\begin{equation}
    S^j = -(\gamma_\mathrm{sink}\Omega_\mathrm{b} \mathrm{d}t) W(r) U^{*j} ,
\end{equation} 
where $j$ indexes the conserved variables (mass, momentum, and energy), $\gamma_\mathrm{sink}=100/\mathcal{M}_a^2$ is the dimensionless sink rate, $\mathrm{d}t$ is the integration timestep and $U^{*j}$ are the torque-free, adiabatic conserved variables (see \citealt{Dempsey2020} and Appendix A in \citealt{Tiede2026}). The window function $W(r)\propto \exp\left[-(r/r_\mathrm{sink})^8\right]$ sharply confines the sink to within $r_\mathrm{sink}=0.05a_\mathrm{b}$. We impose floors of $\Sigma_\mathrm{f}= 10^{-15} M_\mathrm b a_\mathrm b^{-2}$ and $ P_\mathrm f =10^{-20} M_\mathrm b \Omega_\mathrm{b}^2$.\\

An outer boundary buffer acts as both a mass reservoir and damping mechanism to regulate spurious artifacts from the Cartesian boundaries. Inside this region we damp the surface density (at a rate given by the local orbital frequency) to the initial condition surface density ($\Sigma_\mathrm{buffer}$) given in Eq.~\ref{SS73Solution}, the azimuthal velocity to the pressure-corrected sub-Keplerian velocity ($v^2_\phi = GM_\mathrm{b}/r - 3 P/2\Sigma_\mathrm{buffer}$) and the radial velocity to enforce a fixed mass inflow rate $v_r = \dot{M}_{\inf}/2\pi r\Sigma_\mathrm{buffer}$, where $\dot{M}_{\inf}\equiv \lambda_{\mathrm{Edd}} \dot M_\mathrm{Edd}$ is the feeding rate at infinity. 

\begin{figure}
    \centering
    \includegraphics[width=\linewidth]{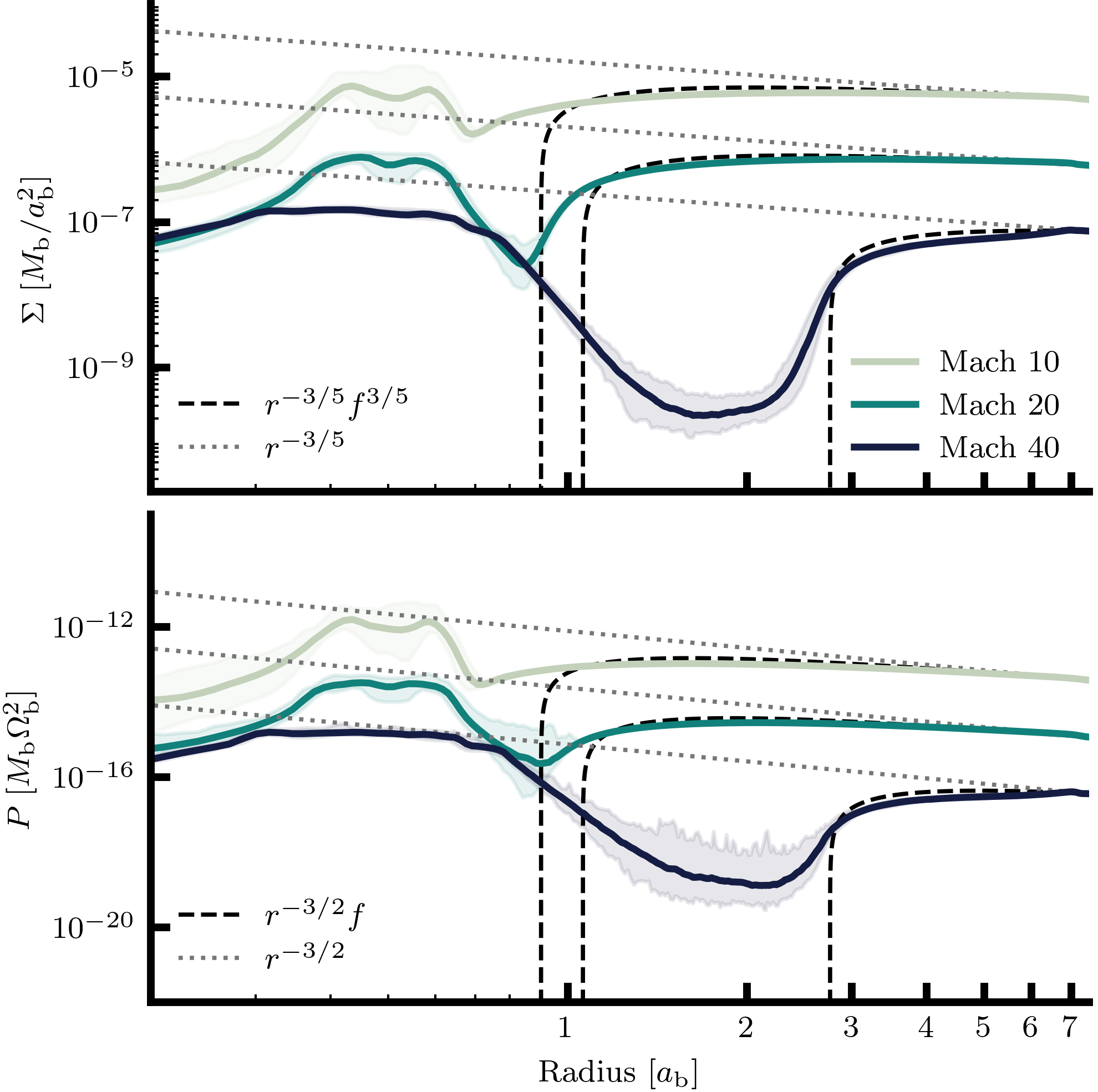}
    \caption{Time and azimuthally averaged surface density profiles (\emph{top}) and vertically-integrated pressure profiles (\emph{bottom}) for circular binaries ($e_\mathrm{b} = 0$). We compare our findings to the initial conditions \citet{SS73} (grey, dotted lines) and Eq.~\ref{TorquedProfiles} (black, dashed lines) using $f = 1+l_0\sqrt{a_\mathrm{b}/r}$ with $l_0= [-0.88,~-1.03,~-1.66]$ for Mach 10, 20 and 40 respectively.\label{fig:CircumbinaryDiskProfile}}
\end{figure}

\begin{figure*}
    \centering
    \includegraphics[width=\linewidth]{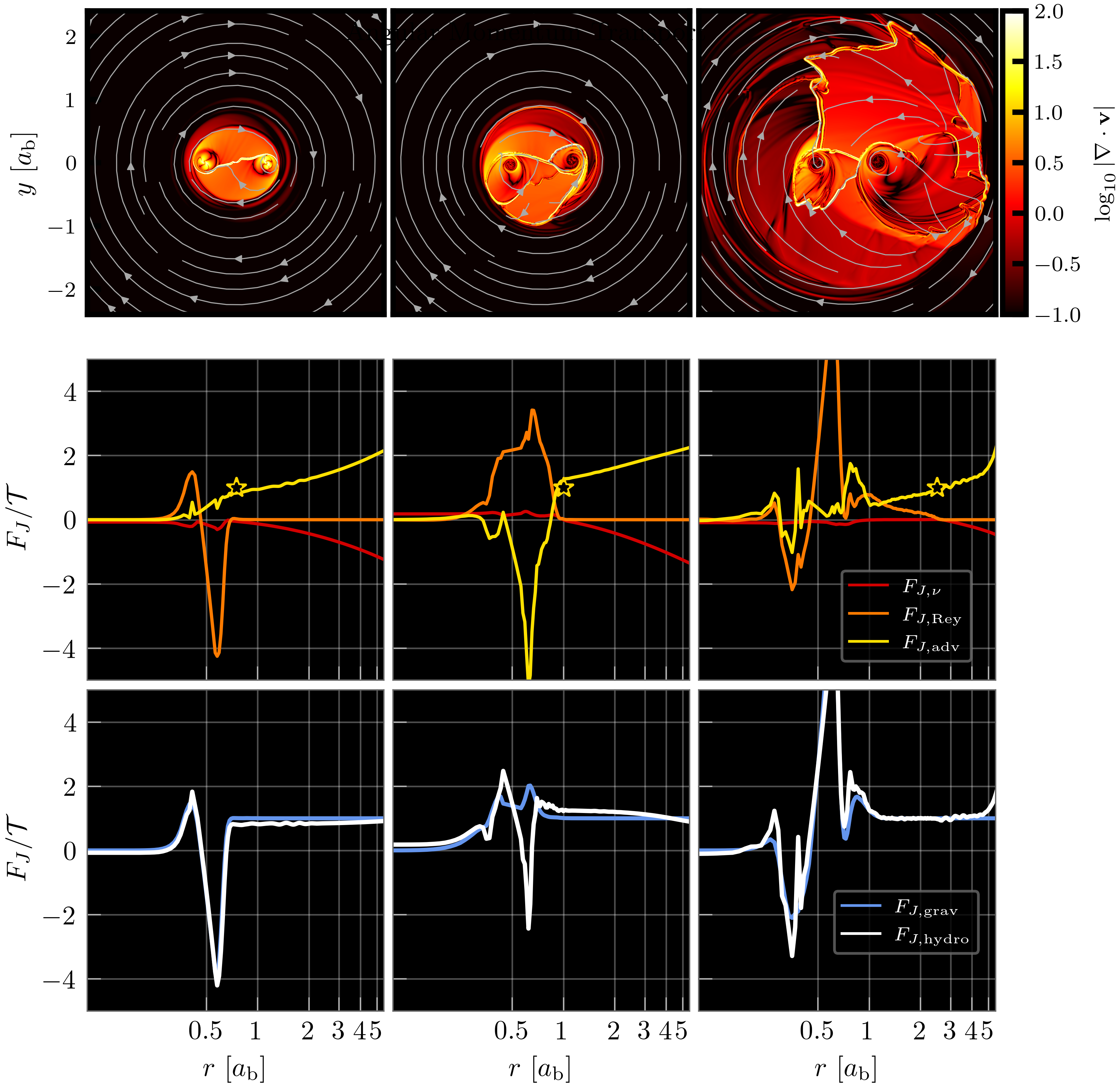}
    \caption{Angular momentum transport for three disks of varying Mach number (left to right: $\mathcal{M}_a = 10,~20,~40$) interacting with a circular binary ($e_\mathrm{b} = 0$). \emph{Top row}: A snapshot of the velocity divergence field clearly showing strong Mach dependence with overlaid streamlines of mass flux $\Sigma\mathbf{v}$ (grey). \emph{Middle row}: Decomposition of the angular momementum flux into it's advective, viscous and Reynolds components ($F_{J,\mathrm{hydro}} = F_{J,\mathrm{adv}} + F_{J,\nu} + F_{J,\mathrm{Rey}}$). \emph{Bottom row}: The cumulative gravitational torque $F_{J,\mathrm{grav}}\equiv \int_0^r dr' 2\pi r'\mathcal{T}(r')$ (blue) compared with the total hydrodynamical flux (white). The time and azimuthally averaged angular momentum flux profiles were averaged over a baseline of $10, 50, 150$ orbits respectively with a cadence of $20$ snapshots per orbit.
    }
    \label{fig:shocks}
\end{figure*}

\section{Results}
\label{sec:Results}
\subsection{Disk Structure}
\label{sec:DiskStruture}
Fig.~\ref{fig:DensityMaps} illustrates a sequence of density snapshots at pericenter with orbital eccentricities $e_\mathrm{b} = [0,~0.3,~0.6]$ (columns) and initial disk Mach numbers $\mathcal{M}_a = 10,~20,~40$ (rows). Across all panels, there is a generic two-component structure comprised of an outer circumbinary disk and inner circumsingle minidisks. Material transitions from the circumbinary disk onto the minidisks through strong shocks, where fluid elements deliver their mass and momentum onto the binary. The intrabinary bridge connecting the two minidisks is a particularly important shock interface where fluid streamlines intersect (see also Fig.~\ref{fig:shocks}). After shocking, material is transported along the bridge with very low angular momentum onto one of the minidisks. The geometry of the bridge and the location at which it attaches to the minidisk sets the sense of rotation of the minidisk.\\

For Mach 10 (\emph{upper row} of Fig.~\ref{fig:DensityMaps}), the bridge connects to the front of each minidisk (ie. the bottom of the left and the top of the right minidisk in panels $i,\, ii,\, iii$ of Fig.~\ref{fig:DensityMaps} 
. Material following this bridge enters the frame of each black hole with negative specific angular momentum, thereby sourcing retrograde minidisks ($\downarrow\downarrow$). For Mach 20 (\emph{middle row} of Fig.~\ref{fig:DensityMaps}), the bridge connects to the back of each minidisk (ie. the top of the left and the bottom of the right minidisk in panels $iv,\, v$ of Fig.~\ref{fig:DensityMaps}), thereby entering the minidisk with positive specific angular momentum and sourcing a prograde configuration ($\uparrow\uparrow$). Finally, for Mach 40 (\emph{lower row} of Fig.~\ref{fig:DensityMaps}), there is an asymmetry between two minidisks, with the bridge attaching to the back of the smaller one and the front of the larger one, leading to one prograde and one retrograde minidisk ($\uparrow\downarrow$). This prograde-retrograde mixed minidisk state is a counter-rotating vortex pair in which the velocity fields of the two minidisks constructively add, rather than cancel in the intrabinary region. This coherent velocity flow pushes the intrabinary bridge into the positive $y$ direction in panel $vii$ of Fig.~\ref{fig:DensityMaps}. We note that the mixed state ($\uparrow\downarrow$) may arise partly because two retrograde minidisks cannot simultaneously extend to their outermost stable streamline radii (Fig.~\ref{fig:StrictlyPeriodicRetrograde}), naturally favouring a configuration in which one minidisk adopts a prograde orientation to occupy a smaller spatial extent. We emphasise that all three states ($\downarrow\downarrow,\,\uparrow\uparrow,\,\uparrow\downarrow$) are stable and persistent over the duration of our simulations. We suggest this stability is self-reinforcing as the rotation of each minidisk is determined by the face to which the bridge attaches, which in turn sets the sign of the angular momentum delivered, sustaining the existing spin state.\\

The states shown in Fig.~\ref{fig:DensityMaps} should be considered representative, and we emphasise that the realized minidisk state is set by the initial conditions. We find no evidence that any state is excluded at a given Mach number or eccentricity. While we have not surveyed the initial-condition space systematically, we find that the initial cavity radius $r_\mathrm{cav,0}$ appears to be a primary factor in selecting the minidisk state: for circular binaries, $r_\mathrm{cav,0}=\,a_\mathrm{b}$ produces $\downarrow\downarrow$ states at both Mach 10 and Mach 20, while $r_\mathrm{init,cav}=0.55 \,a_\mathrm{b}$ produces $\uparrow\downarrow$ at Mach 10 and $\uparrow\uparrow$ at Mach 20.\\

We have verified that minidisks are a stable, persistent feature of circular binaries at Mach 10. With increased spatial resolution, smaller sink regions and faster sink rates, both torque free and acceleration free prescriptions find stable minidisks\footnote{Here, acceleration-free refers to the standard sinks in \cite{Dittmann2021} where mass and momentum are removed in proportion such that the sink applies no acceleration to the gas.}. 
In contrast, minidisks were not found to be persistent features in \cite{ONeill2025} which we attribute to the different viscosity prescription (constant $\nu$ versus constant $\alpha$, see Appendix~\ref{Appendix} for further details).\\ 

Panels $iii$ and $vi$ of Fig.~\ref{fig:DensityMaps} reveal clear two-armed spiral structure in the disk, while panel $vi$ additionally displays a four-armed pattern. As also suggested in \citet{Tiede2024}, these features could arise from retrograde Lindblad resonances excited by the binary potential. The Fourier decomposition of an eccentric binary potential reveals modes $\Psi_{l,m}$ where $m$ and $l$ are the azimuthal and temporal mode numbers respectively. \cite{Nixon2015} show that counter-rotating modes ($l<0$) can excite retrograde Lindblad resonances at discrete radii in the disk, exerting a torque that removes angular momentum from the disk. For an equal-mass binary, the potential is symmetric under rotations by $\pi$, permitting only even $m$ modes. The pattern speed of an $(l,m)$ mode is given by $\Omega_\mathrm{p} = l\Omega_\mathrm{b}/m$ and the corresponding resonant radius in a Keplerian disk follows from $r_{(l, m)} = (m/|l|)^{2/3}a_\mathrm{b}$. The two-armed spiral is consistent with the dominant $(-1,2)$ resonance, at $r_{(-1,2)} \approx 1.587\,a_\mathrm{b}$. This pattern completes a full rotation every two binary orbits, and may be responsible for the power observed in the accretion time series at $0.5\,\Omega_\mathrm{b}$ by \cite{Tiede2024}. The $(-1,4)$ resonance is a four armed spiral structure, although located at $r_{(-1,4)} \approx 2.52\,a_\mathrm{b}$ which lies outside of the spatial domain shown in Fig.~\ref{fig:DensityMaps}. Instead, the four-arm spiral structure seen in panel $(iv)$ of Fig.~\ref{fig:DensityMaps} is consistent with the $(-2,4)$ resonance which lies at radius $r_{(-2,4)} = r_{(-1,2)} \approx 1.587\,a_\mathrm{b}$. Therefore, we suggest that the middle-right panel may appear four-armed because the $m=2$ and $m=4$ components are superimposed on this shared ring, producing four spiral arms of alternating strength rather than a single two-armed pattern.\\

In Fig.~\ref{fig:CircumbinaryDiskProfile} we illustrate the time and azimuthally averaged profiles of disk surface density $\Sigma$ and vertically-integrated pressure ${P}$. We overplot the analytical prediction of Eq.~\ref{TorquedProfiles} (dashed lines), which accounts for the binary torque, alongside the net zero flux \cite{SS73} solution of Eq.~\ref{SS73Solution} (dotted grey lines) as a reference. The torqued analytical profiles in Eq.~\ref{TorquedProfiles} demonstrate good agreement with our simulations in the circumbinary disk, while the net zero flux solution differs significantly, highlighting the importance of the binary torque in shaping the disk structure. Furthermore, the cavity radius $r_\mathrm{cav}$ defined in Section~\ref{sec:CBDTorques} (dashed vertical lines) accurately describes the radius at which the density declines at the cavity edge. Because the circumbinary disk is very close to Keplerian at $r>r_\mathrm{cav}$, the disk pressure in Eq.~\ref{TorquedProfiles} sets the Mach number profile to be
\begin{equation}
    \mathcal{M}(r) = \mathcal{M}_\mathrm{a}\left(\frac{r}{a_\mathrm{b}}\right)^{-1/20} \left(1 + l_0\sqrt{\frac{a_\mathrm{b}}{r}}\right)^{-1/5}.
    \label{MachProfile}
\end{equation}
The binary torque steepens the disk pressure and surface-density profiles (Eq.~\ref{TorquedProfiles}), leading to a steeper Mach-number profile (Eq.~\ref{MachProfile}) relative to the net-zero angular momentum flux \cite{SS73} solution (dashed vs. dotted lines in Fig.~\ref{fig:CircumbinaryDiskProfile}). This is a direct consequence of the $l_0$ torque term, corresponding to a lower sound speed (higher $\mathcal{M}$) in the inner disk.\\

In Fig.~\ref{fig:shocks}, we detail the mechanisms of angular momentum transport for fixed eccentricity binaries ($e=0$) and disks with varying Mach number ($10, 20, 40$, left to right). On the \emph{top} row, we present the velocity divergence map, acting as a proxy for shocks in the disk. Inside the central cavity of each map the velocity divergence is large, while it is weak in the laminar circumbinary disk -- highlighting the importance of Reynolds stresses in the cavity region. In the \emph{middle} row, we decompose all of the hydrodynamical angular momentum fluxes as a function of radius; \ie viscous transport $F_{J,\nu}(r)$ in red, Reynolds stresses $F_{J,\mathrm{Rey}}(r)$ in orange and advected momentum $F_{J,\mathrm{adv}}(r)$ in yellow, each normalised to the total binary torque $T_{\mathrm{b}}$.
This middle row corroborates the fact that Reynolds stresses dominate inside the cavity, before sharply declining at the cavity edge at which point viscous fluxes grow in prominence. We highlight, in particular, the Mach 40 panel (right column middle row) where the large cavity is permeated by shocks (orange) before seamlessly transitioning into viscous stresses (red) near $r=2.3a_\mathrm{b}$. This transition demonstrates that different regions of the disk require different processes to transport the angular momentum. The black star denotes the cavity radius $r_\mathrm{cav}$ in all panels and is the point at which $F_{J,\mathrm{Rey}} =F_{J,\nu} = 0$ which is the boundary between the cavity and the circumbinary disk. At $r_\mathrm{cav}$, the advected flux exactly equals the binary torque as described in Section~\ref{sec:CBDTorques}). On the \emph{bottom} row, we demonstrate that the cumulative gravitational torque (blue) is in agreement with the cumulative total hydrodynamical stresses (white) confirming that the angular momentum deposited by the binary is being transported by the internal stresses in the disk. Together, these results support a consistent picture -- the binary deposits angular momentum into the disk, which is transported outward by hydrodynamical stresses. Decomposing these stresses reveals a clear spatial division between the cavity which is dominated by Reynolds stresses and the circumbinary disk which is dominated by viscous stresses.\\

\begin{figure*}
    \centering
    \includegraphics[width=\linewidth]{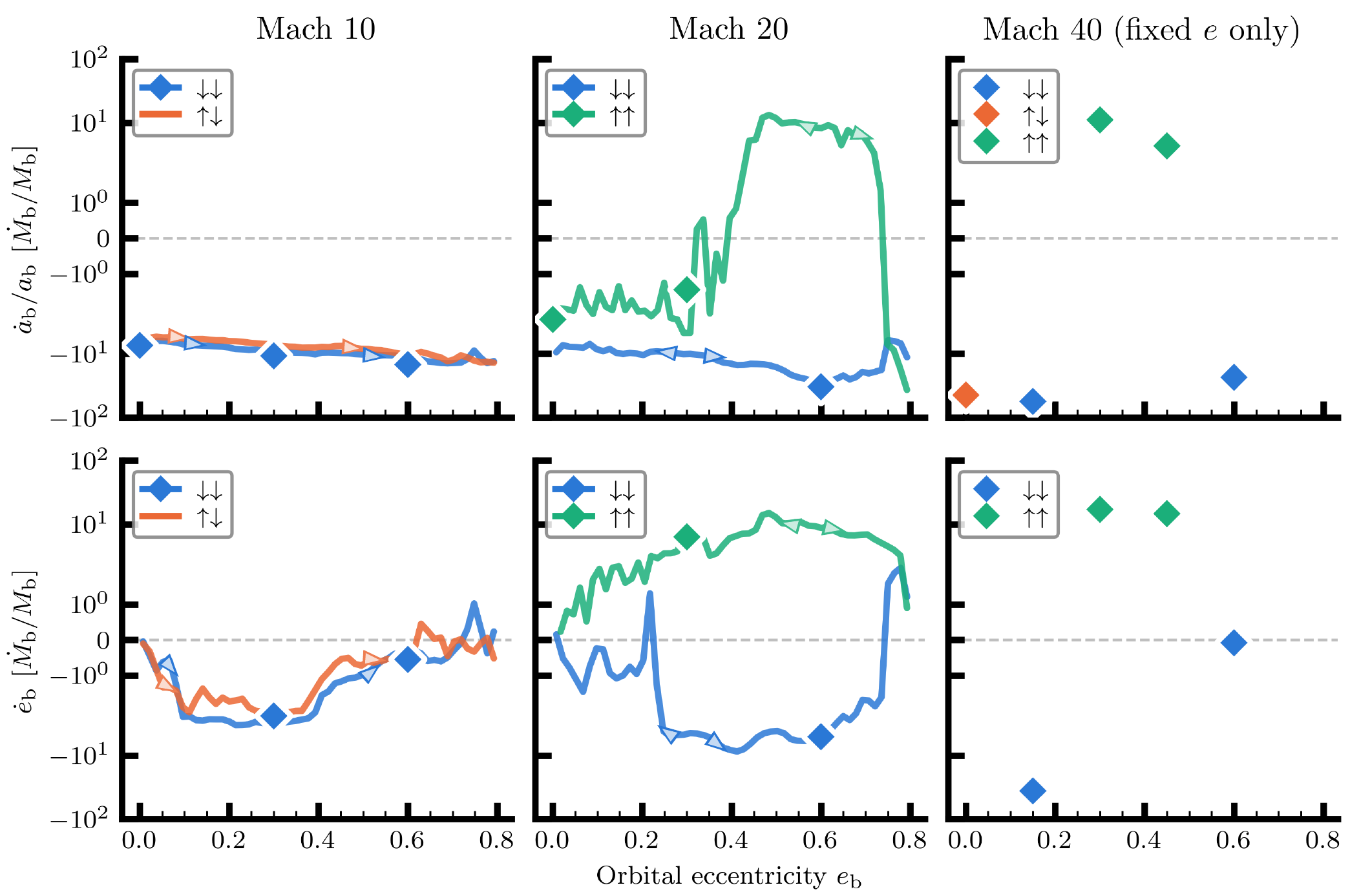}
    \caption{The orbital evolution of equal-mass binaries interacting with retrograde circumbinary disks, as a function of binary eccentricity, for varying Mach numbers. In each panel we show results from fixed-eccentricity simulations (diamonds) and eccentricity sweeps (arrows indicate the sweep direction). For practical reasons, each sweep was initialized at the eccentricity where a given state was most readily obtained, then swept toward both higher and lower eccentricity.}
    \label{fig:OrbitalEvolution}
\end{figure*}

\subsection{Orbital Evolution}
\label{sec:OrbitalEvolution}
The orbital elements of the binary will change in time in response to its interaction with the disk. The force exerted on the binary is given by the sum of both gravitational and accretion contributions\footnote{See section 4.1 of \citealt{Duffell2024} for further details} $\mathbf{F} = \mathbf{F}_g + \mathbf{F}_a$, with external torques ($\mathbf{T}_\mathrm{b}\equiv \mathbf{r}\times\mathbf{F}$), power ($\mathcal{P}\equiv\mathbf{F}\cdot\mathbf{v}$) and accretion ($\dot{M}$) causing secular changes in the binary semi-major axis $\dot{a}_\mathrm{b}$ and eccentricity $\dot{e}_\mathrm{b}$. We directly measure these quantities in our simulations by computing, 
\begin{align}
    \frac{\dot{a}_\mathrm{b}}{a_\mathrm{b}} &= \frac{\dot{M}}{M}\left(1-\frac{2a_\mathrm{b}}{r_\mathrm{b}}\right) - \frac{\mathcal{P}}{E_\mathrm{b}}\\
    \dot{e}_\mathrm{b} & = \left(\frac{1-e^2}{e}\right)\left[\frac{\dot{M}}{M}\left(1-\frac{a_\mathrm{b}}{r_\mathrm{b}}\right) - \frac{\mathcal{P}}{2E_\mathrm{b}} - \frac{\mathbf{T}_\mathrm{b}\cdot\hat{z}}{L_\mathrm{b}}\right],
\end{align}
where, for an equal-mass binary $E_\mathrm{b} = -GM^2/8 a_{\mathrm{b}}$, $L_\mathrm{b} = \sqrt{GM^3a_\mathrm{b}(1-e_\mathrm{b}^2)/16}$, $r_b=a(1-e_\mathrm{b}\cos{\mathcal{E}})$ and $\mathcal{E}$ is the eccentric anomaly of the binary. We illustrate our findings for orbital evolution in Fig.~\ref{fig:OrbitalEvolution}.\\

At Mach 10, (left column of Fig.~\ref{fig:OrbitalEvolution}) we illustrate the rate of semi-major axis evolution (top row) and eccentricity evolution (bottom row) as a function of binary eccentricity for both the $\downarrow\downarrow$ (blue) and $\uparrow\downarrow$ (orange) minidisk states. Both states drive orbital inspiral ($\dot{a}_\mathrm{b}<0$) across all binary eccentricities while typically damping the orbital eccentricity ($\dot e_\mathrm{b}<0$). This is in contrast to the locally isothermal simulations of \cite{Tiede2024} in which $\downarrow\downarrow$ minidisks were found to drive orbital eccentricity. We attribute this discrepancy to different choices of viscosity prescription, \ie constant $\alpha$ in this work and constant $\nu$ in \cite{Tiede2024} which we discuss further in Appendix \ref{Appendix}.\\

\begin{figure*}
    \centering
    \includegraphics[width=0.75\linewidth]{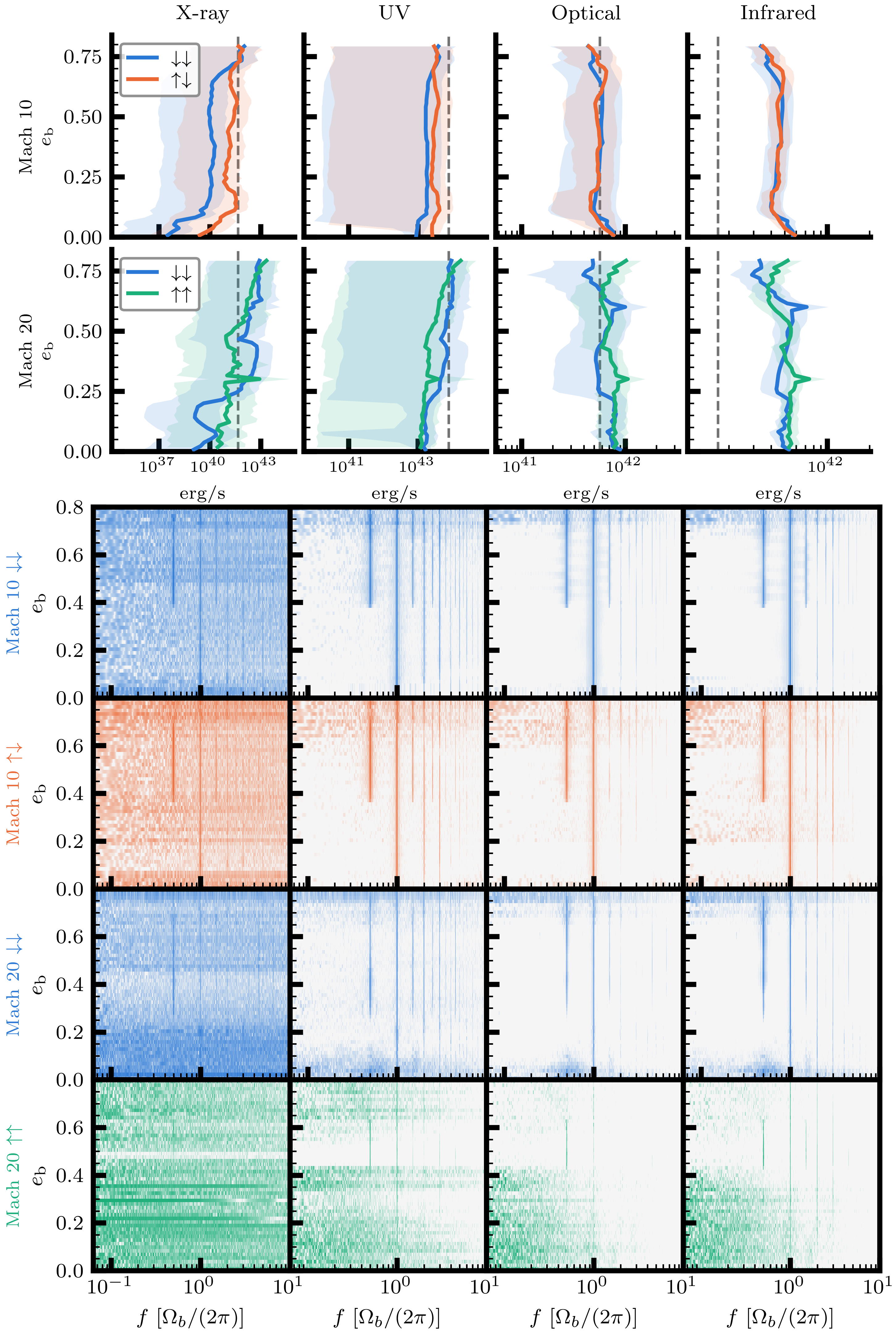}
    \caption{A summary of the electromagnetic emission measured as a function of binary eccentricity. \emph{Top:} Band-specific luminosities, with means and standard deviations computed in eccentricity bins of width $\Delta e = 0.0015$. \emph{Dashed} lines show our estimates of the additional emission from within the sink radius ($r < r_\mathrm{sink}$). \emph{Bottom:} Band-specific Lomb--Scargle periodograms as a function of eccentricity and frequency (in units of the orbital frequency), with each branch's color scale normalized independently and clipped to the upper half of its power range (\ie the top $50\%$ on a linear scale) to emphasize the strongest periodic signals.    }
    \label{fig:DiskEmission}
\end{figure*}

At Mach 20, (middle column of Fig.~\ref{fig:OrbitalEvolution}) the $\downarrow\downarrow$ state (blue) evolves with a decreasing semi-major axis while the eccentricity is typically damped -- a trend very similar to the same $\downarrow\downarrow$ state at Mach 10. However, the $\uparrow\uparrow$ state (green tracks), has an increasing eccentricity throughout, while the semi-major axis undergoes both positive and negative evolution depending on the value of $e_\mathrm{b}$.\footnote{We caution that the sweep timescale ($\Delta e=0.015$ per viscous timescale at $r=a_\mathrm{b}$) introduces uncertainty into the exact eccentricity where $\dot{a}_\mathrm{b}/a_\mathrm{b}$ transitions from negative to positive} When positive, the binary experiences outspiral driven by the disk, although notably with a decreasing pericenter distance $\dot{r}_\mathrm{p}<0$. At high eccentricities ($e_\mathrm{b}\geq0.7$), $\dot{a}_\mathrm{b}$ turns negative and the binary transitions to a full inspiral, ultimately ending in a gravitational wave-driven coalescence.\\

Despite not having completed an eccentricity sweep at Mach 40 (due to computational expenses), our fixed eccentricity simulations reveal similar state behaviour as found at both Mach 10 and Mach 20. Investigation of the velocity maps reveal that our fixed $e_\mathrm{b}=0.3,\,0.45$ simulations are in $\uparrow\uparrow$ states, whereas our $e_\mathrm{b}=0.15, 0.6$ simulations are in $\downarrow\downarrow$ states. We include these fixed eccentricity runs on the right column of Fig.~\ref{fig:OrbitalEvolution}, which further reinforce the trend of circular inspirals for $\downarrow\downarrow$ minidisks and eccentric outspirals for $\uparrow\uparrow$ ones. We note that the magnitude of $\dot{a}_\mathrm{b}/a_\mathrm{b},\,\dot{e}_\mathrm{b}$ are typically largest at Mach 40, suggesting that semi-major axis and eccentricity evolve faster in colder disks.\\

At sufficiently high orbital eccentricities, close pericenter passages truncate the minidisks to progressively smaller radii. As the minidisks shrink, the difference between the prograde and retrograde configurations diminish, so the $\downarrow\downarrow, \, \uparrow\downarrow, \, \uparrow\uparrow$ tracks are expected to converge. This expectation is supported by Fig.~\ref{fig:OrbitalEvolution} where separate branches converge at high eccentricity (near $e_\mathrm{b}=0.8$). At this unifying eccentricity, the sign of $\dot{e}$ is of particular importance, dictating whether the $\uparrow\uparrow$ branch feeds into the $\downarrow\downarrow$ branch (if $\dot{e}<0$) or vice-versa. Fig.~\ref{fig:OrbitalEvolution} suggests that the unification eccentricity has $\dot{e}>0$ for which the $\uparrow\uparrow$ branch continues to drive eccentricity to even higher values. The long-time evolution of binaries implied by these results will be discussed in Section \ref{sec:LISA_Observables} below.\\

\subsection{Electromagnetic Emission}
In this subsection, we summarise the main properties of the electromagnetic emission produced by the disk, choosing to focus on lightcurve properties rather than detectability which we leave for discussion in Section~\ref{sec:EM_Observables}.\\

The band-specific luminosities and periodograms are presented in Fig.~\ref{fig:DiskEmission} as a function of the binary eccentricity for for which we carried out a full sweep. X-ray luminosities are weaker at Mach 10 than at Mach 20, likely reflecting stronger shocks in colder disks producing higher frequency emission. With increasing binary eccentricity, there is a systematic trend of rising X-ray luminosity with the brightest emission produced during the pericenter passage. Our choice of assigning emission only to cells with $\tau_\mathrm{eff}>1$ excludes optically-thin regions and therefore any non-thermal emission. The high-frequency luminosities reported here should therefore be regarded as conservative lower bounds. The optical and infrared bands all have quite steady emission with changing binary eccentricity in Fig.~\ref{fig:DiskEmission} because the bulk of lower frequency emission originates in outer regions of the disk, which are less strongly affected by binary dynamics.\\ 

The periodograms look qualitatively similar across all frequency bands for a given sweep branch. The orbital frequency $\Omega_\mathrm{b}$, is the most powerful harmonic throughout, although power appears at $0.5~\Omega_\mathrm{b}$ from the $(-1,2)$ resonance once the binary becomes sufficiently eccentric. The eccentricity threshold for this activation is both Mach-number and branch dependent, reflecting the competition between resonant forcing, viscosity, and pressure support. At Mach 10 we find power in the $f = 0.5~\Omega_\mathrm{b}$ harmonic at eccentricities greater than $e_\mathrm{b}=0.4$, whereas the locally isothermal simulations of \cite{Tiede2024} find a higher threshold of $e_\mathrm{b}=0.55$. At Mach 20, this activation threshold is $e_\mathrm{b}=0.25$ for the $\downarrow\downarrow$ branch and $e_\mathrm{b}=0.45$ for the $\uparrow\uparrow$ branch. More broadly, the $\downarrow\downarrow$ periodogram is richer in harmonic structure than the $\uparrow\uparrow$ periodogram, further emphasising the physical distinction between the two branches in their time series behaviour.\\ 

\section{Discussion}
\label{sec:Discussion}
\subsection{LISA Observables}
\label{sec:LISA_Observables}
Retrograde circumbinary disks can host both prograde and retrograde minidisks, giving rise to distinct dynamical states. In this section we assess the detectability of binary eccentricity in the LISA band, assuming an evolution dictated by the Mach 20 $\uparrow\uparrow$ branch, and we discuss the observational implications for the effective spins of the merging binary.\\

By combining the circumbinary disk torques (see Fig.~\ref{fig:OrbitalEvolution}) with gravitational-wave radiation reaction, we integrate the coupled ordinary differential equations for the binary semi-major axis $\dot{a}$, eccentricity $\dot{e}$ and mass $\dot{M}$ for two binaries with initial masses $8\times10^6 \mathrm{M}_\odot$ (panels a and c) and $8\times10^5 \mathrm{M}_\odot$ (panels b and d) at redshift $z=1$, starting from an initial separation of $5\times10^{-3},\mathrm{pc}$ and evolving until merger. We choose these initial conditions to ensure that the subsequent evolution remains within the eccentricity bounds of our parameter space ($e_\mathrm{max}=0.8$), and to avoid over-extrapolating our results beyond their realm of validity.\\

\begin{figure}
    \centering
    \includegraphics[width=\linewidth]{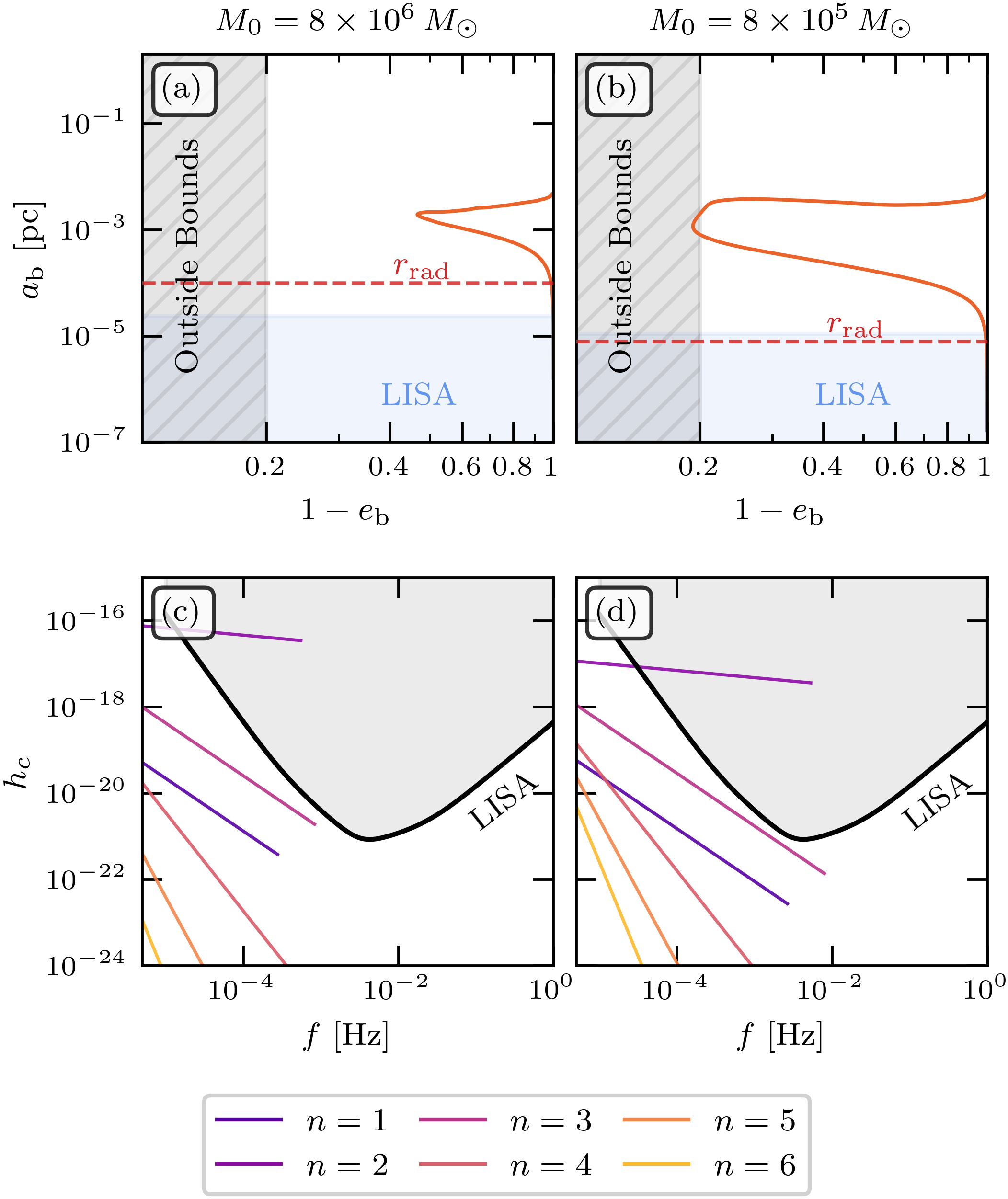}
    \caption{Massive binary evolution integrated according to circumbinary disk evolution (on the Mach 20 $\uparrow\uparrow$ branch), mass accretion and gravitational wave-emission \citep{Peters1964}. We initialise both binaries at $a_\mathrm{i} = 5\times10^{-3}\,\mathrm{pc}$ with $e_\mathrm{i}=0.0$ and we constrain the orbital eccentricity by $e_\mathrm{max}\sim 0.8$ in Fig.~\ref{fig:OrbitalEvolution}. \emph{Top}: semi-major axis and eccentricity tracks. In \emph{red} we denote the radiation-pressure dominated region and in \emph{blue} the separation at which a circular binary will enter the LISA sensitivity band. \emph{Bottom}: higher harmonics in the LISA band using the sensitivity curve given by \citet{Robson2019}.}
    \label{fig:IntegratedTrajectories}
\end{figure}

\begin{figure*}
    \centering
    \includegraphics[width=\linewidth]{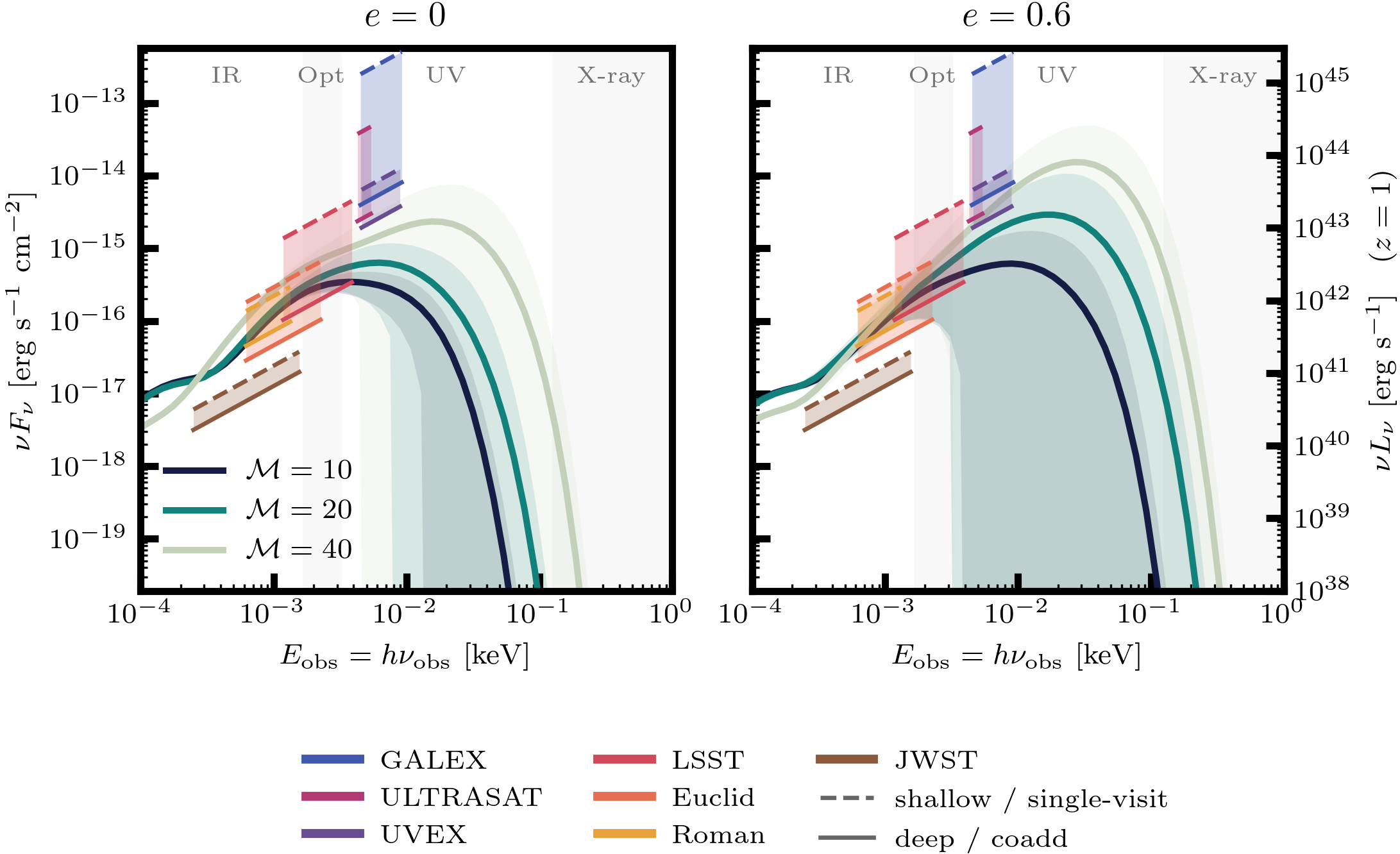}
    \caption{Detectability analysis for a catalogue of current and upcoming EM instruments. We consider the thermal emission produced by a face-on retrograde circumbinary disks at redshift $z=1$, (neglecting the subdomain and superdomain contributions of the disk).
    }
    \label{fig:SEDs}
\end{figure*}

Eccentric binaries emit gravitational waves across multiple harmonics $n$. Because a circular binary radiates solely at $n=2$, the presence of power in other harmonics is a signature of an eccentric source. The $n^\mathrm{th}$ harmonic has observer-frame frequency 
\begin{equation}
    f_n = \frac{n\Omega_\mathrm{b}}{2\pi(1+z)}.
\end{equation}
For a given harmonic, the power emitted in gravitational waves is given by
\begin{equation}
    \dot{E}_n = \frac{32}{5}\frac{G^4 M_1^2 M_2^2 (M_1+M_2)}{c^5 a^5}\,g(n,e),
\end{equation}
where $g(n,e)$ is the \cite{PetersMathews1963} enhancement factor, which distributes power across harmonics. The occupancy time at each frequency is set by the rate of change of the orbital frequency, which evolves through both semi-major axis decay and mass accretion,
\begin{equation}
    \dot{f}_n = f_n\left|-\frac{3}{2}\frac{\dot{a}}{a} + \frac{1}{2}\frac{\dot M}{M}\right|.
\end{equation}
The characteristic strain measures the power radiated in gravitational waves weighted by the occupancy time at that frequency. We compute it at each harmonic as
\begin{equation}
    h^2_{c,n} = \frac{2G}{\pi^2 c^3}\frac{\dot{E}_n}{D_\mathrm{L}^2\,\dot f_n},
\end{equation}
where $D_\mathrm{L}$ is the luminosity distance for a source at redshift $z$. We present our results in Fig.~\ref{fig:IntegratedTrajectories}.\\

In Fig.~\ref{fig:IntegratedTrajectories} we find that higher-order harmonics $n \geq 2$ are just below the detection threshold. Recall, however, that our evolution is limited to a ceiling eccentricity of $e_\mathrm{b}=0.8$. If eccentricity driving persists beyond $e_\mathrm{b}=0.8$, residual binary eccentricity may be retained within the LISA band. Regarding integration timescales, both binaries merge around $t_\mathrm{integ} = 8.2\times10^8$ years which is comparable to the total lifetime of active AGN phases \citep{Martini2004, Marconi2004}. However, over these timescales retrograde circumbinary disks may be unstable, as discussed in Section~\ref{sec:Instability}.\\

The black hole spins $\mathbf{S}_{1,2}$ are expected to align with the angular momenta of their respective minidisks through accretion, so that the three branches drive distinct spin configurations at merger. In the aligned, maximally-spinning limit, our results in Section~\ref{sec:OrbitalEvolution} predict three subpopulations of merging binaries: a $\downarrow\downarrow$ population with circular inspirals and $\chi_\mathrm{eff}=-1$ \citep{Garg2024} an $\uparrow\downarrow$ population also with circular inspirals but $\chi_\mathrm{eff}=0$ and a $\uparrow\uparrow$ population population with eccentric inspirals and $\chi_\mathrm{eff}=+1$.\\

Whether these limiting spin values are reached depends on how long the spin configuration persists relative to the accretion spin-up time. For aligned black-hole spin and minidisk angular momenta, we can estimate spin-up timescale using spin-up efficiencies from Novikov–Thorne thin-disk theory, consistent with 3D GRMHD simulations \citep{Penna2010}. Provided the average accretion rate onto the black holes is greater than $\sim 1\%$ of the Eddington accretion rate (at the initial mass), the black holes can reach high spin within $\sim 1 \, \mathrm{Gyr}$, independent of their initial spin magnitude and initial binary mass. This estimate assumes matter reaches the minidisk ISCO with maximal (circular) angular momentum. If, however the minidisks are strongly perturbed and material crosses the ISCO on significantly eccentric orbits, the spin-up timescale would lengthen.\\

\subsection{Electromagnetic Observables}
\label{sec:EM_Observables}
Accretion onto the binary powers bright electromagnetic emission with potential observability for current and upcoming surveys. In this section, we compile a list of EM instruments and assess their capabilities for detecting retrograde circumbinary disks in Fig.~\ref{fig:SEDs}.\\

In Fig.~\ref{fig:SEDs} we show time-averaged SEDs for circular ($e_\mathrm{b}=0$) and eccentric ($e_\mathrm{b}=0.6$
) binaries interacting with disks of Mach number $\mathcal{M}=10,\,20,\,40$. Solid lines show the mean SED and shaded regions the standard deviation. We overplot current and upcoming EM instruments, with their shallow (single-visit) and deep (multi-visit coadd) magnitudes, to assess which are capable of detecting the disk emission. We consider a face-on disk at redshift $z=1$, and neglect both the subdomain and superdomain emission (the unresolved emission from within the sink region and from outside the spatial domain, respectively), making our luminosities conservative lower bounds. We do not include any relativistic effects such as Doppler boosting or binary self-lensing (see later discussion) that would alter the luminosities in an observer dependent way. We expect that JWST \citep{JWST}, ROMAN \citep{ROMAN}, EUCLID \citep{EUCLID}, LSST \citep{LSST} and ULTRASAT \citep{Ultrasat} are/will be capable of detecting these binaries, which we would expect will have periodic variability on orbital timescales as illustrated in Fig.~\ref{fig:DiskEmission}.\\

We briefly discuss the prospects for observing binary self-lensing in retrograde circumbinary disks. For binaries viewed near edge-on, each black hole periodically lenses light from its companion's minidisk, producing two flares per binary orbit \citep{Dorazio2018}. Because each flare magnifies the minidisk's Doppler-boosted emission \citep{Dorazio2015, Krauth2024}, its lightcurve is not time-symmetric: as the finite-sized minidisk is lensed by the companion, its approaching (beamed, brighter) and receding (dimmed) sides are lensed at slightly different times, breaking the flare's symmetry. This asymmetry offers a potential discriminant between prograde and retrograde minidisk orientations. We expect the $\uparrow\uparrow$ state to closely resemble the prograde case. Meanwhile, the $\downarrow\downarrow$ state, with both minidisks retrograde, may instead produce flares whose shape is the mirror image of the prograde case. Finally, the $\uparrow\downarrow$ state is the most distinctive: oppositely-rotating minidisks imprint opposite relativistic beaming patterns on successive flares, breaking the flare-to-flare symmetry with no prograde analogue.\\

A second, independent discriminant arises from the minidisk sizes. As discussed in Section~\ref{sec:MinidiskStability} retrograde minidisks can exist to much larger separations than their prograde counterparts. Because binary self-lensing magnification is suppressed by finite-source effects, these extended retrograde minidisks should produce systematically weaker and broader lensing flares. Their lower outer-edge orbital velocities likewise weaken the Doppler-boost amplitude. These size-related effects may distinguish retrograde from prograde configurations independently of the sense of rotation. We highlight the importance of inclination, minidisk obscuration and optical depths, all of which may alter the self-lensing signatures which we leave for future work.

\subsection{Instability Analysis}
\label{sec:Instability}
There are two stability criteria relevant for retrograde binary disk-interactions. When one of these criteria fails, the binary will reorient itself into a prograde configuration and the subsequent orbital evolution will not be as described by Section~\ref{sec:OrbitalEvolution}. In this subsection we provide a brief discussion on these criteria relevant for binaries retaining high eccentricity during the gravitational wave-driven inspiral.\\

The first condition is a simple `flip', in which the binary's orbital angular momentum $L=\frac{1}{4} M\Omega_\mathrm{b}a_\mathrm{b}^2\sqrt{1-e_\mathrm{b}^2}$ is driven through zero and reverses sign. Since $L$ decreases as the binary evolves toward higher eccentricity, we estimate the threshold for reversal by relating the per-orbit extraction $\Delta L=2\pi\Omega^{-1}l_0\dot{M}a^2\Omega$ with the binary orbital angular momentum. We consider the retrograde state to be stable when $L>|\Delta L|$, giving,
\begin{equation}
    1-e^2> 7.8\times10^{-15}\left(\frac{a}{a_\mathrm{b}}\right)^3 \left(\frac{M}{M_\mathrm{b}}\right)^{-1} l_0^2.
    \label{FlipStable}
\end{equation}
If we assume that the torque parameter $l_0$ is of order unity (although potentially larger for highly eccentric binaries and/or higher Mach disks, \citealt{Tiede2025}), Eq.~\ref{FlipStable} predicts stable retrograde binary-disk interactions up to extremely large orbital eccentricities, \eg when $e\geq 1-10^{-7}$. The second instability is when the disk's angular momentum dominates the binary's angular momentum \citep[$L_\mathrm{b}\leq0.5 L_\mathrm{d}$, see][]{King2005, Lodato2009, NixonKingPringle2011, Nixon12_StableRetrograde, Garg2024}, at which point the retrograde state is no longer stable. The misaligned disk undergoes differential precession, which builds a warp. Viscous dissipation of this warp drives both the disk and, through back-reaction, the binary toward their common angular-momentum axis. Once the disk dominates the angular-momentum budget, that axis lies on the prograde side, so the binary reorients into a prograde configuration. For an $\alpha$-viscosity disk, the viscosity $\nu$ can be given as, 
\begin{equation}
    \nu = \frac{\alpha a_\mathrm{b}^2\Omega_\mathrm{b}}{\mathcal{M}_\mathrm{a}^2}\left(\frac{r}{a_\mathrm{b}}\right)^{3/5}.
\end{equation}
We define the time-dependent warp radius $r_\mathrm{w}$ as the extent over which the disk has viscously evolved by time $t$, obtained by integrating the viscous radial velocity $dr/dt=-3\nu/2r$ over the range $r\in [0,r_\mathrm{w}]$, giving
\begin{equation}
    r_\mathrm{w} = \left(\frac{21}{10}\frac{\alpha\gamma a_\mathrm{b}^{9/10} t}{\sqrt{GM}}\frac{P}{\Sigma}\right)^{5/7},
    \label{WarpRadius}
\end{equation}
where we have assumed that the binary mass does not change appreciably during the propagation of the warp, negating the need for a time-dependent \cite{SS73} disk model. For simplicity, we assume that the vertically-acting kinematic viscosity is comparable to the lateral viscosity $\nu_z\sim\nu$ as appropriate for a large-amplitude warp, although in the small-warp regime the vertical viscosity may be enhanced up to $\nu_z\sim \nu/2\alpha^2$\citep{Papaloizou1983, Ogilvie1999}. For simplicity, we will assume that the binary's change in angular momentum is primarily due to its increasing eccentricity $a_\mathrm{f}/a_\mathrm{i},~M_\mathrm{i}/M_\mathrm{f}\ll (1-e_\mathrm{f}^2)^{-1/2}$. The total disk angular momentum contained within the warp radius is given by,
\begin{align}
    \frac{L_\mathrm{d}}{Ma^2\Omega} \approx 3.2\times10^{-11}&\left(\frac{\alpha}{0.1}\right)^\frac{2}{7}\left(\frac{\gamma}{5/3}\right)^{\frac{19}{14}}\left(\frac{M}{M_\mathrm{b}}\right)^{-\frac{35}{84}}\nonumber\\
    \times & \left(\frac{\lambda_\mathrm{Edd}}{0.1}\right)^\frac{8}{7}\left(\frac{t}{\mathrm{yr}}\right)^\frac{19}{14}\left(\frac{a}{a_\mathrm{b}}\right)^{\frac{129}{84}}.
\end{align}
By invoking the tilt criterion, $L_\mathrm{b} < 0.5L_\mathrm{d}$ \citep{NixonKingPringle2011}, we find that the limiting eccentricity is given by,
\begin{align}
    1-e^2 > 0.68&\left(\frac{t}{10^7\mathrm{yr}}\right)^\frac{19}{7}\left(\frac{\alpha}{0.1}\right)^\frac{4}{7}\left(\frac{\gamma}{5/3}\right)^{\frac{19}{7}} \\ \times&\left(\frac{M}{M_\mathrm{b}}\right)^{-\frac{35}{42}}
     \left(\frac{\lambda_\mathrm{Edd}}{0.1}\right)^\frac{16}{7}\left(\frac{a}{a_\mathrm{b}}\right)^{\frac{129}{42}}.
\end{align}
Thus after $10^7$ years, the angular momentum of the disk enclosed within the warp radius will exceed $2\, L_\mathrm{b}$ if the binary has eccentricity above $e\geq 0.56$. Consequently, the system will re-orientate into a prograde configuration after this timescale.\\

For an inspiralling binary, tilting is inevitable as $L_\mathrm{b}\to~0$. However, for the binary to reach large eccentricities prior to merger, it may be favourable for tilting to occur at as small a binary semi-major axis as possible. To achieve this, the binary must evolve quickly ($|\dot{a}/a|, |\dot{e}| \gg 1$) to avoid the warp radius propagating quickly and dominating the angular momentum budget at early times. We propose high Mach number disks may achieve this quick orbital evolution \citep{Tiede2025, Dittmann2026, Tiede2026}, enabling them to stay stable for longer and reach large eccentricities prior to merger, but we reserve such an analysis for future studies.\\

We note that similar alignment processes apply to individual minidisks around a spinning black hole. \citet{King2005} show that the minidisk can either align or counter-align with the black hole spin depending on the ratio of disk to spin angular momentum, mediated by Lense-Thirring precession rather than the binary-induced precession considered above. This is directly relevant to the binary reorientation discussed in Section~\ref{sec:LISA_Observables}.

\subsection{Caveats}
The results presented here rest on a number of simplifying assumptions. Our simulated disk parameters imply hyper-Eddington accretion rates ($\dot{m}/\dot{m}_\mathrm{Edd} \gg 0.1$), and to obtain physically motivated electromagnetic emission estimates we apply an approximate disk remapping, which likely misrepresents contributions from shocks and tidal heating. We also neglect the tilting instability \cite{Nixon12_StableRetrograde} to which retrograde circumbinary disks are prone; we estimate the tilting timescale to be $\sim 10^7$~yr, shortening at higher binary eccentricities, which is particularly relevant for the eccentric inspiralling LISA sources identified in Section~\ref{sec:Results}. In our simulations we model both masses as softened Newtonian potentials, justified by binary separations far exceeding the gravitational radius ($a_\mathrm{b} = 2530\,r_\mathrm{G}$). Strong-field gravity effects are likely confined within the minidisks, with negligible influence on the large-scale circumbinary disk dynamics. The disk itself is treated as an ideal monatomic gas with an $\alpha$ viscosity prescription; radiation pressure, magnetic fields, and turbulence are all neglected, each of which could modify the effective viscosity, shock structure, and cavity thermodynamics. Finally, by solving the vertically integrated equations we suppress all out-of-plane dynamics, including disk warping. \cite{Roedig2014} and \cite{Overton2024} find that retrograde minidisks may be unstable to out-of-plane tilting in 3D, suggesting this is a potentially significant limitation of our 2D framework.

\subsection{Outlook}
The results presented in this study hold potential astrophysical applications for SMBHBs interacting with circumbinary accretion disks. We consider a number of future research directions building from the results found here.\\

The initial conditions of the disk are the primary factor determining which spin state is realised, although we have not explored this systematically. The mixed $\uparrow\downarrow$ state also remains uncharacterised, due to the computational demands of eccentricity sweeps at Mach 40. Fully mapping the conditions under which each state is common (and the associated orbital evolution) is one of the most astrophysically relevant questions arising from this work. Dedicated simulations spanning a range of mass ratios, Mach numbers, and initial disk conditions would clarify whether the $\uparrow\uparrow$ and $\downarrow\downarrow$ states are equally accessible, or whether one is preferred under realistic astrophysical conditions. Additionally, for unequal mass binaries (where the two minidisks are distinguishable), the number of spin states might be expected to extend to $4$, offering richer binary-disk dynamics.\\

Simulations of retrograde disks involving astrophysically realistic Mach numbers is a clear priority for several reasons. First, the race between orbital inspiral and disk tilting will dictate whether or not binaries will reach the gravitational wave regime via retrograde evolution alone, or whether a flip to prograde orientation intervenes before gravitational wave-driven inspiral. Second, higher-Mach simulations will be required to capture the thin-stream collisions in the disk, which are expected to dominate the high-frequency (UV and X-ray) emission. This emission may have important observable implications, potentially connecting to nuclear transients such as quasi-periodic eruptions \citep{Miniutti2019} or quasi-periodic oscillations \citep[e.g.][for the analogous stellar-mass phenomenon]{Remillard2006}.\\

Upcoming multimessenger searches promise the first direct detections of SMBHBs through combined electromagnetic and gravitational wave observations. Discerning the distinct predictions of prograde and retrograde disk populations will aid in the identification of binary systems and offer insight into the landscape of environmental interactions experienced by massive black hole binaries. Incorporating retrograde disk-driven evolution into SMBHB population synthesis models would also be a valuable step towards interpreting the nHz gravitational wave background detected by current pulsar timing arrays \citep{Agazie2023, EPTA2023, Reardon2023}.

\section{Conclusions}
\label{sec:Conclusions}
In this paper we present hydrodynamical simulations of retrograde circumbinary disks around eccentric SMBHBs. We summarise our main findings below.\\

\textbf{(1) \emph{Minidisk Spin States}}: We identify multiple stable disk configurations at fixed Mach number and binary eccentricity, distinguished by the spins of their minidisks: both retrograde ($\downarrow\downarrow$), both prograde ($\uparrow\uparrow$) and a mixed state ($\uparrow\downarrow$). For an equal-mass binary these form a triad of solutions selected by the initial conditions. Each imparts a different torque on the binary, leading to distinct orbital evolution (Fig.~\ref{fig:OrbitalEvolution}), electromagnetic emission (Fig.~\ref{fig:DiskEmission}), and potentially different black-hole spin configurations.

\pagebreak

\textbf{(2) \emph{Gravitational Wave Observables}}: Binaries in the $\downarrow\downarrow$ and $\uparrow\downarrow$ states evolve toward circular inspiral, whereas those in the $\uparrow\uparrow$ state gain eccentricity until gravitational-wave emission drives coalescence. Integrating these tracks, we find residual eccentricity may persist into the LISA band (Fig.~\ref{fig:IntegratedTrajectories}). The minidisk spin states may further imprint distinct effective spin $\chi_\mathrm{eff}$ on the binary.\\ 

\textbf{(3) \emph{Electromagnetic Observables}}: Under conservative assumptions, the emission from retrograde disks is detectable by current and upcoming surveys (\eg JWST, Roman, Euclid, LSST, and ULTRASAT). Across the rest-frame X-ray, UV, optical and Infrared bands, we predict lightcurve periodicities at $\Omega_\mathrm{b}$ and $0.5\Omega_\mathrm{b}$ -- likely a signature of retrograde circumbinary disks. We further highlight potentially distinctive self-lensing signatures.\\

\textbf{(4) \emph{Angular Momentum Transport}}: The time- and azimuthally-averaged surface density and pressure profiles agree well with the 1D model of \cite{Rafikov2013}. For negative torque parameter, this yields a relation between the cavity radius and the binary torque. Within the cavity, angular momentum transport is dominated by Reynolds stresses rather than viscosity.\\

\section*{Acknowledgements}

We thank Henry Whitehead and Kevin Park for useful discussions and feedback on the manuscript. The Tycho supercomputer hosted at the SCIENCE HPC center at the University of Copenhagen was used in this work. This research was supported by the Scientific Service Units (SSU) of IST Austria through resources provided by Scientific Computing (SciComp), the Life Science Facility (LSF) and the XYZ Facility (xxx). The Anvil supercomputer \cite{Anvil} hosted at Purdue University was used in this work.

Large language models (LLMs) were used to assist with manuscript formatting and presentation. The authors take full responsibility for the content and conclusions of this work.

This work was supported by the European Union’s Horizon 2023 research and innovation program under Marie Sklodowska-Curie grant agreement No. 101148364, and by Sapere Aude Starting grant No. 121587 through the Danish Independent Research Fund. It was additionally supported by NSF AAG No. 2511544...

This work made use of the following software packages: \texttt{matplotlib} \citep{Hunter:2007}, \texttt{numpy} \citep{numpy}, \texttt{python} \citep{python}, and \texttt{scipy} \citep{2020SciPy-NMeth,scipy_20383145}. Software citation information aggregated using \texttt{\href{https://www.tomwagg.com/software-citation-station/}{The Software Citation Station}} \citep{software-citation-station-paper,software-citation-station-zenodo}.

\section*{Data Availability}
The data underlying this article will be shared on reasonable request to the corresponding author.

\bibliographystyle{mnras}
\bibliography{bibliography}

@software{Zrake24,
       author = {{Zrake}, Jonathan and {MacFadyen}, Andrew},
        title = "{Sailfish: GPU-accelerated grid-based astrophysics gas dynamics code}",
 howpublished = {Astrophysics Source Code Library, record ascl:2408.004},
         year = 2024,
        month = aug,
          eid = {ascl:2408.004},
archivePrefix = {ascl},
       eprint = {2408.004},
       adsurl = {https://ui.adsabs.harvard.edu/abs/2024ascl.soft08004Z}
}

@ARTICLE{DOrazio2021,
       author = {{D'Orazio}, Daniel J. and {Duffell}, Paul C.},
        title = "{Orbital Evolution of Equal-mass Eccentric Binaries due to a Gas Disk: Eccentric Inspirals and Circular Outspirals}",
      journal = {\apjl},
         year = 2021,
        month = jun,
       volume = {914},
       number = {1},
          eid = {L21},
        pages = {L21},
          doi = {10.3847/2041-8213/ac0621},
archivePrefix = {arXiv},
       eprint = {2103.09251},
 primaryClass = {astro-ph.HE},
       adsurl = {https://ui.adsabs.harvard.edu/abs/2021ApJ...914L..21D}
}

@ARTICLE{Tiede2024,
       author = {{Tiede}, Christopher and {D'Orazio}, Daniel J.},
        title = "{Eccentric binaries in retrograde discs}",
      journal = {\mnras},
         year = 2024,
        month = jan,
       volume = {527},
       number = {3},
        pages = {6021-6037},
          doi = {10.1093/mnras/stad3551},
archivePrefix = {arXiv},
       eprint = {2307.03775},
 primaryClass = {astro-ph.GA},
       adsurl = {https://ui.adsabs.harvard.edu/abs/2024MNRAS.527.6021T}
}

@ARTICLE{AmaroSeoane23,
       author = {{Amaro-Seoane}, Pau and {Andrews}, Jeff and {Arca Sedda}, Manuel and {Askar}, Abbas and {Baghi}, Quentin and {Balasov}, Razvan and {Bartos}, Imre and {Bavera}, Simone S. and {Bellovary}, Jillian and {Berry}, Christopher P.~L. and {Berti}, Emanuele and {Bianchi}, Stefano and {Blecha}, Laura and {Blondin}, St{\'e}phane and {Bogdanovi{\'c}}, Tamara and {Boissier}, Samuel and {Bonetti}, Matteo and {Bonoli}, Silvia and {Bortolas}, Elisa and {Breivik}, Katelyn and {Capelo}, Pedro R. and {Caramete}, Laurentiu and {Cattorini}, Federico and {Charisi}, Maria and {Chaty}, Sylvain and {Chen}, Xian and {Chru{\'s}li{\'n}ska}, Martyna and {Chua}, Alvin J.~K. and {Church}, Ross and {Colpi}, Monica and {D'Orazio}, Daniel and {Danielski}, Camilla and {Davies}, Melvyn B. and {Dayal}, Pratika and {De Rosa}, Alessandra and {Derdzinski}, Andrea and {Destounis}, Kyriakos and {Dotti}, Massimo and {Du{\c{t}}an}, Ioana and {Dvorkin}, Irina and {Fabj}, Gaia and {Foglizzo}, Thierry and {Ford}, Saavik and {Fouvry}, Jean-Baptiste and {Franchini}, Alessia and {Fragos}, Tassos and {Fryer}, Chris and {Gaspari}, Massimo and {Gerosa}, Davide and {Graziani}, Luca and {Groot}, Paul and {Habouzit}, Melanie and {Haggard}, Daryl and {Haiman}, Zoltan and {Han}, Wen-Biao and {Istrate}, Alina and {Johansson}, Peter H. and {Khan}, Fazeel Mahmood and {Kimpson}, Tomas and {Kokkotas}, Kostas and {Kong}, Albert and {Korol}, Valeriya and {Kremer}, Kyle and {Kupfer}, Thomas and {Lamberts}, Astrid and {Larson}, Shane and {Lau}, Mike and {Liu}, Dongliang and {Lloyd-Ronning}, Nicole and {Lodato}, Giuseppe and {Lupi}, Alessandro and {Ma}, Chung-Pei and {Maccarone}, Tomas and {Mandel}, Ilya and {Mangiagli}, Alberto and {Mapelli}, Michela and {Mathis}, St{\'e}phane and {Mayer}, Lucio and {McGee}, Sean and {McKernan}, Berry and {Miller}, M. Coleman and {Mota}, David F. and {Mumpower}, Matthew and {Nasim}, Syeda S. and {Nelemans}, Gijs and {Noble}, Scott and {Pacucci}, Fabio and {Panessa}, Francesca and {Paschalidis}, Vasileios and {Pfister}, Hugo and {Porquet}, Delphine and {Quenby}, John and {Ricarte}, Angelo and {R{\"o}pke}, Friedrich K. and {Regan}, John and {Rosswog}, Stephan and {Ruiter}, Ashley and {Ruiz}, Milton and {Runnoe}, Jessie and {Schneider}, Raffaella and {Schnittman}, Jeremy and {Secunda}, Amy and {Sesana}, Alberto and {Seto}, Naoki and {Shao}, Lijing and {Shapiro}, Stuart and {Sopuerta}, Carlos and {Stone}, Nicholas C. and {Suvorov}, Arthur and {Tamanini}, Nicola and {Tamfal}, Tomas and {Tauris}, Thomas and {Temmink}, Karel and {Tomsick}, John and {Toonen}, Silvia and {Torres-Orjuela}, Alejandro and {Toscani}, Martina and {Tsokaros}, Antonios and {Unal}, Caner and {V{\'a}zquez-Aceves}, Ver{\'o}nica and {Valiante}, Rosa and {van Putten}, Maurice and {van Roestel}, Jan and {Vignali}, Christian and {Volonteri}, Marta and {Wu}, Kinwah and {Younsi}, Ziri and {Yu}, Shenghua and {Zane}, Silvia and {Zwick}, Lorenz and {Antonini}, Fabio and {Baibhav}, Vishal and {Barausse}, Enrico and {Bonilla Rivera}, Alexander and {Branchesi}, Marica and {Branduardi-Raymont}, Graziella and {Burdge}, Kevin and {Chakraborty}, Srija and {Cuadra}, Jorge and {Dage}, Kristen and {Davis}, Benjamin and {de Mink}, Selma E. and {Decarli}, Roberto and {Doneva}, Daniela and {Escoffier}, Stephanie and {Gandhi}, Poshak and {Haardt}, Francesco and {Lousto}, Carlos O. and {Nissanke}, Samaya and {Nordhaus}, Jason and {O'Shaughnessy}, Richard and {Portegies Zwart}, Simon and {Pound}, Adam and {Schussler}, Fabian and {Sergijenko}, Olga and {Spallicci}, Alessandro and {Vernieri}, Daniele and {Vigna-G{\'o}mez}, Alejandro},
        title = "{Astrophysics with the Laser Interferometer Space Antenna}",
      journal = {Living Reviews in Relativity},
         year = 2023,
        month = dec,
       volume = {26},
       number = {1},
          eid = {2},
        pages = {2},
          doi = {10.1007/s41114-022-00041-y},
archivePrefix = {arXiv},
       eprint = {2203.06016},
 primaryClass = {gr-qc},
       adsurl = {https://ui.adsabs.harvard.edu/abs/2023LRR....26....2A}
}

@ARTICLE{TinQian2021,
       author = {{Gong}, Yungui and {Luo}, Jun and {Wang}, Bin},
        title = "{Concepts and status of Chinese space gravitational wave detection projects}",
      journal = {Nature Astronomy},
         year = 2021,
        month = sep,
       volume = {5},
        pages = {881-889},
          doi = {10.1038/s41550-021-01480-3},
archivePrefix = {arXiv},
       eprint = {2109.07442},
 primaryClass = {astro-ph.IM},
       adsurl = {https://ui.adsabs.harvard.edu/abs/2021NatAs...5..881G}
}

@ARTICLE{NanoGrav2023,
       author = {{Agazie}, Gabriella and {Anumarlapudi}, Akash and {Archibald}, Anne M. and {Arzoumanian}, Zaven and {Baker}, Paul T. and {B{\'e}csy}, Bence and {Blecha}, Laura and {Brazier}, Adam and {Brook}, Paul R. and {Burke-Spolaor}, Sarah and {Burnette}, Rand and {Case}, Robin and {Charisi}, Maria and {Chatterjee}, Shami and {Chatziioannou}, Katerina and {Cheeseboro}, Belinda D. and {Chen}, Siyuan and {Cohen}, Tyler and {Cordes}, James M. and {Cornish}, Neil J. and {Crawford}, Fronefield and {Cromartie}, H. Thankful and {Crowter}, Kathryn and {Cutler}, Curt J. and {Decesar}, Megan E. and {Degan}, Dallas and {Demorest}, Paul B. and {Deng}, Heling and {Dolch}, Timothy and {Drachler}, Brendan and {Ellis}, Justin A. and {Ferrara}, Elizabeth C. and {Fiore}, William and {Fonseca}, Emmanuel and {Freedman}, Gabriel E. and {Garver-Daniels}, Nate and {Gentile}, Peter A. and {Gersbach}, Kyle A. and {Glaser}, Joseph and {Good}, Deborah C. and {G{\"u}ltekin}, Kayhan and {Hazboun}, Jeffrey S. and {Hourihane}, Sophie and {Islo}, Kristina and {Jennings}, Ross J. and {Johnson}, Aaron D. and {Jones}, Megan L. and {Kaiser}, Andrew R. and {Kaplan}, David L. and {Kelley}, Luke Zoltan and {Kerr}, Matthew and {Key}, Joey S. and {Klein}, Tonia C. and {Laal}, Nima and {Lam}, Michael T. and {Lamb}, William G. and {Lazio}, T. Joseph W. and {Lewandowska}, Natalia and {Littenberg}, Tyson B. and {Liu}, Tingting and {Lommen}, Andrea and {Lorimer}, Duncan R. and {Luo}, Jing and {Lynch}, Ryan S. and {Ma}, Chung-Pei and {Madison}, Dustin R. and {Mattson}, Margaret A. and {McEwen}, Alexander and {McKee}, James W. and {McLaughlin}, Maura A. and {McMann}, Natasha and {Meyers}, Bradley W. and {Meyers}, Patrick M. and {Mingarelli}, Chiara M.~F. and {Mitridate}, Andrea and {Natarajan}, Priyamvada and {Ng}, Cherry and {Nice}, David J. and {Ocker}, Stella Koch and {Olum}, Ken D. and {Pennucci}, Timothy T. and {Perera}, Benetge B.~P. and {Petrov}, Polina and {Pol}, Nihan S. and {Radovan}, Henri A. and {Ransom}, Scott M. and {Ray}, Paul S. and {Romano}, Joseph D. and {Sardesai}, Shashwat C. and {Schmiedekamp}, Ann and {Schmiedekamp}, Carl and {Schmitz}, Kai and {Schult}, Levi and {Shapiro-Albert}, Brent J. and {Siemens}, Xavier and {Simon}, Joseph and {Siwek}, Magdalena S. and {Stairs}, Ingrid H. and {Stinebring}, Daniel R. and {Stovall}, Kevin and {Sun}, Jerry P. and {Susobhanan}, Abhimanyu and {Swiggum}, Joseph K. and {Taylor}, Jacob and {Taylor}, Stephen R. and {Turner}, Jacob E. and {Unal}, Caner and {Vallisneri}, Michele and {van Haasteren}, Rutger and {Vigeland}, Sarah J. and {Wahl}, Haley M. and {Wang}, Qiaohong and {Witt}, Caitlin A. and {Young}, Olivia and {Nanograv Collaboration}},
        title = "{The NANOGrav 15 yr Data Set: Evidence for a Gravitational-wave Background}",
      journal = {\apjl},
         year = 2023,
        month = jul,
       volume = {951},
       number = {1},
          eid = {L8},
        pages = {L8},
          doi = {10.3847/2041-8213/acdac6},
archivePrefix = {arXiv},
       eprint = {2306.16213},
 primaryClass = {astro-ph.HE},
       adsurl = {https://ui.adsabs.harvard.edu/abs/2023ApJ...951L...8A}
}

@ARTICLE{LSST19,
       author = {{Ivezi{\'c}}, {\v{Z}}eljko and {Kahn}, Steven M. and {Tyson}, J. Anthony and {Abel}, Bob and {Acosta}, Emily and {Allsman}, Robyn and {Alonso}, David and {AlSayyad}, Yusra and {Anderson}, Scott F. and {Andrew}, John and {Angel}, James Roger P. and {Angeli}, George Z. and {Ansari}, Reza and {Antilogus}, Pierre and {Araujo}, Constanza and {Armstrong}, Robert and {Arndt}, Kirk T. and {Astier}, Pierre and {Aubourg}, {\'E}ric and {Auza}, Nicole and {Axelrod}, Tim S. and {Bard}, Deborah J. and {Barr}, Jeff D. and {Barrau}, Aurelian and {Bartlett}, James G. and {Bauer}, Amanda E. and {Bauman}, Brian J. and {Baumont}, Sylvain and {Bechtol}, Ellen and {Bechtol}, Keith and {Becker}, Andrew C. and {Becla}, Jacek and {Beldica}, Cristina and {Bellavia}, Steve and {Bianco}, Federica B. and {Biswas}, Rahul and {Blanc}, Guillaume and {Blazek}, Jonathan and {Blandford}, Roger D. and {Bloom}, Josh S. and {Bogart}, Joanne and {Bond}, Tim W. and {Booth}, Michael T. and {Borgland}, Anders W. and {Borne}, Kirk and {Bosch}, James F. and {Boutigny}, Dominique and {Brackett}, Craig A. and {Bradshaw}, Andrew and {Brandt}, William Nielsen and {Brown}, Michael E. and {Bullock}, James S. and {Burchat}, Patricia and {Burke}, David L. and {Cagnoli}, Gianpietro and {Calabrese}, Daniel and {Callahan}, Shawn and {Callen}, Alice L. and {Carlin}, Jeffrey L. and {Carlson}, Erin L. and {Chandrasekharan}, Srinivasan and {Charles-Emerson}, Glenaver and {Chesley}, Steve and {Cheu}, Elliott C. and {Chiang}, Hsin-Fang and {Chiang}, James and {Chirino}, Carol and {Chow}, Derek and {Ciardi}, David R. and {Claver}, Charles F. and {Cohen-Tanugi}, Johann and {Cockrum}, Joseph J. and {Coles}, Rebecca and {Connolly}, Andrew J. and {Cook}, Kem H. and {Cooray}, Asantha and {Covey}, Kevin R. and {Cribbs}, Chris and {Cui}, Wei and {Cutri}, Roc and {Daly}, Philip N. and {Daniel}, Scott F. and {Daruich}, Felipe and {Daubard}, Guillaume and {Daues}, Greg and {Dawson}, William and {Delgado}, Francisco and {Dellapenna}, Alfred and {de Peyster}, Robert and {de Val-Borro}, Miguel and {Digel}, Seth W. and {Doherty}, Peter and {Dubois}, Richard and {Dubois-Felsmann}, Gregory P. and {Durech}, Josef and {Economou}, Frossie and {Eifler}, Tim and {Eracleous}, Michael and {Emmons}, Benjamin L. and {Fausti Neto}, Angelo and {Ferguson}, Henry and {Figueroa}, Enrique and {Fisher-Levine}, Merlin and {Focke}, Warren and {Foss}, Michael D. and {Frank}, James and {Freemon}, Michael D. and {Gangler}, Emmanuel and {Gawiser}, Eric and {Geary}, John C. and {Gee}, Perry and {Geha}, Marla and {Gessner}, Charles J.~B. and {Gibson}, Robert R. and {Gilmore}, D. Kirk and {Glanzman}, Thomas and {Glick}, William and {Goldina}, Tatiana and {Goldstein}, Daniel A. and {Goodenow}, Iain and {Graham}, Melissa L. and {Gressler}, William J. and {Gris}, Philippe and {Guy}, Leanne P. and {Guyonnet}, Augustin and {Haller}, Gunther and {Harris}, Ron and {Hascall}, Patrick A. and {Haupt}, Justine and {Hernandez}, Fabio and {Herrmann}, Sven and {Hileman}, Edward and {Hoblitt}, Joshua and {Hodgson}, John A. and {Hogan}, Craig and {Howard}, James D. and {Huang}, Dajun and {Huffer}, Michael E. and {Ingraham}, Patrick and {Innes}, Walter R. and {Jacoby}, Suzanne H. and {Jain}, Bhuvnesh and {Jammes}, Fabrice and {Jee}, M. James and {Jenness}, Tim and {Jernigan}, Garrett and {Jevremovi{\'c}}, Darko and {Johns}, Kenneth and {Johnson}, Anthony S. and {Johnson}, Margaret W.~G. and {Jones}, R. Lynne and {Juramy-Gilles}, Claire and {Juri{\'c}}, Mario and {Kalirai}, Jason S. and {Kallivayalil}, Nitya J. and {Kalmbach}, Bryce and {Kantor}, Jeffrey P. and {Karst}, Pierre and {Kasliwal}, Mansi M. and {Kelly}, Heather and {Kessler}, Richard and {Kinnison}, Veronica and {Kirkby}, David and {Knox}, Lloyd and {Kotov}, Ivan V. and {Krabbendam}, Victor L. and {Krughoff}, K. Simon and {Kub{\'a}nek}, Petr and {Kuczewski}, John and {Kulkarni}, Shri and {Ku}, John and {Kurita}, Nadine R. and {Lage}, Craig S. and {Lambert}, Ron and {Lange}, Travis and {Langton}, J. Brian and {Le Guillou}, Laurent and {Levine}, Deborah and {Liang}, Ming and {Lim}, Kian-Tat and {Lintott}, Chris J. and {Long}, Kevin E. and {Lopez}, Margaux and {Lotz}, Paul J. and {Lupton}, Robert H. and {Lust}, Nate B. and {MacArthur}, Lauren A. and {Mahabal}, Ashish and {Mandelbaum}, Rachel and {Markiewicz}, Thomas W. and {Marsh}, Darren S. and {Marshall}, Philip J. and {Marshall}, Stuart and {May}, Morgan and {McKercher}, Robert and {McQueen}, Michelle and {Meyers}, Joshua and {Migliore}, Myriam and {Miller}, Michelle and {Mills}, David J.},
        title = "{LSST: From Science Drivers to Reference Design and Anticipated Data Products}",
      journal = {\apj},
         year = 2019,
        month = mar,
       volume = {873},
       number = {2},
          eid = {111},
        pages = {111},
          doi = {10.3847/1538-4357/ab042c},
archivePrefix = {arXiv},
       eprint = {0805.2366},
 primaryClass = {astro-ph},
       adsurl = {https://ui.adsabs.harvard.edu/abs/2019ApJ...873..111I}
}

@ARTICLE{ZTF2019,
       author = {{Bellm}, Eric C. and {Kulkarni}, Shrinivas R. and {Graham}, Matthew J. and {Dekany}, Richard and {Smith}, Roger M. and {Riddle}, Reed and {Masci}, Frank J. and {Helou}, George and {Prince}, Thomas A. and {Adams}, Scott M. and {Barbarino}, C. and {Barlow}, Tom and {Bauer}, James and {Beck}, Ron and {Belicki}, Justin and {Biswas}, Rahul and {Blagorodnova}, Nadejda and {Bodewits}, Dennis and {Bolin}, Bryce and {Brinnel}, Valery and {Brooke}, Tim and {Bue}, Brian and {Bulla}, Mattia and {Burruss}, Rick and {Cenko}, S. Bradley and {Chang}, Chan-Kao and {Connolly}, Andrew and {Coughlin}, Michael and {Cromer}, John and {Cunningham}, Virginia and {De}, Kishalay and {Delacroix}, Alex and {Desai}, Vandana and {Duev}, Dmitry A. and {Eadie}, Gwendolyn and {Farnham}, Tony L. and {Feeney}, Michael and {Feindt}, Ulrich and {Flynn}, David and {Franckowiak}, Anna and {Frederick}, S. and {Fremling}, C. and {Gal-Yam}, Avishay and {Gezari}, Suvi and {Giomi}, Matteo and {Goldstein}, Daniel A. and {Golkhou}, V. Zach and {Goobar}, Ariel and {Groom}, Steven and {Hacopians}, Eugean and {Hale}, David and {Henning}, John and {Ho}, Anna Y.~Q. and {Hover}, David and {Howell}, Justin and {Hung}, Tiara and {Huppenkothen}, Daniela and {Imel}, David and {Ip}, Wing-Huen and {Ivezi{\'c}}, {\v{Z}}eljko and {Jackson}, Edward and {Jones}, Lynne and {Juric}, Mario and {Kasliwal}, Mansi M. and {Kaspi}, S. and {Kaye}, Stephen and {Kelley}, Michael S.~P. and {Kowalski}, Marek and {Kramer}, Emily and {Kupfer}, Thomas and {Landry}, Walter and {Laher}, Russ R. and {Lee}, Chien-De and {Lin}, Hsing Wen and {Lin}, Zhong-Yi and {Lunnan}, Ragnhild and {Giomi}, Matteo and {Mahabal}, Ashish and {Mao}, Peter and {Miller}, Adam A. and {Monkewitz}, Serge and {Murphy}, Patrick and {Ngeow}, Chow-Choong and {Nordin}, Jakob and {Nugent}, Peter and {Ofek}, Eran and {Patterson}, Maria T. and {Penprase}, Bryan and {Porter}, Michael and {Rauch}, Ludwig and {Rebbapragada}, Umaa and {Reiley}, Dan and {Rigault}, Mickael and {Rodriguez}, Hector and {van Roestel}, Jan and {Rusholme}, Ben and {van Santen}, Jakob and {Schulze}, S. and {Shupe}, David L. and {Singer}, Leo P. and {Soumagnac}, Maayane T. and {Stein}, Robert and {Surace}, Jason and {Sollerman}, Jesper and {Szkody}, Paula and {Taddia}, F. and {Terek}, Scott and {Van Sistine}, Angela and {van Velzen}, Sjoert and {Vestrand}, W. Thomas and {Walters}, Richard and {Ward}, Charlotte and {Ye}, Quan-Zhi and {Yu}, Po-Chieh and {Yan}, Lin and {Zolkower}, Jeffry},
        title = "{The Zwicky Transient Facility: System Overview, Performance, and First Results}",
      journal = {\pasp},
         year = 2019,
        month = jan,
       volume = {131},
       number = {995},
        pages = {018002},
          doi = {10.1088/1538-3873/aaecbe},
archivePrefix = {arXiv},
       eprint = {1902.01932},
 primaryClass = {astro-ph.IM},
       adsurl = {https://ui.adsabs.harvard.edu/abs/2019PASP..131a8002B}
}

@ARTICLE{Nixon12_StableRetrograde,
       author = {{Nixon}, Christopher J.},
        title = "{Stable counteralignment of a circumbinary disc}",
      journal = {\mnras},
         year = 2012,
        month = jul,
       volume = {423},
       number = {3},
        pages = {2597-2600},
          doi = {10.1111/j.1365-2966.2012.21072.x},
archivePrefix = {arXiv},
       eprint = {1204.4185},
 primaryClass = {astro-ph.HE},
       adsurl = {https://ui.adsabs.harvard.edu/abs/2012MNRAS.423.2597N}
}

@ARTICLE{Martin2017,
       author = {{Martin}, Rebecca G. and {Lubow}, Stephen H.},
        title = "{Polar Alignment of a Protoplanetary Disk around an Eccentric Binary}",
      journal = {\apjl},
         year = 2017,
        month = feb,
       volume = {835},
       number = {2},
          eid = {L28},
        pages = {L28},
          doi = {10.3847/2041-8213/835/2/L28},
archivePrefix = {arXiv},
       eprint = {1702.00545},
 primaryClass = {astro-ph.EP},
       adsurl = {https://ui.adsabs.harvard.edu/abs/2017ApJ...835L..28M}
}

@ARTICLE{Lubow2018,
       author = {{Lubow}, Stephen H. and {Martin}, Rebecca G.},
        title = "{Linear analysis of the evolution of nearly polar low-mass circumbinary discs}",
      journal = {\mnras},
         year = 2018,
        month = jan,
       volume = {473},
       number = {3},
        pages = {3733-3746},
          doi = {10.1093/mnras/stx2643},
archivePrefix = {arXiv},
       eprint = {1710.02233},
 primaryClass = {astro-ph.SR},
       adsurl = {https://ui.adsabs.harvard.edu/abs/2018MNRAS.473.3733L}
}

@ARTICLE{King2005,
       author = {{King}, A.~R. and {Lubow}, S.~H. and {Ogilvie}, G.~I. and {Pringle}, J.~E.},
        title = "{Aligning spinning black holes and accretion discs}",
      journal = {\mnras},
         year = 2005,
        month = oct,
       volume = {363},
       number = {1},
        pages = {49-56},
          doi = {10.1111/j.1365-2966.2005.09378.x},
archivePrefix = {arXiv},
       eprint = {astro-ph/0507098},
 primaryClass = {astro-ph},
       adsurl = {https://ui.adsabs.harvard.edu/abs/2005MNRAS.363...49K}
}

@ARTICLE{King2006,
       author = {{King}, A.~R. and {Pringle}, J.~E.},
        title = "{Growing supermassive black holes by chaotic accretion}",
      journal = {\mnras},
         year = 2006,
        month = nov,
       volume = {373},
       number = {1},
        pages = {L90-L92},
          doi = {10.1111/j.1745-3933.2006.00249.x},
archivePrefix = {arXiv},
       eprint = {astro-ph/0609598},
 primaryClass = {astro-ph},
       adsurl = {https://ui.adsabs.harvard.edu/abs/2006MNRAS.373L..90K}
}

@ARTICLE{Mayer2007,
       author = {{Mayer}, L. and {Kazantzidis}, S. and {Madau}, P. and {Colpi}, M. and {Quinn}, T. and {Wadsley}, J.},
        title = "{Rapid Formation of Supermassive Black Hole Binaries in Galaxy Mergers with Gas}",
      journal = {Science},
         year = 2007,
        month = jun,
       volume = {316},
       number = {5833},
        pages = {1874},
          doi = {10.1126/science.1141858},
archivePrefix = {arXiv},
       eprint = {0706.1562},
 primaryClass = {astro-ph},
       adsurl = {https://ui.adsabs.harvard.edu/abs/2007Sci...316.1874M}
}

@ARTICLE{Chapon2013,
       author = {{Chapon}, Damien and {Mayer}, Lucio and {Teyssier}, Romain},
        title = "{Hydrodynamics of galaxy mergers with supermassive black holes: is there a last parsec problem?}",
      journal = {\mnras},
         year = 2013,
        month = mar,
       volume = {429},
       number = {4},
        pages = {3114-3122},
          doi = {10.1093/mnras/sts568},
archivePrefix = {arXiv},
       eprint = {1110.6086},
 primaryClass = {astro-ph.GA},
       adsurl = {https://ui.adsabs.harvard.edu/abs/2013MNRAS.429.3114C}
}

@ARTICLE{Bennert2026,
       author = {{Bennert}, Vardha N. and {Winkel}, Nico and {Treu}, Tommaso and {Ding}, Xuheng and {U}, Vivian and {Remigio}, Raymond P. and {Barth}, Aaron J. and {Malkan}, Matthew A. and {Villafa{\~n}a}, Lizvette and {Allen}, Samantha and {Johnson}, Ellie and {Contreras}, Sebastian and {Kim}, Minjin and {Birrer}, Simon and {Jahnke}, Knud and {Zheng}, Shaoping},
        title = "{The Host Galaxies of Active Galactic Nuclei with Direct Black Hole Mass Measurements}",
      journal = {\apj},
         year = 2026,
        month = mar,
       volume = {1000},
       number = {1},
          eid = {48},
        pages = {48},
          doi = {10.3847/1538-4357/ae41ad},
archivePrefix = {arXiv},
       eprint = {2602.06116},
 primaryClass = {astro-ph.GA},
       adsurl = {https://ui.adsabs.harvard.edu/abs/2026ApJ..1000...48B}
}

@ARTICLE{Hopkins2011,
       author = {{Hopkins}, Philip F. and {Quataert}, Eliot},
        title = "{An analytic model of angular momentum transport by gravitational torques: from galaxies to massive black holes}",
      journal = {\mnras},
         year = 2011,
        month = aug,
       volume = {415},
       number = {2},
        pages = {1027-1050},
          doi = {10.1111/j.1365-2966.2011.18542.x},
archivePrefix = {arXiv},
       eprint = {1007.2647},
 primaryClass = {astro-ph.CO},
       adsurl = {https://ui.adsabs.harvard.edu/abs/2011MNRAS.415.1027H}
}

@ARTICLE{Levine2010,
       author = {{Levine}, Robyn and {Gnedin}, Nickolay Y. and {Hamilton}, Andrew J.~S.},
        title = "{Measuring Gas Accretion and Angular Momentum Near Simulated Supermassive Black Holes}",
      journal = {\apj},
         year = 2010,
        month = jun,
       volume = {716},
       number = {2},
        pages = {1386-1396},
          doi = {10.1088/0004-637X/716/2/1386},
archivePrefix = {arXiv},
       eprint = {1004.3785},
 primaryClass = {astro-ph.CO},
       adsurl = {https://ui.adsabs.harvard.edu/abs/2010ApJ...716.1386L}
}

@ARTICLE{MacFadyen2008,
       author = {{MacFadyen}, Andrew I. and {Milosavljevi{\'c}}, Milo{\v{s}}},
        title = "{An Eccentric Circumbinary Accretion Disk and the Detection of Binary Massive Black Holes}",
      journal = {\apj},
         year = 2008,
        month = jan,
       volume = {672},
       number = {1},
        pages = {83-93},
          doi = {10.1086/523869},
archivePrefix = {arXiv},
       eprint = {astro-ph/0607467},
 primaryClass = {astro-ph},
       adsurl = {https://ui.adsabs.harvard.edu/abs/2008ApJ...672...83M}
}

@ARTICLE{Cuadra2009,
       author = {{Cuadra}, J. and {Armitage}, P.~J. and {Alexander}, R.~D. and {Begelman}, M.~C.},
        title = "{Massive black hole binary mergers within subparsec scale gas discs}",
      journal = {\mnras},
         year = 2009,
        month = mar,
       volume = {393},
       number = {4},
        pages = {1423-1432},
          doi = {10.1111/j.1365-2966.2008.14147.x},
archivePrefix = {arXiv},
       eprint = {0809.0311},
 primaryClass = {astro-ph},
       adsurl = {https://ui.adsabs.harvard.edu/abs/2009MNRAS.393.1423C}
}

@ARTICLE{Shi2012,
       author = {{Shi}, Ji-Ming and {Krolik}, Julian H. and {Lubow}, Stephen H. and {Hawley}, John F.},
        title = "{Three-dimensional Magnetohydrodynamic Simulations of Circumbinary Accretion Disks: Disk Structures and Angular Momentum Transport}",
      journal = {\apj},
         year = 2012,
        month = apr,
       volume = {749},
       number = {2},
          eid = {118},
        pages = {118},
          doi = {10.1088/0004-637X/749/2/118},
archivePrefix = {arXiv},
       eprint = {1110.4866},
 primaryClass = {astro-ph.HE},
       adsurl = {https://ui.adsabs.harvard.edu/abs/2012ApJ...749..118S}
}

@ARTICLE{Shi2015,
       author = {{Shi}, Ji-Ming and {Krolik}, Julian H.},
        title = "{Three-dimensional MHD Simulation of Circumbinary Accretion Disks. II. Net Accretion Rate}",
      journal = {\apj},
         year = 2015,
        month = jul,
       volume = {807},
       number = {2},
          eid = {131},
        pages = {131},
          doi = {10.1088/0004-637X/807/2/131},
archivePrefix = {arXiv},
       eprint = {1503.05561},
 primaryClass = {astro-ph.HE},
       adsurl = {https://ui.adsabs.harvard.edu/abs/2015ApJ...807..131S}
}

@ARTICLE{Moody2019,
       author = {{Moody}, Mackenzie S.~L. and {Shi}, Ji-Ming and {Stone}, James M.},
        title = "{Hydrodynamic Torques in Circumbinary Accretion Disks}",
      journal = {\apj},
         year = 2019,
        month = apr,
       volume = {875},
       number = {1},
          eid = {66},
        pages = {66},
          doi = {10.3847/1538-4357/ab09ee},
archivePrefix = {arXiv},
       eprint = {1903.00008},
 primaryClass = {astro-ph.HE},
       adsurl = {https://ui.adsabs.harvard.edu/abs/2019ApJ...875...66M}
}

@ARTICLE{RyanWS2021,
       author = {{Westernacher-Schneider}, John Ryan and {Zrake}, Jonathan and {MacFadyen}, Andrew and {Haiman}, Zolt{\'a}n},
        title = "{Multi-band light curves from eccentric accreting supermassive black hole binaries}",
      journal = {arXiv e-prints},
         year = 2021,
        month = nov,
          eid = {arXiv:2111.06882},
        pages = {arXiv:2111.06882},
archivePrefix = {arXiv},
       eprint = {2111.06882},
 primaryClass = {astro-ph.HE},
       adsurl = {https://ui.adsabs.harvard.edu/abs/2021arXiv211106882W}
}

@article{Zrake2020,
   title={Equilibrium Eccentricity of Accreting Binaries},
   volume={909},
   ISSN={2041-8213},
   url={http://dx.doi.org/10.3847/2041-8213/abdd1c},
   DOI={10.3847/2041-8213/abdd1c},
   number={1},
   journal={The Astrophysical Journal Letters},
   publisher={American Astronomical Society},
   author={Zrake, Jonathan and Tiede, Christopher and MacFadyen, Andrew and Haiman, Zoltán},
   year={2021},
   month={Mar},
   pages={L13}
}

@article{Farris15,
author = {Farris, Brian D. and Duffell, Paul and MacFadyen, Andrew I. and Haiman, Zoltán},
title = {Binary black hole accretion during inspiral and merger},
journal = {Monthly Notices of the Royal Astronomical Society: Letters},
volume = {447},
number = {1},
pages = {L80},
year = {2015},
doi = {10.1093/mnrasl/slu184},
URL = { + http://dx.doi.org/10.1093/mnrasl/slu184},
eprint = {/oup/backfile/content_public/journal/mnrasl/447/1/10.1093/mnrasl/slu184/2/slu184.pdf}
}

@ARTICLE{Miranda2017,
   author = {{Miranda}, R. and {Mu{\~n}oz}, D.~J. and {Lai}, D.},
    title = "{Viscous hydrodynamics simulations of circumbinary accretion discs: variability, quasi-steady state and angular momentum transfer}",
  journal = {\mnras},
archivePrefix = "arXiv",
   eprint = {1610.07263},
 primaryClass = "astro-ph.SR",
     year = 2017,
    month = apr,
   volume = 466,
    pages = {1170-1191},
      doi = {10.1093/mnras/stw3189},
   adsurl = {http://adsabs.harvard.edu/abs/2017MNRAS.466.1170M}
}

@ARTICLE{Betancourt2026,
       author = {{Betancourt}, Leonardo and {MacFadyen}, Andrew and {Zrake}, Jonathan},
        title = "{Eccentric Disks from Circumbinary Rings}",
      journal = {arXiv e-prints},
         year = 2026,
        month = jan,
          eid = {arXiv:2601.00741},
        pages = {arXiv:2601.00741},
archivePrefix = {arXiv},
       eprint = {2601.00741},
 primaryClass = {astro-ph.HE},
       adsurl = {https://ui.adsabs.harvard.edu/abs/2026arXiv260100741B}
}

@ARTICLE{Tiede2020,
       author = {{Tiede}, Christopher and {Zrake}, Jonathan and {MacFadyen}, Andrew and {Haiman}, Zoltan},
        title = "{Gas-driven Inspiral of Binaries in Thin Accretion Disks}",
      journal = {\apj},
         year = 2020,
        month = sep,
       volume = {900},
       number = {1},
          eid = {43},
        pages = {43},
          doi = {10.3847/1538-4357/aba432},
archivePrefix = {arXiv},
       eprint = {2005.09555},
 primaryClass = {astro-ph.GA},
       adsurl = {https://ui.adsabs.harvard.edu/abs/2020ApJ...900...43T}
}

@ARTICLE{Tiede2025,
       author = {{Tiede}, Christopher and {Zrake}, Jonathan and {MacFadyen}, Andrew and {Haiman}, Zolt{\'a}n},
        title = "{Suppressed Accretion onto Massive Black Hole Binaries Surrounded by Thin Disks}",
      journal = {\apj},
         year = 2025,
        month = may,
       volume = {984},
       number = {2},
          eid = {144},
        pages = {144},
          doi = {10.3847/1538-4357/adc727},
archivePrefix = {arXiv},
       eprint = {2410.03830},
 primaryClass = {astro-ph.GA},
       adsurl = {https://ui.adsabs.harvard.edu/abs/2025ApJ...984..144T}
}

@ARTICLE{Dittmann2023,
       author = {{Dittmann}, Alexander J. and {Ryan}, Geoffrey and {Miller}, M. Coleman},
        title = "{The Decoupling of Binaries from Their Circumbinary Disks}",
      journal = {\apjl},
         year = 2023,
        month = jun,
       volume = {949},
       number = {2},
          eid = {L30},
        pages = {L30},
          doi = {10.3847/2041-8213/acd183},
archivePrefix = {arXiv},
       eprint = {2303.16204},
 primaryClass = {astro-ph.HE},
       adsurl = {https://ui.adsabs.harvard.edu/abs/2023ApJ...949L..30D}
}

@ARTICLE{Dittmann2022,
       author = {{Dittmann}, Alexander J. and {Ryan}, Geoffrey},
        title = "{A survey of disc thickness and viscosity in circumbinary accretion: Binary evolution, variability, and disc morphology}",
      journal = {\mnras},
         year = 2022,
        month = jul,
       volume = {513},
       number = {4},
        pages = {6158-6176},
          doi = {10.1093/mnras/stac935},
archivePrefix = {arXiv},
       eprint = {2201.07816},
 primaryClass = {astro-ph.HE},
       adsurl = {https://ui.adsabs.harvard.edu/abs/2022MNRAS.513.6158D}
}

@ARTICLE{Bourne2024,
       author = {{Bourne}, Martin A. and {Fiacconi}, Davide and {Sijacki}, Debora and {Piotrowska}, Joanna M. and {Koudmani}, Sophie},
        title = "{Dynamics and spin alignment in massive, gravito-turbulent circumbinary discs around supermassive black hole binaries}",
      journal = {\mnras},
         year = 2024,
        month = nov,
       volume = {534},
       number = {4},
        pages = {3448-3477},
          doi = {10.1093/mnras/stae2143},
archivePrefix = {arXiv},
       eprint = {2311.17144},
 primaryClass = {astro-ph.HE},
       adsurl = {https://ui.adsabs.harvard.edu/abs/2024MNRAS.534.3448B}
}

@ARTICLE{NixonKingPringle2011,
       author = {{Nixon}, C.~J. and {King}, A.~R. and {Pringle}, J.~E.},
        title = "{The final parsec problem: aligning a binary with an external accretion disc}",
      journal = {\mnras},
         year = 2011,
        month = oct,
       volume = {417},
       number = {1},
        pages = {L66-L69},
          doi = {10.1111/j.1745-3933.2011.01121.x},
archivePrefix = {arXiv},
       eprint = {1107.5056},
 primaryClass = {astro-ph.GA},
       adsurl = {https://ui.adsabs.harvard.edu/abs/2011MNRAS.417L..66N}
}

@ARTICLE{Nixon2011,
       author = {{Nixon}, C.~J. and {Cossins}, P.~J. and {King}, A.~R. and {Pringle}, J.~E.},
        title = "{Retrograde accretion and merging supermassive black holes}",
      journal = {\mnras},
         year = 2011,
        month = apr,
       volume = {412},
       number = {3},
        pages = {1591-1598},
          doi = {10.1111/j.1365-2966.2010.17952.x},
archivePrefix = {arXiv},
       eprint = {1011.1914},
 primaryClass = {astro-ph.HE},
       adsurl = {https://ui.adsabs.harvard.edu/abs/2011MNRAS.412.1591N}
}

@ARTICLE{Roedig2014,
       author = {{Roedig}, Constanze and {Sesana}, Alberto},
        title = "{Migration of massive black hole binaries in self-gravitating discs: retrograde versus prograde}",
      journal = {\mnras},
         year = 2014,
        month = apr,
       volume = {439},
       number = {4},
        pages = {3476-3489},
          doi = {10.1093/mnras/stu194},
archivePrefix = {arXiv},
       eprint = {1307.6283},
 primaryClass = {astro-ph.HE},
       adsurl = {https://ui.adsabs.harvard.edu/abs/2014MNRAS.439.3476R}
}

@ARTICLE{Bankert2015,
       author = {{Bankert}, Justin and {Krolik}, Julian H. and {Shi}, Jiming},
        title = "{Structure of Retrograde Circumbinary Accretion Disks}",
      journal = {\apj},
         year = 2015,
        month = mar,
       volume = {801},
       number = {2},
          eid = {114},
        pages = {114},
          doi = {10.1088/0004-637X/801/2/114},
       adsurl = {https://ui.adsabs.harvard.edu/abs/2015ApJ...801..114B}
}

@ARTICLE{Schnittman2015,
       author = {{Schnittman}, Jeremy D. and {Krolik}, Julian H.},
        title = "{Evolution of a Binary Black Hole with a Retrograde Circumbinary Accretion Disk}",
      journal = {\apj},
         year = 2015,
        month = jun,
       volume = {806},
       number = {1},
          eid = {88},
        pages = {88},
          doi = {10.1088/0004-637X/806/1/88},
archivePrefix = {arXiv},
       eprint = {1504.00311},
 primaryClass = {astro-ph.HE},
       adsurl = {https://ui.adsabs.harvard.edu/abs/2015ApJ...806...88S}
}

@ARTICLE{AmaroSeoane2016,
       author = {{Amaro-Seoane}, Pau and {Maureira-Fredes}, Cristi{\'a}n and {Dotti}, Massimo and {Colpi}, Monica},
        title = "{Retrograde binaries of massive black holes in circumbinary accretion discs}",
      journal = {\aap},
         year = 2016,
        month = jun,
       volume = {591},
          eid = {A114},
        pages = {A114},
          doi = {10.1051/0004-6361/201526172},
archivePrefix = {arXiv},
       eprint = {1604.01392},
 primaryClass = {astro-ph.GA},
       adsurl = {https://ui.adsabs.harvard.edu/abs/2016A&A...591A.114A}
}

@ARTICLE{Hobbs2011,
       author = {{Hobbs}, Alexander and {Nayakshin}, Sergei and {Power}, Chris and {King}, Andrew},
        title = "{Feeding supermassive black holes through supersonic turbulence and ballistic accretion}",
      journal = {\mnras},
         year = 2011,
        month = jun,
       volume = {413},
       number = {4},
        pages = {2633-2650},
          doi = {10.1111/j.1365-2966.2011.18333.x},
archivePrefix = {arXiv},
       eprint = {1001.3883},
 primaryClass = {astro-ph.HE},
       adsurl = {https://ui.adsabs.harvard.edu/abs/2011MNRAS.413.2633H}
}

@ARTICLE{ONeill2025,
       author = {{O'Neill}, David and {Tiede}, Christopher and {D'Orazio}, Daniel J. and {Haiman}, Zolt{\'a}n and {MacFadyen}, Andrew},
        title = "{Gravitational Wave Decoupling in Retrograde Circumbinary Disks}",
      journal = {\apj},
         year = 2025,
        month = nov,
       volume = {993},
       number = {2},
          eid = {206},
        pages = {206},
          doi = {10.3847/1538-4357/ae0ca8},
archivePrefix = {arXiv},
       eprint = {2501.11679},
 primaryClass = {astro-ph.HE},
       adsurl = {https://ui.adsabs.harvard.edu/abs/2025ApJ...993..206O}
}

@ARTICLE{Nixon2015,
       author = {{Nixon}, Chris and {Lubow}, Stephen H.},
        title = "{Resonances in retrograde circumbinary discs}",
      journal = {\mnras},
         year = 2015,
        month = apr,
       volume = {448},
       number = {4},
        pages = {3472-3483},
          doi = {10.1093/mnras/stv166},
archivePrefix = {arXiv},
       eprint = {1501.07277},
 primaryClass = {astro-ph.HE},
       adsurl = {https://ui.adsabs.harvard.edu/abs/2015MNRAS.448.3472N}
}

@ARTICLE{ERosita2021,
       author = {{Predehl}, P. and {Andritschke}, R. and {Arefiev}, V. and {Babyshkin}, V. and {Batanov}, O. and {Becker}, W. and {B{\"o}hringer}, H. and {Bogomolov}, A. and {Boller}, T. and {Borm}, K. and {Bornemann}, W. and {Br{\"a}uninger}, H. and {Br{\"u}ggen}, M. and {Brunner}, H. and {Brusa}, M. and {Bulbul}, E. and {Buntov}, M. and {Burwitz}, V. and {Burkert}, W. and {Clerc}, N. and {Churazov}, E. and {Coutinho}, D. and {Dauser}, T. and {Dennerl}, K. and {Doroshenko}, V. and {Eder}, J. and {Emberger}, V. and {Eraerds}, T. and {Finoguenov}, A. and {Freyberg}, M. and {Friedrich}, P. and {Friedrich}, S. and {F{\"u}rmetz}, M. and {Georgakakis}, A. and {Gilfanov}, M. and {Granato}, S. and {Grossberger}, C. and {Gueguen}, A. and {Gureev}, P. and {Haberl}, F. and {H{\"a}lker}, O. and {Hartner}, G. and {Hasinger}, G. and {Huber}, H. and {Ji}, L. and {Kienlin}, A. v. and {Kink}, W. and {Korotkov}, F. and {Kreykenbohm}, I. and {Lamer}, G. and {Lomakin}, I. and {Lapshov}, I. and {Liu}, T. and {Maitra}, C. and {Meidinger}, N. and {Menz}, B. and {Merloni}, A. and {Mernik}, T. and {Mican}, B. and {Mohr}, J. and {M{\"u}ller}, S. and {Nandra}, K. and {Nazarov}, V. and {Pacaud}, F. and {Pavlinsky}, M. and {Perinati}, E. and {Pfeffermann}, E. and {Pietschner}, D. and {Ramos-Ceja}, M.~E. and {Rau}, A. and {Reiffers}, J. and {Reiprich}, T.~H. and {Robrade}, J. and {Salvato}, M. and {Sanders}, J. and {Santangelo}, A. and {Sasaki}, M. and {Scheuerle}, H. and {Schmid}, C. and {Schmitt}, J. and {Schwope}, A. and {Shirshakov}, A. and {Steinmetz}, M. and {Stewart}, I. and {Str{\"u}der}, L. and {Sunyaev}, R. and {Tenzer}, C. and {Tiedemann}, L. and {Tr{\"u}mper}, J. and {Voron}, V. and {Weber}, P. and {Wilms}, J. and {Yaroshenko}, V.},
        title = "{The eROSITA X-ray telescope on SRG}",
      journal = {\aap},
         year = 2021,
        month = mar,
       volume = {647},
          eid = {A1},
        pages = {A1},
          doi = {10.1051/0004-6361/202039313},
archivePrefix = {arXiv},
       eprint = {2010.03477},
 primaryClass = {astro-ph.HE},
       adsurl = {https://ui.adsabs.harvard.edu/abs/2021A&A...647A...1P}
}

@article{InPTA_EPTA,
   title={The second data release from the European Pulsar Timing Array: III. Search for gravitational wave signals},
   volume={678},
   ISSN={1432-0746},
   url={http://dx.doi.org/10.1051/0004-6361/202346844},
   DOI={10.1051/0004-6361/202346844},
   journal={Astronomy \&; Astrophysics},
   publisher={EDP Sciences},
   author={Antoniadis, J. and Arumugam, P. and Arumugam, S. and Babak, S. and Bagchi, M. and Bak Nielsen, A.-S. and Bassa, C. G. and Bathula, A. and Berthereau, A. and Bonetti, M. and Bortolas, E. and Brook, P. R. and Burgay, M. and Caballero, R. N. and Chalumeau, A. and Champion, D. J. and Chanlaridis, S. and Chen, S. and Cognard, I. and Dandapat, S. and Deb, D. and Desai, S. and Desvignes, G. and Dhanda-Batra, N. and Dwivedi, C. and Falxa, M. and Ferdman, R. D. and Franchini, A. and Gair, J. R. and Goncharov, B. and Gopakumar, A. and Graikou, E. and Grießmeier, J.-M. and Guillemot, L. and Guo, Y. J. and Gupta, Y. and Hisano, S. and Hu, H. and Iraci, F. and Izquierdo-Villalba, D. and Jang, J. and Jawor, J. and Janssen, G. H. and Jessner, A. and Joshi, B. C. and Kareem, F. and Karuppusamy, R. and Keane, E. F. and Keith, M. J. and Kharbanda, D. and Kikunaga, T. and Kolhe, N. and Kramer, M. and Krishnakumar, M. A. and Lackeos, K. and Lee, K. J. and Liu, K. and Liu, Y. and Lyne, A. G. and McKee, J. W. and Maan, Y. and Main, R. A. and Mickaliger, M. B. and Niţu, I. C. and Nobleson, K. and Paladi, A. K. and Parthasarathy, A. and Perera, B. B. P. and Perrodin, D. and Petiteau, A. and Porayko, N. K. and Possenti, A. and Prabu, T. and Quelquejay Leclere, H. and Rana, P. and Samajdar, A. and Sanidas, S. A. and Sesana, A. and Shaifullah, G. and Singha, J. and Speri, L. and Spiewak, R. and Srivastava, A. and Stappers, B. W. and Surnis, M. and Susarla, S. C. and Susobhanan, A. and Takahashi, K. and Tarafdar, P. and Theureau, G. and Tiburzi, C. and van der Wateren, E. and Vecchio, A. and Venkatraman Krishnan, V. and Verbiest, J. P. W. and Wang, J. and Wang, L. and Wu, Z.},
   year={2023},
   month=oct, pages={A50} }

@article{ParkesPTA,
   title={Search for an Isotropic Gravitational-wave Background with the Parkes Pulsar Timing Array},
   volume={951},
   ISSN={2041-8213},
   url={http://dx.doi.org/10.3847/2041-8213/acdd02},
   DOI={10.3847/2041-8213/acdd02},
   number={1},
   journal={The Astrophysical Journal Letters},
   publisher={American Astronomical Society},
   author={Reardon, Daniel J. and Zic, Andrew and Shannon, Ryan M. and Hobbs, George B. and Bailes, Matthew and Di Marco, Valentina and Kapur, Agastya and Rogers, Axl F. and Thrane, Eric and Askew, Jacob and Bhat, N. D. Ramesh and Cameron, Andrew and Curyło, Małgorzata and Coles, William A. and Dai, Shi and Goncharov, Boris and Kerr, Matthew and Kulkarni, Atharva and Levin, Yuri and Lower, Marcus E. and Manchester, Richard N. and Mandow, Rami and Miles, Matthew T. and Nathan, Rowina S. and Osłowski, Stefan and Russell, Christopher J. and Spiewak, Renée and Zhang, Songbo and Zhu, Xing-Jiang},
   year={2023},
   month=jun, pages={L6} }

@ARTICLE{Kinney2000, 
       author = {{Kinney}, A.~L. and {Schmitt}, H.~R. and {Clarke}, C.~J. and {Pringle}, J.~E. and {Ulvestad}, J.~S. and {Antonucci}, R.~R.~J.},
        title = "{Jet Directions in Seyfert Galaxies}",
      journal = {\apj},
         year = 2000,
        month = jul,
       volume = {537},
       number = {1},
        pages = {152-177},
          doi = {10.1086/309016},
archivePrefix = {arXiv},
       eprint = {astro-ph/0002131},
 primaryClass = {astro-ph},
       adsurl = {https://ui.adsabs.harvard.edu/abs/2000ApJ...537..152K}
}

@ARTICLE{Schmitt2002,
       author = {{Schmitt}, H.~R. and {Pringle}, J.~E. and {Clarke}, C.~J. and {Kinney}, A.~L.},
        title = "{The Orientation of Jets Relative to Dust Disks in Radio Galaxies}",
      journal = {\apj},
         year = 2002,
        month = aug,
       volume = {575},
       number = {1},
        pages = {150-155},
          doi = {10.1086/341211},
archivePrefix = {arXiv},
       eprint = {astro-ph/0204247},
 primaryClass = {astro-ph},
       adsurl = {https://ui.adsabs.harvard.edu/abs/2002ApJ...575..150S}
}

@ARTICLE{Dorazio2015,
       author = {{D'Orazio}, Daniel J. and {Haiman}, Zolt{\'a}n and {Schiminovich}, David},
        title = "{Relativistic boost as the cause of periodicity in a massive black-hole binary candidate}",
      journal = {\nat},
         year = 2015,
        month = sep,
       volume = {525},
       number = {7569},
        pages = {351-353},
          doi = {10.1038/nature15262},
archivePrefix = {arXiv},
       eprint = {1509.04301},
 primaryClass = {astro-ph.HE},
       adsurl = {https://ui.adsabs.harvard.edu/abs/2015Natur.525..351D}
}

@ARTICLE{RyanMacFadyen2017,
       author = {{Ryan}, Geoffrey and {MacFadyen}, Andrew},
        title = "{Minidisks in Binary Black Hole Accretion}",
      journal = {\apj},
         year = 2017,
        month = feb,
       volume = {835},
       number = {2},
          eid = {199},
        pages = {199},
          doi = {10.3847/1538-4357/835/2/199},
archivePrefix = {arXiv},
       eprint = {1611.00341},
 primaryClass = {astro-ph.HE},
       adsurl = {https://ui.adsabs.harvard.edu/abs/2017ApJ...835..199R}
}

@ARTICLE{Duffell2024,
       author = {{Duffell}, Paul C. and {Dittmann}, Alexander J. and {D'Orazio}, Daniel J. and {Franchini}, Alessia and {Kratter}, Kaitlin M. and {Penzlin}, Anna B.~T. and {Ragusa}, Enrico and {Siwek}, Magdalena and {Tiede}, Christopher and {Wang}, Haiyang and {Zrake}, Jonathan and {Dempsey}, Adam M. and {Haiman}, Zoltan and {Lupi}, Alessandro and {Pirog}, Michal and {Ryan}, Geoffrey},
        title = "{The Santa Barbara Binary‑disk Code Comparison}",
      journal = {\apj},
         year = 2024,
        month = aug,
       volume = {970},
       number = {2},
          eid = {156},
        pages = {156},
          doi = {10.3847/1538-4357/ad5a7e},
archivePrefix = {arXiv},
       eprint = {2402.13039},
 primaryClass = {astro-ph.SR},
       adsurl = {https://ui.adsabs.harvard.edu/abs/2024ApJ...970..156D}
}

@ARTICLE{Haiman2009,
       author = {{Haiman}, Zolt{\'a}n and {Kocsis}, Bence and {Menou}, Kristen},
        title = "{The Population of Viscosity- and Gravitational Wave-driven Supermassive Black Hole Binaries Among Luminous Active Galactic Nuclei}",
      journal = {\apj},
         year = 2009,
        month = aug,
       volume = {700},
       number = {2},
        pages = {1952-1969},
          doi = {10.1088/0004-637X/700/2/1952},
archivePrefix = {arXiv},
       eprint = {0904.1383},
 primaryClass = {astro-ph.CO},
       adsurl = {https://ui.adsabs.harvard.edu/abs/2009ApJ...700.1952H}
}

@ARTICLE{SS73,
       author = {{Shakura}, N.~I. and {Sunyaev}, R.~A.},
        title = "{Black holes in binary systems. Observational appearance.}",
      journal = {\aap},
         year = 1973,
        month = jan,
       volume = {24},
        pages = {337-355},
       adsurl = {https://ui.adsabs.harvard.edu/abs/1973A&A....24..337S}
}

@ARTICLE{Goodman2003,
       author = {{Goodman}, Jeremy},
        title = "{Self-gravity and quasi-stellar object discs}",
      journal = {\mnras},
         year = 2003,
        month = mar,
       volume = {339},
       number = {4},
        pages = {937-948},
          doi = {10.1046/j.1365-8711.2003.06241.x},
archivePrefix = {arXiv},
       eprint = {astro-ph/0201001},
 primaryClass = {astro-ph},
       adsurl = {https://ui.adsabs.harvard.edu/abs/2003MNRAS.339..937G}
}

@BOOK{FrankKingBook2002,
       author = {{Frank}, Juhan and {King}, Andrew and {Raine}, Derek J.},
        title = "{Accretion Power in Astrophysics: Third Edition}",
         year = 2002,
       adsurl = {https://ui.adsabs.harvard.edu/abs/2002apa..book.....F}
}

@ARTICLE{Rafikov2013,
       author = {{Rafikov}, Roman R.},
        title = "{Structure and Evolution of Circumbinary Disks around Supermassive Black Hole Binaries}",
      journal = {\apj},
         year = 2013,
        month = sep,
       volume = {774},
       number = {2},
          eid = {144},
        pages = {144},
          doi = {10.1088/0004-637X/774/2/144},
archivePrefix = {arXiv},
       eprint = {1205.5017},
 primaryClass = {astro-ph.GA},
       adsurl = {https://ui.adsabs.harvard.edu/abs/2013ApJ...774..144R}
}

@ARTICLE{BalbusHawley91,
       author = {{Balbus}, Steven A. and {Hawley}, John F.},
        title = "{A Powerful Local Shear Instability in Weakly Magnetized Disks. I. Linear Analysis}",
      journal = {\apj},
         year = 1991,
        month = jul,
       volume = {376},
        pages = {214},
          doi = {10.1086/170270},
       adsurl = {https://ui.adsabs.harvard.edu/abs/1991ApJ...376..214B}
}

@ARTICLE{Schulze2010,
       author = {{Schulze}, A. and {Wisotzki}, L.},
        title = "{Low redshift AGN in the Hamburg/ESO Survey . II. The active black hole mass function and the distribution function of Eddington ratios}",
      journal = {\aap},
         year = 2010,
        month = jun,
       volume = {516},
          eid = {A87},
        pages = {A87},
          doi = {10.1051/0004-6361/201014193},
archivePrefix = {arXiv},
       eprint = {1004.2671},
 primaryClass = {astro-ph.CO},
       adsurl = {https://ui.adsabs.harvard.edu/abs/2010A&A...516A..87S}
}

@ARTICLE{Ananna2022,
       author = {{Ananna}, Tonima Tasnim and {Urry}, C. Megan and {Ricci}, Claudio and {Natarajan}, Priyamvada and {Hickox}, Ryan C. and {Trakhtenbrot}, Benny and {Treister}, Ezequiel and {Weigel}, Anna K. and {Ueda}, Yoshihiro and {Koss}, Michael J. and {Bauer}, F.~E. and {Temple}, Matthew J. and {Balokovi{\'c}}, Mislav and {Mushotzky}, Richard and {Auge}, Connor and {Sanders}, David B. and {Kakkad}, Darshan and {Sartori}, Lia F. and {Marchesi}, Stefano and {Harrison}, Fiona and {Stern}, Daniel and {Oh}, Kyuseok and {Caglar}, Turgay and {Powell}, Meredith C. and {Podjed}, Stephanie A. and {Mej{\'\i}a-Restrepo}, Julian E.},
        title = "{Probing the Structure and Evolution of BASS Active Galactic Nuclei through Eddington Ratios}",
      journal = {\apjl},
         year = 2022,
        month = nov,
       volume = {939},
       number = {1},
          eid = {L13},
        pages = {L13},
          doi = {10.3847/2041-8213/ac9979},
archivePrefix = {arXiv},
       eprint = {2210.08401},
 primaryClass = {astro-ph.GA},
       adsurl = {https://ui.adsabs.harvard.edu/abs/2022ApJ...939L..13A}
}

@ARTICLE{Abramowicz1988,
       author = {{Abramowicz}, M.~A. and {Czerny}, B. and {Lasota}, J.~P. and {Szuszkiewicz}, E.},
        title = "{Slim Accretion Disks}",
      journal = {\apj},
         year = 1988,
        month = sep,
       volume = {332},
        pages = {646},
          doi = {10.1086/166683},
       adsurl = {https://ui.adsabs.harvard.edu/abs/1988ApJ...332..646A}
}

@ARTICLE{Paczynski1977,
       author = {{Paczynski}, B.},
        title = "{A model of accretion disks in close binaries.}",
      journal = {\apj},
         year = 1977,
        month = sep,
       volume = {216},
        pages = {822-826},
          doi = {10.1086/155526},
       adsurl = {https://ui.adsabs.harvard.edu/abs/1977ApJ...216..822P}
}

@ARTICLE{Overton2024,
       author = {{Overton}, Madeline and {Martin}, Rebecca G. and {Lubow}, Stephen H. and {Lepp}, Stephen},
        title = "{Retrograde discs around one component of a binary are unstable to tilting}",
      journal = {\mnras},
         year = 2024,
        month = feb,
       volume = {528},
       number = {1},
        pages = {L106-L111},
          doi = {10.1093/mnrasl/slad172},
archivePrefix = {arXiv},
       eprint = {2311.10864},
 primaryClass = {astro-ph.SR},
       adsurl = {https://ui.adsabs.harvard.edu/abs/2024MNRAS.528L.106O}
}

@ARTICLE{Artymowicz1994,
       author = {{Artymowicz}, Pawel and {Lubow}, Stephen H.},
        title = "{Dynamics of Binary-Disk Interaction. I. Resonances and Disk Gap Sizes}",
      journal = {\apj},
         year = 1994,
        month = feb,
       volume = {421},
        pages = {651},
          doi = {10.1086/173679},
       adsurl = {https://ui.adsabs.harvard.edu/abs/1994ApJ...421..651A}
}

@ARTICLE{Overton2025,
       author = {{Overton}, Madeline and {Martin}, Rebecca G. and {Lubow}, Stephen H. and {Lepp}, Stephen},
        title = "{Disc breaking through forced eccentricity growth}",
      journal = {\mnras},
         year = 2025,
        month = jun,
       volume = {540},
       number = {1},
        pages = {L41-L47},
          doi = {10.1093/mnrasl/slaf029},
archivePrefix = {arXiv},
       eprint = {2503.22586},
 primaryClass = {astro-ph.SR},
       adsurl = {https://ui.adsabs.harvard.edu/abs/2025MNRAS.540L..41O}
}

@ARTICLE{Papaloizou1977,
       author = {{Papaloizou}, J. and {Pringle}, J.~E.},
        title = "{Tidal torques on accretion discs in close binary systems.}",
      journal = {\mnras},
         year = 1977,
        month = nov,
       volume = {181},
        pages = {441-454},
          doi = {10.1093/mnras/181.3.441},
       adsurl = {https://ui.adsabs.harvard.edu/abs/1977MNRAS.181..441P}
}

@ARTICLE{Morais2012,
       author = {{Morais}, M.~H.~M. and {Giuppone}, C.~A.},
        title = "{Stability of prograde and retrograde planets in circular binary systems}",
      journal = {\mnras},
         year = 2012,
        month = jul,
       volume = {424},
       number = {1},
        pages = {52-64},
          doi = {10.1111/j.1365-2966.2012.21151.x},
archivePrefix = {arXiv},
       eprint = {1204.4718},
 primaryClass = {astro-ph.EP},
       adsurl = {https://ui.adsabs.harvard.edu/abs/2012MNRAS.424...52M}
}

@ARTICLE{Tiede2026,
       author = {{Tiede}, Christopher and {O'Neill}, David and {D'Orazio}, Daniel J.},
        title = "{Dynamics and detectability of long-lived non-accretion phases for massive black hole binaries in cold, thermally regulating disks}",
      journal = {arXiv e-prints},
         year = 2026,
        month = jun,
          eid = {arXiv:2606.04082},
        pages = {arXiv:2606.04082},
archivePrefix = {arXiv},
       eprint = {2606.04082},
 primaryClass = {astro-ph.HE},
       adsurl = {https://ui.adsabs.harvard.edu/abs/2026arXiv260604082T}
}

@ARTICLE{Peters1964,
       author = {{Peters}, P.~C.},
        title = "{Gravitational Radiation and the Motion of Two Point Masses}",
      journal = {Physical Review},
         year = 1964,
        month = nov,
       volume = {136},
       number = {4B},
        pages = {1224-1232},
          doi = {10.1103/PhysRev.136.B1224},
       adsurl = {https://ui.adsabs.harvard.edu/abs/1964PhRv..136.1224P}
}

@article{Hunter:2007,
  author        = {Hunter, J. D.},
  title         = {Matplotlib: A 2D graphics environment},
  journal       = {Computing in Science \& Engineering},
  volume        = {9},
  number        = {3},
  pages         = {90--95},
  publisher     = {IEEE COMPUTER SOC},
  doi           = {10.1109/MCSE.2007.55},
  year          = 2007
}

@article{numpy,
  title         = {Array programming with {NumPy}},
  author        = {Charles R. Harris and K. Jarrod Millman and St{\'{e}}fan J. van der Walt and Ralf Gommers and Pauli Virtanen and David Cournapeau and Eric Wieser and Julian Taylor and Sebastian Berg and Nathaniel J. Smith and Robert Kern and Matti Picus and Stephan Hoyer and Marten H. van Kerkwijk and Matthew Brett and Allan Haldane and Jaime Fern{\'{a}}ndez del R{\'{i}}o and Mark Wiebe and Pearu Peterson and Pierre G{\'{e}}rard-Marchant and Kevin Sheppard and Tyler Reddy and Warren Weckesser and Hameer Abbasi and Christoph Gohlke and Travis E. Oliphant},
  year          = {2020},
  month         = sep,
  journal       = {Nature},
  volume        = {585},
  number        = {7825},
  pages         = {357--362},
  doi           = {10.1038/s41586-020-2649-2},
  publisher     = {Springer Science and Business Media {LLC}},
  url           = {https://doi.org/10.1038/s41586-020-2649-2}
}

@book{python,
  author        = {Van Rossum, Guido and Drake, Fred L.},
  title         = {Python 3 Reference Manual},
  year          = {2009},
  isbn          = {1441412697},
  publisher     = {CreateSpace},
  address       = {Scotts Valley, CA}
}

@software{scipy_20383145,
  author       = {Ralf Gommers and
                  Pauli Virtanen and
                  Matt Haberland and
                  Evgeni Burovski and
                  Tyler Reddy and
                  Warren Weckesser and
                  Andrew Nelson and
                  Travis E. Oliphant and
                  David Cournapeau and
                  Ilhan Polat and
                  alexbrc and
                  Pamphile Roy and
                  Pearu Peterson and
                  Lucas Colley and
                  Josh Wilson and
                  endolith and
                  Jake Bowhay and
                  Nikolay Mayorov and
                  Albert Steppi and
                  Stefan van der Walt and
                  Matthew Brett and
                  Denis Laxalde and
                  Eric Larson and
                  Atsushi Sakai and
                  Jarrod Millman and
                  Lars and
                  peterbell10 and
                  CJ Carey and
                  Paul van Mulbregt and
                  Daniel Schmitz},
  title        = {scipy/scipy: SciPy 1.18.0rc1},
  month        = may,
  year         = 2026,
  publisher    = {Zenodo},
  version      = {v1.18.0rc1},
  doi          = {10.5281/zenodo.20383145},
  url          = {https://doi.org/10.5281/zenodo.20383145},
}

@article{2020SciPy-NMeth,
  author        = {Virtanen, Pauli and Gommers, Ralf and Oliphant, Travis E. and Haberland, Matt and Reddy, Tyler and Cournapeau, David and Burovski, Evgeni and Peterson, Pearu and Weckesser, Warren and Bright, Jonathan and {van der Walt}, St{\'e}fan J. and Brett, Matthew and Wilson, Joshua and Millman, K. Jarrod and Mayorov, Nikolay and Nelson, Andrew R. J. and Jones, Eric and Kern, Robert and Larson, Eric and Carey, C J and Polat, {\.I}lhan and Feng, Yu and Moore, Eric W. and {VanderPlas}, Jake and Laxalde, Denis and Perktold, Josef and Cimrman, Robert and Henriksen, Ian and Quintero, E. A. and Harris, Charles R. and Archibald, Anne M. and Ribeiro, Ant{\^o}nio H. and Pedregosa, Fabian and {van Mulbregt}, Paul and {SciPy 1.0 Contributors}},
  title         = {{{SciPy} 1.0: Fundamental Algorithms for Scientific Computing in Python}},
  journal       = {Nature Methods},
  year          = {2020},
  volume        = {17},
  pages         = {261--272},
  adsurl        = {https://rdcu.be/b08Wh},
  doi           = {10.1038/s41592-019-0686-2}
}

@article{software-citation-station-paper,
  author        = {{Wagg}, Tom and {Broekgaarden}, Floor S.},
  title         = "{Streamlining and standardizing software citations with The Software Citation Station}",
  journal       = {arXiv e-prints},
  year          = 2024,
  month         = jun,
  eid           = {arXiv:2406.04405},
  pages         = {arXiv:2406.04405},
  archiveprefix = {arXiv},
  eprint        = {2406.04405},
  primaryclass  = {astro-ph.IM},
  adsurl        = {https://ui.adsabs.harvard.edu/abs/2024arXiv240604405W}
}

@software{software-citation-station-zenodo,
  author       = {Tom Wagg and
                  Floor Broekgaarden and
                  Phil Van-Lane and
                  Kai Wu and
                  Kayhan Gültekin},
  title        = {TomWagg/software-citation-station: v1.4},
  month        = nov,
  year         = 2025,
  publisher    = {Zenodo},
  version      = {v1.4},
  doi          = {10.5281/zenodo.17654855},
  url          = {https://doi.org/10.5281/zenodo.17654855},
  swhid        = {swh:1:dir:9a009430037c791424a572f542e9a5d5c1fb44ff
                   ;origin=https://doi.org/10.5281/zenodo.13225526;vi
                   sit=swh:1:snp:ef11f058d718d691f0661c9445c2251328cb
                   ac95;anchor=swh:1:rel:84cde4e532032537b7f014c4467c
                   102430a53fa2;path=TomWagg-software-citation-
                   station-61a588a
                  },
}

@ARTICLE{Garg2024,
       author = {{Garg}, Mudit and {Tiede}, Christopher and {D'Orazio}, Daniel J.},
        title = "{Accretion-mediated spin-eccentricity correlations in LISA massive black hole binaries}",
      journal = {\mnras},
         year = 2024,
        month = nov,
       volume = {534},
       number = {4},
        pages = {3705-3712},
          doi = {10.1093/mnras/stae2357},
archivePrefix = {arXiv},
       eprint = {2405.04411},
 primaryClass = {astro-ph.HE},
       adsurl = {https://ui.adsabs.harvard.edu/abs/2024MNRAS.534.3705G}
}

@ARTICLE{Agazie2023,
       author = {{Agazie}, Gabriella and {Anumarlapudi}, Akash and {Archibald}, Anne M. and {Arzoumanian}, Zaven and {Baker}, Paul T. and {B{\'e}csy}, Bence and {Blecha}, Laura and {Brazier}, Adam and {Brook}, Paul R. and {Burke-Spolaor}, Sarah and {Burnette}, Rand and {Case}, Robin and {Charisi}, Maria and {Chatterjee}, Shami and {Chatziioannou}, Katerina and {Cheeseboro}, Belinda D. and {Chen}, Siyuan and {Cohen}, Tyler and {Cordes}, James M. and {Cornish}, Neil J. and {Crawford}, Fronefield and {Cromartie}, H. Thankful and {Crowter}, Kathryn and {Cutler}, Curt J. and {Decesar}, Megan E. and {Degan}, Dallas and {Demorest}, Paul B. and {Deng}, Heling and {Dolch}, Timothy and {Drachler}, Brendan and {Ellis}, Justin A. and {Ferrara}, Elizabeth C. and {Fiore}, William and {Fonseca}, Emmanuel and {Freedman}, Gabriel E. and {Garver-Daniels}, Nate and {Gentile}, Peter A. and {Gersbach}, Kyle A. and {Glaser}, Joseph and {Good}, Deborah C. and {G{\"u}ltekin}, Kayhan and {Hazboun}, Jeffrey S. and {Hourihane}, Sophie and {Islo}, Kristina and {Jennings}, Ross J. and {Johnson}, Aaron D. and {Jones}, Megan L. and {Kaiser}, Andrew R. and {Kaplan}, David L. and {Kelley}, Luke Zoltan and {Kerr}, Matthew and {Key}, Joey S. and {Klein}, Tonia C. and {Laal}, Nima and {Lam}, Michael T. and {Lamb}, William G. and {Lazio}, T. Joseph W. and {Lewandowska}, Natalia and {Littenberg}, Tyson B. and {Liu}, Tingting and {Lommen}, Andrea and {Lorimer}, Duncan R. and {Luo}, Jing and {Lynch}, Ryan S. and {Ma}, Chung-Pei and {Madison}, Dustin R. and {Mattson}, Margaret A. and {McEwen}, Alexander and {McKee}, James W. and {McLaughlin}, Maura A. and {McMann}, Natasha and {Meyers}, Bradley W. and {Meyers}, Patrick M. and {Mingarelli}, Chiara M.~F. and {Mitridate}, Andrea and {Natarajan}, Priyamvada and {Ng}, Cherry and {Nice}, David J. and {Ocker}, Stella Koch and {Olum}, Ken D. and {Pennucci}, Timothy T. and {Perera}, Benetge B.~P. and {Petrov}, Polina and {Pol}, Nihan S. and {Radovan}, Henri A. and {Ransom}, Scott M. and {Ray}, Paul S. and {Romano}, Joseph D. and {Sardesai}, Shashwat C. and {Schmiedekamp}, Ann and {Schmiedekamp}, Carl and {Schmitz}, Kai and {Schult}, Levi and {Shapiro-Albert}, Brent J. and {Siemens}, Xavier and {Simon}, Joseph and {Siwek}, Magdalena S. and {Stairs}, Ingrid H. and {Stinebring}, Daniel R. and {Stovall}, Kevin and {Sun}, Jerry P. and {Susobhanan}, Abhimanyu and {Swiggum}, Joseph K. and {Taylor}, Jacob and {Taylor}, Stephen R. and {Turner}, Jacob E. and {Unal}, Caner and {Vallisneri}, Michele and {van Haasteren}, Rutger and {Vigeland}, Sarah J. and {Wahl}, Haley M. and {Wang}, Qiaohong and {Witt}, Caitlin A. and {Young}, Olivia and {Nanograv Collaboration}},
        title = "{The NANOGrav 15 yr Data Set: Evidence for a Gravitational-wave Background}",
      journal = {\apjl},
         year = 2023,
        month = jul,
       volume = {951},
       number = {1},
          eid = {L8},
        pages = {L8},
          doi = {10.3847/2041-8213/acdac6},
archivePrefix = {arXiv},
       eprint = {2306.16213},
 primaryClass = {astro-ph.HE},
       adsurl = {https://ui.adsabs.harvard.edu/abs/2023ApJ...951L...8A}
}

@ARTICLE{EPTA2023,
       author = {{EPTA Collaboration} and {InPTA Collaboration} and {Antoniadis}, J. and {Arumugam}, P. and {Arumugam}, S. and {Babak}, S. and {Bagchi}, M. and {Bak Nielsen}, A.-S. and {Bassa}, C.~G. and {Bathula}, A. and {Berthereau}, A. and {Bonetti}, M. and {Bortolas}, E. and {Brook}, P.~R. and {Burgay}, M. and {Caballero}, R.~N. and {Chalumeau}, A. and {Champion}, D.~J. and {Chanlaridis}, S. and {Chen}, S. and {Cognard}, I. and {Dandapat}, S. and {Deb}, D. and {Desai}, S. and {Desvignes}, G. and {Dhanda-Batra}, N. and {Dwivedi}, C. and {Falxa}, M. and {Ferdman}, R.~D. and {Franchini}, A. and {Gair}, J.~R. and {Goncharov}, B. and {Gopakumar}, A. and {Graikou}, E. and {Grie{\ss}meier}, J.-M. and {Guillemot}, L. and {Guo}, Y.~J. and {Gupta}, Y. and {Hisano}, S. and {Hu}, H. and {Iraci}, F. and {Izquierdo-Villalba}, D. and {Jang}, J. and {Jawor}, J. and {Janssen}, G.~H. and {Jessner}, A. and {Joshi}, B.~C. and {Kareem}, F. and {Karuppusamy}, R. and {Keane}, E.~F. and {Keith}, M.~J. and {Kharbanda}, D. and {Kikunaga}, T. and {Kolhe}, N. and {Kramer}, M. and {Krishnakumar}, M.~A. and {Lackeos}, K. and {Lee}, K.~J. and {Liu}, K. and {Liu}, Y. and {Lyne}, A.~G. and {McKee}, J.~W. and {Maan}, Y. and {Main}, R.~A. and {Mickaliger}, M.~B. and {Ni{\c{t}}u}, I.~C. and {Nobleson}, K. and {Paladi}, A.~K. and {Parthasarathy}, A. and {Perera}, B.~B.~P. and {Perrodin}, D. and {Petiteau}, A. and {Porayko}, N.~K. and {Possenti}, A. and {Prabu}, T. and {Quelquejay Leclere}, H. and {Rana}, P. and {Samajdar}, A. and {Sanidas}, S.~A. and {Sesana}, A. and {Shaifullah}, G. and {Singha}, J. and {Speri}, L. and {Spiewak}, R. and {Srivastava}, A. and {Stappers}, B.~W. and {Surnis}, M. and {Susarla}, S.~C. and {Susobhanan}, A. and {Takahashi}, K. and {Tarafdar}, P. and {Theureau}, G. and {Tiburzi}, C. and {van der Wateren}, E. and {Vecchio}, A. and {Venkatraman Krishnan}, V. and {Verbiest}, J.~P.~W. and {Wang}, J. and {Wang}, L. and {Wu}, Z.},
        title = "{The second data release from the European Pulsar Timing Array. III. Search for gravitational wave signals}",
      journal = {\aap},
         year = 2023,
        month = oct,
       volume = {678},
          eid = {A50},
        pages = {A50},
          doi = {10.1051/0004-6361/202346844},
archivePrefix = {arXiv},
       eprint = {2306.16214},
 primaryClass = {astro-ph.HE},
       adsurl = {https://ui.adsabs.harvard.edu/abs/2023A&A...678A..50E}
}

@ARTICLE{Reardon2023,
       author = {{Reardon}, Daniel J. and {Zic}, Andrew and {Shannon}, Ryan M. and {Hobbs}, George B. and {Bailes}, Matthew and {Di Marco}, Valentina and {Kapur}, Agastya and {Rogers}, Axl F. and {Thrane}, Eric and {Askew}, Jacob and {Bhat}, N.~D. Ramesh and {Cameron}, Andrew and {Cury{\l}o}, Ma{\l}gorzata and {Coles}, William A. and {Dai}, Shi and {Goncharov}, Boris and {Kerr}, Matthew and {Kulkarni}, Atharva and {Levin}, Yuri and {Lower}, Marcus E. and {Manchester}, Richard N. and {Mandow}, Rami and {Miles}, Matthew T. and {Nathan}, Rowina S. and {Os{\l}owski}, Stefan and {Russell}, Christopher J. and {Spiewak}, Ren{\'e}e and {Zhang}, Songbo and {Zhu}, Xing-Jiang},
        title = "{Search for an Isotropic Gravitational-wave Background with the Parkes Pulsar Timing Array}",
      journal = {\apjl},
         year = 2023,
        month = jul,
       volume = {951},
       number = {1},
          eid = {L6},
        pages = {L6},
          doi = {10.3847/2041-8213/acdd02},
archivePrefix = {arXiv},
       eprint = {2306.16215},
 primaryClass = {astro-ph.HE},
       adsurl = {https://ui.adsabs.harvard.edu/abs/2023ApJ...951L...6R}
}

@ARTICLE{Gigi2026,
       author = {{Gigi}, Loren and {Gaskell}, C. Martin},
        title = "{The direction of rotation of supermassive black holes is unrelated to the direction of rotation of the host galaxy}",
      journal = {arXiv e-prints},
         year = 2026,
        month = jul,
          eid = {arXiv:2607.06902},
        pages = {arXiv:2607.06902},
          doi = {10.48550/arXiv.2607.06902},
archivePrefix = {arXiv},
       eprint = {2607.06902},
 primaryClass = {astro-ph.GA},
       adsurl = {https://ui.adsabs.harvard.edu/abs/2026arXiv260706902G}
}

@ARTICLE{Fischer2013,
       author = {{Fischer}, T.~C. and {Crenshaw}, D.~M. and {Kraemer}, S.~B. and {Schmitt}, H.~R.},
        title = "{Determining Inclinations of Active Galactic Nuclei via their Narrow-line Region Kinematics. I. Observational Results}",
      journal = {\apjs},
         year = 2013,
        month = nov,
       volume = {209},
       number = {1},
          eid = {1},
        pages = {1},
          doi = {10.1088/0067-0049/209/1/1},
archivePrefix = {arXiv},
       eprint = {1308.4129},
 primaryClass = {astro-ph.CO},
       adsurl = {https://ui.adsabs.harvard.edu/abs/2013ApJS..209....1F}
}

@ARTICLE{Fischer2014,
       author = {{Fischer}, T.~C. and {Crenshaw}, D.~M. and {Kraemer}, S.~B. and {Schmitt}, H.~R. and {Turner}, T.~J.},
        title = "{Determining Inclinations of Active Galactic Nuclei via Their Narrow-line Region Kinematics. II. Correlation with Observed Properties}",
      journal = {\apj},
         year = 2014,
        month = apr,
       volume = {785},
       number = {1},
          eid = {25},
        pages = {25},
          doi = {10.1088/0004-637X/785/1/25},
archivePrefix = {arXiv},
       eprint = {1402.3509},
 primaryClass = {astro-ph.GA},
       adsurl = {https://ui.adsabs.harvard.edu/abs/2014ApJ...785...25F}
}

@ARTICLE{DeLaurentiis2025,
       author = {{DeLaurentiis}, Stanislav and {Haiman}, Zolt{\'a}n and {Westernacher-Schneider}, John Ryan and {Krauth}, Luke Major and {Davelaar}, Jordy and {Zrake}, Jonathan and {MacFadyen}, Andrew},
        title = "{Relativistic Binary Precession: Impact on Eccentric Massive Binary Black Hole Accretion and Hydrodynamics}",
      journal = {\apj},
         year = 2025,
        month = feb,
       volume = {980},
       number = {1},
          eid = {55},
        pages = {55},
          doi = {10.3847/1538-4357/ada612},
archivePrefix = {arXiv},
       eprint = {2405.07897},
 primaryClass = {astro-ph.HE},
       adsurl = {https://ui.adsabs.harvard.edu/abs/2025ApJ...980...55D}
}

@ARTICLE{Ragusa2016,
   author = {{Ragusa}, E. and {Lodato}, G. and {Price}, D.~J.},
    title = "{Suppression of the accretion rate in thin discs around binary black holes}",
  journal = {\mnras},
archivePrefix = "arXiv",
   eprint = {1605.01730},
 primaryClass = "astro-ph.HE",
     year = 2016,
    month = aug,
   volume = 460,
    pages = {1243-1253},
      doi = {10.1093/mnras/stw1081},
   adsurl = {http://adsabs.harvard.edu/abs/2016MNRAS.460.1243R}
}

@ARTICLE{Tanaka2010,
       author = {{Tanaka}, Takamitsu and {Haiman}, Zolt{\'a}n and {Menou}, Kristen},
        title = "{Witnessing the Birth of a Quasar}",
      journal = {\aj},
         year = 2010,
        month = aug,
       volume = {140},
       number = {2},
        pages = {642-651},
          doi = {10.1088/0004-6256/140/2/642},
archivePrefix = {arXiv},
       eprint = {1004.5411},
 primaryClass = {astro-ph.CO},
       adsurl = {https://ui.adsabs.harvard.edu/abs/2010AJ....140..642T}
}

@ARTICLE{DOrazio2013,
       author = {{D'Orazio}, Daniel J. and {Haiman}, Zolt{\'a}n and {MacFadyen}, Andrew},
        title = "{Accretion into the central cavity of a circumbinary disc}",
      journal = {\mnras},
         year = 2013,
        month = dec,
       volume = {436},
       number = {4},
        pages = {2997-3020},
          doi = {10.1093/mnras/stt1787},
archivePrefix = {arXiv},
       eprint = {1210.0536},
 primaryClass = {astro-ph.GA},
       adsurl = {https://ui.adsabs.harvard.edu/abs/2013MNRAS.436.2997D}
}

@ARTICLE{Pichardo2005,
       author = {{Pichardo}, Barbara and {Sparke}, Linda S. and {Aguilar}, Luis A.},
        title = "{Circumstellar and circumbinary discs in eccentric stellar binaries}",
      journal = {\mnras},
         year = 2005,
        month = may,
       volume = {359},
       number = {2},
        pages = {521-530},
          doi = {10.1111/j.1365-2966.2005.08905.x},
archivePrefix = {arXiv},
       eprint = {astro-ph/0501244},
 primaryClass = {astro-ph},
       adsurl = {https://ui.adsabs.harvard.edu/abs/2005MNRAS.359..521P}
}

@ARTICLE{Ivanov2015,
       author = {{Ivanov}, P.~B. and {Papaloizou}, J.~C.~B. and {Paardekooper}, S.-J. and {Polnarev}, A.~G.},
        title = "{The evolution of a binary in a retrograde circular orbit embedded in an accretion disk}",
      journal = {\aap},
         year = 2015,
        month = apr,
       volume = {576},
          eid = {A29},
        pages = {A29},
          doi = {10.1051/0004-6361/201424359},
archivePrefix = {arXiv},
       eprint = {1410.3250},
 primaryClass = {astro-ph.HE},
       adsurl = {https://ui.adsabs.harvard.edu/abs/2015A&A...576A..29I}
}

@BOOK{RybickiLightman1986,
       author = {{Rybicki}, George B. and {Lightman}, Alan P.},
        title = "{Radiative Processes in Astrophysics}",
         year = 1986,
       adsurl = {https://ui.adsabs.harvard.edu/abs/1986rpa..book.....R}
}

@ARTICLE{Dempsey2020,
       author = {{Dempsey}, Adam M. and {Mu{\~n}oz}, Diego and {Lithwick}, Yoram},
        title = "{Inner Boundary Condition in Quasi-Lagrangian Simulations of Accretion Disks}",
      journal = {\apjl},
         year = 2020,
        month = apr,
       volume = {892},
       number = {2},
          eid = {L29},
        pages = {L29},
          doi = {10.3847/2041-8213/ab800e},
archivePrefix = {arXiv},
       eprint = {2002.05164},
 primaryClass = {astro-ph.EP},
       adsurl = {https://ui.adsabs.harvard.edu/abs/2020ApJ...892L..29D}
}

@ARTICLE{Ultrasat,
       author = {{Shvartzvald}, Y. and {Waxman}, E. and {Gal-Yam}, A. and {Ofek}, E.~O. and {Ben-Ami}, S. and {Berge}, D. and {Kowalski}, M. and {B{\"u}hler}, R. and {Worm}, S. and {Rhoads}, J.~E. and {Arcavi}, I. and {Maoz}, D. and {Polishook}, D. and {Stone}, N. and {Trakhtenbrot}, B. and {Ackermann}, M. and {Aharonson}, O. and {Birnholtz}, O. and {Chelouche}, D. and {Guetta}, D. and {Hallakoun}, N. and {Horesh}, A. and {Kushnir}, D. and {Mazeh}, T. and {Nordin}, J. and {Ofir}, A. and {Ohm}, S. and {Parsons}, D. and {Pe'er}, A. and {Perets}, H.~B. and {Perdelwitz}, V. and {Poznanski}, D. and {Sadeh}, I. and {Sagiv}, I. and {Shahaf}, S. and {Soumagnac}, M. and {Tal-Or}, L. and {Santen}, J. Van and {Zackay}, B. and {Guttman}, O. and {Rekhi}, P. and {Townsend}, A. and {Weinstein}, A. and {Wold}, I.},
        title = "{ULTRASAT: A Wide-field Time-domain UV Space Telescope}",
      journal = {\apj},
         year = 2024,
        month = mar,
       volume = {964},
       number = {1},
          eid = {74},
        pages = {74},
          doi = {10.3847/1538-4357/ad2704},
archivePrefix = {arXiv},
       eprint = {2304.14482},
 primaryClass = {astro-ph.IM},
       adsurl = {https://ui.adsabs.harvard.edu/abs/2024ApJ...964...74S}
}

@ARTICLE{LSST,
       author = {{Ivezi{\'c}}, {\v{Z}}eljko and {Kahn}, Steven M. and {Tyson}, J. Anthony and {Abel}, Bob and {Acosta}, Emily and {Allsman}, Robyn and {Alonso}, David and {AlSayyad}, Yusra and {Anderson}, Scott F. and {Andrew}, John and {Angel}, James Roger P. and {Angeli}, George Z. and {Ansari}, Reza and {Antilogus}, Pierre and {Araujo}, Constanza and {Armstrong}, Robert and {Arndt}, Kirk T. and {Astier}, Pierre and {Aubourg}, {\'E}ric and {Auza}, Nicole and {Axelrod}, Tim S. and {Bard}, Deborah J. and {Barr}, Jeff D. and {Barrau}, Aurelian and {Bartlett}, James G. and {Bauer}, Amanda E. and {Bauman}, Brian J. and {Baumont}, Sylvain and {Bechtol}, Ellen and {Bechtol}, Keith and {Becker}, Andrew C. and {Becla}, Jacek and {Beldica}, Cristina and {Bellavia}, Steve and {Bianco}, Federica B. and {Biswas}, Rahul and {Blanc}, Guillaume and {Blazek}, Jonathan and {Blandford}, Roger D. and {Bloom}, Josh S. and {Bogart}, Joanne and {Bond}, Tim W. and {Booth}, Michael T. and {Borgland}, Anders W. and {Borne}, Kirk and {Bosch}, James F. and {Boutigny}, Dominique and {Brackett}, Craig A. and {Bradshaw}, Andrew and {Brandt}, William Nielsen and {Brown}, Michael E. and {Bullock}, James S. and {Burchat}, Patricia and {Burke}, David L. and {Cagnoli}, Gianpietro and {Calabrese}, Daniel and {Callahan}, Shawn and {Callen}, Alice L. and {Carlin}, Jeffrey L. and {Carlson}, Erin L. and {Chandrasekharan}, Srinivasan and {Charles-Emerson}, Glenaver and {Chesley}, Steve and {Cheu}, Elliott C. and {Chiang}, Hsin-Fang and {Chiang}, James and {Chirino}, Carol and {Chow}, Derek and {Ciardi}, David R. and {Claver}, Charles F. and {Cohen-Tanugi}, Johann and {Cockrum}, Joseph J. and {Coles}, Rebecca and {Connolly}, Andrew J. and {Cook}, Kem H. and {Cooray}, Asantha and {Covey}, Kevin R. and {Cribbs}, Chris and {Cui}, Wei and {Cutri}, Roc and {Daly}, Philip N. and {Daniel}, Scott F. and {Daruich}, Felipe and {Daubard}, Guillaume and {Daues}, Greg and {Dawson}, William and {Delgado}, Francisco and {Dellapenna}, Alfred and {de Peyster}, Robert and {de Val-Borro}, Miguel and {Digel}, Seth W. and {Doherty}, Peter and {Dubois}, Richard and {Dubois-Felsmann}, Gregory P. and {Durech}, Josef and {Economou}, Frossie and {Eifler}, Tim and {Eracleous}, Michael and {Emmons}, Benjamin L. and {Fausti Neto}, Angelo and {Ferguson}, Henry and {Figueroa}, Enrique and {Fisher-Levine}, Merlin and {Focke}, Warren and {Foss}, Michael D. and {Frank}, James and {Freemon}, Michael D. and {Gangler}, Emmanuel and {Gawiser}, Eric and {Geary}, John C. and {Gee}, Perry and {Geha}, Marla and {Gessner}, Charles J.~B. and {Gibson}, Robert R. and {Gilmore}, D. Kirk and {Glanzman}, Thomas and {Glick}, William and {Goldina}, Tatiana and {Goldstein}, Daniel A. and {Goodenow}, Iain and {Graham}, Melissa L. and {Gressler}, William J. and {Gris}, Philippe and {Guy}, Leanne P. and {Guyonnet}, Augustin and {Haller}, Gunther and {Harris}, Ron and {Hascall}, Patrick A. and {Haupt}, Justine and {Hernandez}, Fabio and {Herrmann}, Sven and {Hileman}, Edward and {Hoblitt}, Joshua and {Hodgson}, John A. and {Hogan}, Craig and {Howard}, James D. and {Huang}, Dajun and {Huffer}, Michael E. and {Ingraham}, Patrick and {Innes}, Walter R. and {Jacoby}, Suzanne H. and {Jain}, Bhuvnesh and {Jammes}, Fabrice and {Jee}, M. James and {Jenness}, Tim and {Jernigan}, Garrett and {Jevremovi{\'c}}, Darko and {Johns}, Kenneth and {Johnson}, Anthony S. and {Johnson}, Margaret W.~G. and {Jones}, R. Lynne and {Juramy-Gilles}, Claire and {Juri{\'c}}, Mario and {Kalirai}, Jason S. and {Kallivayalil}, Nitya J. and {Kalmbach}, Bryce and {Kantor}, Jeffrey P. and {Karst}, Pierre and {Kasliwal}, Mansi M. and {Kelly}, Heather and {Kessler}, Richard and {Kinnison}, Veronica and {Kirkby}, David and {Knox}, Lloyd and {Kotov}, Ivan V. and {Krabbendam}, Victor L. and {Krughoff}, K. Simon and {Kub{\'a}nek}, Petr and {Kuczewski}, John and {Kulkarni}, Shri and {Ku}, John and {Kurita}, Nadine R. and {Lage}, Craig S. and {Lambert}, Ron and {Lange}, Travis and {Langton}, J. Brian and {Le Guillou}, Laurent and {Levine}, Deborah and {Liang}, Ming and {Lim}, Kian-Tat and {Lintott}, Chris J. and {Long}, Kevin E. and {Lopez}, Margaux and {Lotz}, Paul J. and {Lupton}, Robert H. and {Lust}, Nate B. and {MacArthur}, Lauren A. and {Mahabal}, Ashish and {Mandelbaum}, Rachel and {Markiewicz}, Thomas W. and {Marsh}, Darren S. and {Marshall}, Philip J. and {Marshall}, Stuart and {May}, Morgan and {McKercher}, Robert and {McQueen}, Michelle and {Meyers}, Joshua and {Migliore}, Myriam and {Miller}, Michelle and {Mills}, David J.},
        title = "{LSST: From Science Drivers to Reference Design and Anticipated Data Products}",
      journal = {\apj},
         year = 2019,
        month = mar,
       volume = {873},
       number = {2},
          eid = {111},
        pages = {111},
          doi = {10.3847/1538-4357/ab042c},
archivePrefix = {arXiv},
       eprint = {0805.2366},
 primaryClass = {astro-ph},
       adsurl = {https://ui.adsabs.harvard.edu/abs/2019ApJ...873..111I}
}

@ARTICLE{ROMAN,
       author = {{Observations Time Allocation Committee}, Roman and {Community Survey Definition Committees}, Core},
        title = "{Roman Observations Time Allocation Committee: Final Report and Recommendations}",
      journal = {arXiv e-prints},
         year = 2025,
        month = may,
          eid = {arXiv:2505.10574},
        pages = {arXiv:2505.10574},
          doi = {10.48550/arXiv.2505.10574},
archivePrefix = {arXiv},
       eprint = {2505.10574},
 primaryClass = {astro-ph.IM},
       adsurl = {https://ui.adsabs.harvard.edu/abs/2025arXiv250510574O}
}

@ARTICLE{Mahesh2024,
       author = {{Mahesh}, Siddharth and {McWilliams}, Sean T. and {Pirog}, Michal},
        title = "{Analytical and Numerical Analysis of Circumbinary Disk Dynamics. I. Coplanar Systems}",
      journal = {\apj},
         year = 2024,
        month = sep,
       volume = {973},
       number = {1},
          eid = {18},
        pages = {18},
          doi = {10.3847/1538-4357/ad6149},
archivePrefix = {arXiv},
       eprint = {2305.01533},
 primaryClass = {astro-ph.SR},
       adsurl = {https://ui.adsabs.harvard.edu/abs/2024ApJ...973...18M}
}

\appendix
\section{Viscosity Prescription and Assumed Thermodynamics}
\label{Appendix}

\begin{figure}
    \centering
    \includegraphics[width=\linewidth]{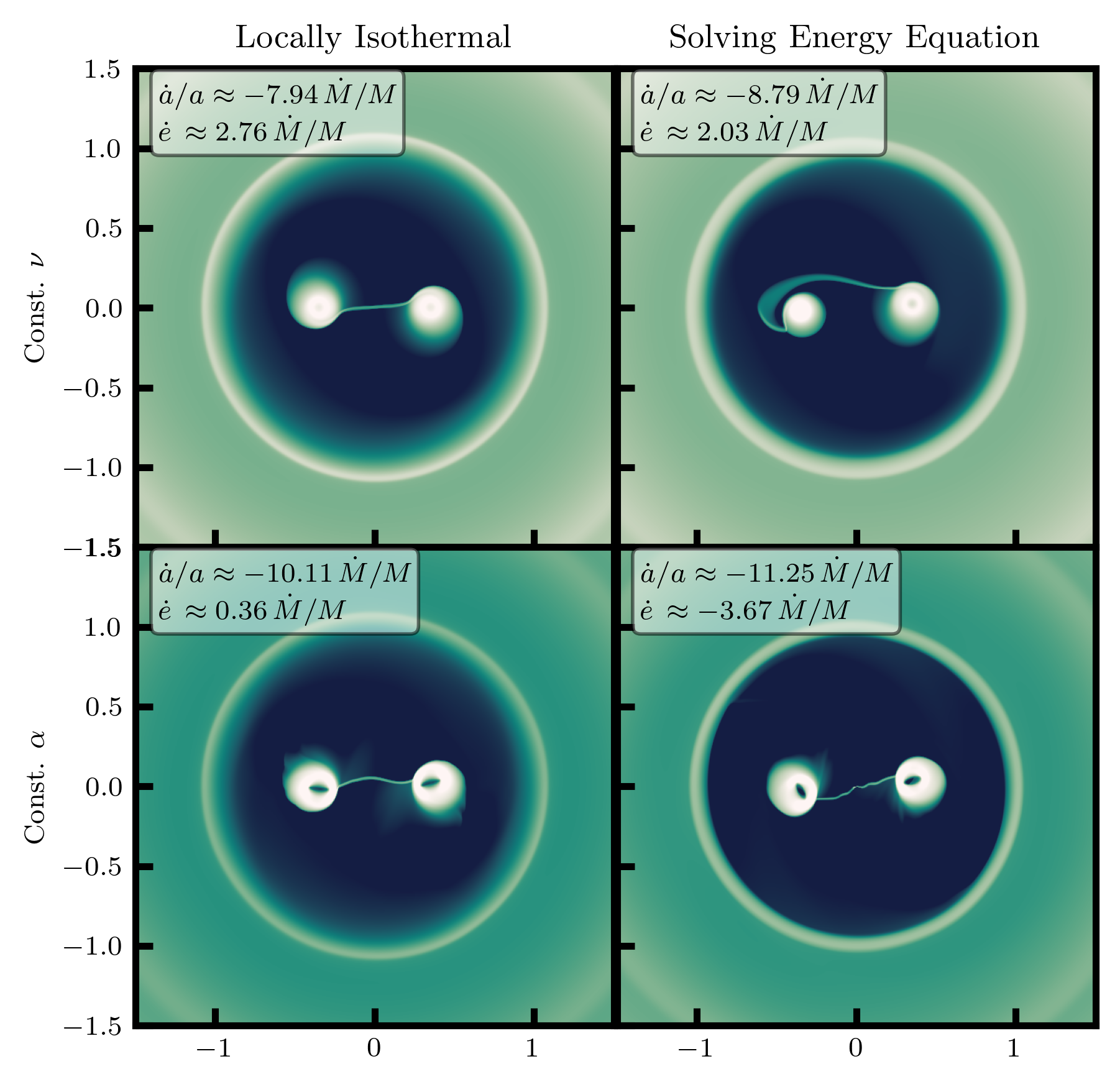}
    \caption{A comparison of different Mach 10 retrograde disks around eccentric binaries of $e_\mathrm{b}=0.3$. \emph{Top row}: Constant kinematic viscosity $\nu =\mathrm{const.}$, \emph{Bottom row}: Constant $\alpha$ disks with $\nu\propto r^{3/5}$. \emph{Left column}: locally isothermal simulations with $c_\mathrm{s}^2\propto r^2\Omega^2$, \emph{Right column}: self-consistently solving the energy equation to balance viscous and shock heating with blackbody cooling. The cat-eye features at the minidisks in the constant-$\alpha$ panels are associated with the torque-free treatment of the sinks.}
    \label{fig:AppendixFig}
\end{figure}

In this appendix, we briefly compare the sensitivity of our results to different choices of viscosity (constant $\alpha$ or consant $\nu$) and disk thermodynamics (locally isothermal or by solving an energy equation). We compare our findings with previous studies to highlight qualitative differences between these choices.\\

In Fig.~\ref{fig:AppendixFig} we illustrate simulations of Mach 10 disks around binaries with eccentricity $e_\mathrm{b}=0.3$, using four different combinations of viscosity prescriptions and disk thermodynamics. We restrict our analysis to simulations that are in the $\downarrow\downarrow$ state -- although we have verified that the $\uparrow\uparrow$ state also exist for locally isothermal, constant $\nu$ disks when initialised with $r_\mathrm{cav} = 0.3a_\mathrm{b}$. Furthermore, in the limit of a small (spatially resolved) sink with a fast removal rate $\gamma_\mathrm{sink}\gg 100/\mathcal{M}_a^2$, we have verified that minidisks exist when an $\alpha$ prescription is solved together with a consistent energy equation, in contrast to the locally isothermal, constant-$\nu$ simulations of \cite{ONeill2025}.\\

We find excellent agreement with \cite{Tiede2024} in the morphological features and orbital evolution shown in the top-left panel of Fig.~\ref{fig:AppendixFig}. We argue that the choice of viscosity prescription is more impactful than the thermodynamics, with both panels in the top row showing similar $\dot{a}/a$ and $\dot{e}$ values. In contrast, the bottom two panels have larger $\dot{a}/a$ and weaker or negative $\dot{e}$ evolution.

\end{document}